\documentclass[graybox,natbib,nosecnum]{svmult}
\bibpunct{(}{)}{;}{a}{}{,} 

\pdfoutput=1   

\usepackage{mathptmx}       
\usepackage{helvet}         
\usepackage{courier}        
\usepackage{type1cm}        

\usepackage{makeidx}         
\usepackage[T1]{fontenc}
\usepackage[utf8]{inputenc}
\usepackage{xcolor}
\usepackage{graphicx}        
\usepackage{multicol}        
\usepackage[bottom]{footmisc}
\usepackage[normalem]{ulem}	
\usepackage{hyperref}  
\usepackage[export]{adjustbox} 
\usepackage[margin=10pt]{subfig}
\usepackage{soul}   
\usepackage{textcomp}
\usepackage[T1]{fontenc}

\newcommand{\ignore}[1]{}

\makeindex

\begin{document}

\title*{Connecting Planetary Composition with Formation: a New Paradigm Emerges}
\author{Ralph E. Pudritz, Alex J. Cridland, Julie Inglis \& Mathew Alessi}
\institute{Ralph E. Pudritz \at Department of Physics and Astronomy, McMaster University, Hamilton, Ontario, Canada, L8S 4K1; Origins Institute, McMaster University, Hamilton, Ontario, Canada, L8S 4E8,
\email{pudritz@mcmaster.ca}
\and Alex J. Cridland \at  Universit\"ats-Sternwarte M\"unchen, Ludwig-Maximilians-Universit\"at, Scheinerstr. 1, 81679 M\"unchen, Deutschland  \email{A.Cridland@lmu.de  }
\and Julie Inglis \at Division of Geological and Planetary Sciences, California Institute of Technology, Pasadena, CA, 91125 \email{jinglis@caltech.edu} 
\and Matthew Alessi \at Hamilton Health Sciences, Greater Hamilton Metropolitan Area, Hamilton, Ontario, Canada,    \email{matthew.alessi3@gmail.com}
}

\maketitle
\abstract{ Extensive ground and space based surveys have now characterized the properties of thousands of exoplanets; their radii, masses, orbits around their host stars, and the beginnings of accurate measurements of the chemical compositions of their atmospheres and cores.  How are these properties linked to their formation in physically and chemically evolving protoplanetary disks wherein they accrete pebbles, planetesimals and gas as they undergo migration?  To address this challenge, our review assembles a large and varied body of exoplanet observations as well as recent Atacama Large Millimeter Array (ALMA) and James Webb Space Telescope (JWST) observations of disk structure, chemistry, kinematics, and winds.  The latest advances in theory and MHD simulations that bear on these issues are also reviewed and compared with the observations. Taken together, this review argues that a new dynamic paradigm for planet formation is emerging wherein MHD disk winds and not disk turbulence play a central role in disk evolution and planet formation including: angular momentum transport, gap and ring formation.  disk astrochemistry, and planet formation and migration.  These processes leave their mark on the resulting atmospheric composition, radii, and orbital characteristics of exoplanet populations, offering the possibility of future observational tests.    
     }


\graphicspath{{./IMAGES/}{./SOURCE_FILES/IMAGES/}{./IMAGES_edition_2/}}

\section{Introduction}

Planet formation is widely understood to take place in disks of dust and gas that are ubiquitous around forming stars of all masses \citep{Miotello2023, Drazkowska2023}. Most of the exoplanets we observe today are far removed from their birth in these environments billions of years ago.  Nevertheless, their composition, structure, and to some degree, their orbital properties encode key elements of their formation histories. This review attempts to summarize some of the major advances in our understanding of the composition and other related physical properties of observed exoplanet populations as it connects with new insights as to how planets formed.  We begin with a thumbnail sketch of the basic topics covered in the review.

Early classic theories of planet formation argued that disks form as a consequence of angular momentum conservation during the collapse of isolated, rotating, dense molecular cloud cores \citep{Terebey+1984}.  However, observational advances over the last 15 years by the Herschel observatory \citep{Andre2010}, the Atacama Large Millimeter Array \citep[ALMA,][]{ALMA2015}, and the James Webb Space Telescope \citep[JWST,][]{Watkins2023} reveal that star formation is a highly dynamical process that takes place within the turbulent, magnetized, filaments that make up molecular clouds \citep{Andre2014}.   Recent observations show that ongoing inflow along these filaments continuously feeds disks with fresh gas, dust, and angular momentum well into the so-called Class II phase of protostellar evolution when planetary systems are observed to be forming \citep{Pineda2023}. This physical picture was predicted by sophisticated numerical magnetohydrodynamical (MHD) simulations of turbulent, magnetized, self-gravitating clouds, made possible by the enormous advances in computing power and code design. A new generation of simulations showed that disks are formed by converging filamentary flows and streamers \citep{Seifried2014, Bate2018, Kuffmeier2020, Kuffmeier2024}. Disk formation is also accompanied by strong disk torques exerted by the magnetic fields threading the infalling gas. There is now excellent evidence from VLBA, ALMA, and JWST observations that the jets and outflows that accompany the formation of stars of all masses are magnetohydrodynamic (MHD) disk winds that brake the disk and carry off its angular momentum, driving accretion flow onto the forming star \citep{Moscadelli2022,Pascucci2023,Pascucci2025,Bacciotti2025}.

During the subsequent main phase of gaseous protoplanetary disks (PPDs) evolution,  protostars and their planetary systems are forged out of the same accreted gas and dust.  However, the evolution of gas and solid materials bifurcates.  Over the 3-10 Myrs lifetime of PPDs, initially micron size dust grains settle to the disk midplane, undergo radial drift, accumulate into rings at gas pressure maxima, collide, agglomerate, and ultimately become a  mixture of pebbles and planetesimals out of which rocky planets or the cores of gas giants are built (see Chapter by Ormel;   The refractory and volatile species that these solids accumulate in their journeys are some of the construction materials of the growing planets that determine their chemical compositions and structure. In the  core accretion picture of planet formation \citep{Pollack1996}, giant planets accrete their gas very rapidly (within a few $10^5  $ yrs) by gravitational instability  once their rocky cores grow beyond 10-20 Earth masses, provided that their disks still contain sufficient gas  (see Youdin and Zhu - this volume). Theory, simulations, and recent observations show that outflows efficiently remove disk angular momentum to drive accretion flow onto the forming star and the forming planets.  The gravitational interaction between disks and planets, as well as the gravitational planet-planet interaction lead to planetary migration.  In this way, forming planets may accrete building materials from different physical regions in their disks as defined, for example, by the various ice lines of volatile species (CO, CO$_2$, H$_2$O). The gaseous, disk dominated era of planet formation ends as a consequence of FUV and X-ray driven photoevaporative flows. 

ALMA has revolutionized our understanding of disk structure in having resolved exquisite ring and gapped structure as well as asymmetries in dust emission \citep{Andrews2018, vanderMarel2021, Bae2023}. Although these structures are generally thought to arise from planet formation and/or planet-disk interaction, MHD processes have been shown to produce such structure in the absence of planets. 
Most recently, the JWST, MIRI Mid-INfrared Disk Survey (MINDS) has probed for the first time the chemistry of the inner, planet-forming region (10 AU) of PPDs \citep{Kamp2023,Henning_MINDS2024}. This survey has so far detected abundant CO$_2$ \citep{Grant2023}, water and a varying C/O ratio across a PPD \citep{Gasman2023}, mid IR emission from atomic and molecular hydrogen \citep{Franceschi2024}, and even hydrocarbon molecules consistent with high C/O gas chemistry \citep{Kanwar2024}.
Meanwhile, major advances in theory and simulations include dust evolution and the growth of planets by some combination of pebble and planetesimal accretion and the demonstration that MHD disk winds and not turbulence may be the main transporter of disk angular momentum \citep{BaiStone2013,Gressel2015,Aoyama_Bai2023}   

In the
post-gaseous disk phase, rapid dynamical evolution of the planetary system may occur.  The recent discovery of numerous sub-Neptunian mass planets arranged in compact multi-planet systems with regularly spaced orbits inside 1 AU ("peas in a pod") raises the question of how these arise from a chaotic formation process \citep{Weiss2023}. The final assembly of a variety of left-over, small bodies into rocky planets is driven by gravitational perturbations of the remaining planetesimal and cometary populations by the giant planets.  The large eccentricities that planetesimals acquire mix material from large regions of the former PPD, that once again can alter planetary composition.

There are several likely enduring imprints of these processes on planet composition. As an example, the ratios of elemental abundances in their atmospheres (C/O, C/N, etc) may leave a fingerprint of some kind as to where in the disk a planet accreted most of its mass \citep{Oberg2011a, ObergBergin2021}.  However, planet migration and pebble drift complicate this picture depending on how angular momentum is moved or lost from disks, the details of planet-disk interaction leading to planet migration, and how and where pebbles may be trapped.  The internal structure of the planets also depends on the composition of materials accreted during their formation.  The latter are divided roughly into iron species in the core, the silicates that comprise the mantel, as well as ices, particularly water ice that can account for a half of the planet's mass, as is suggested for mini-Neptunes.  Once again, the range of disk radii over which these materials are accreted and how they are sequestered within the forming planet will affect the planetary radii \citep{Alessi_Pudritz2022, Haldemann2024, Burn2024, Luo2024}. At later stages planetary atmospheres evolve by photoevaporation or secondary outgassing to degrees that depend on their mass, metallicity, relative elemental abundances,  stellar type of the host stars, and the radiative flux they intercept from same.

The TESS observatory has contributed a substantial number of planets to the total confirmed list  which currently stands at over 5830 exoplanets, with thousands more yet to be confirmed  \footnote{See NASA's Exoplanet Archive: https://exoplanetarchive.ipac.caltech.edu/ }.  Most exoplanets have been discovered by transit observations coupled  with radial velocity (RV) observations  \citep{Mayor1995,Queloz2000,Pepe2004,UdrySantos2007,Howard2010,Howard2012,Batalha2014,Bowler2016}.  The statistical properties of at least four planetary populations (hot and warm Jupiters, mini-Neptunes, and super-Earths) have been much better characterized.  

Arguably, one of the most significant recent discoveries in this regard is the "radius valley": the dearth of planets at 1.3-1.5 Earth radii that separates the radii of rocky, super-Earth planets from the larger mini-Neptunes \citep{Fulton2017}.  Explaining its origin is, as we will see, an important test for various planet formation models. Among the most important theoretical tools that we have to make these predictions is population synthesis wherein the best available physical and astrochemistry models  (often based on detailed simulations) of planet formation and migration in gaseous disks are input into large Monte-Carlo simulations in which populations are computed by drawing on the initial distributions of various model parameters such as initial disk mass and turbulence amplitude.

\begin{figure*}
\centering
\includegraphics[width=\textwidth]{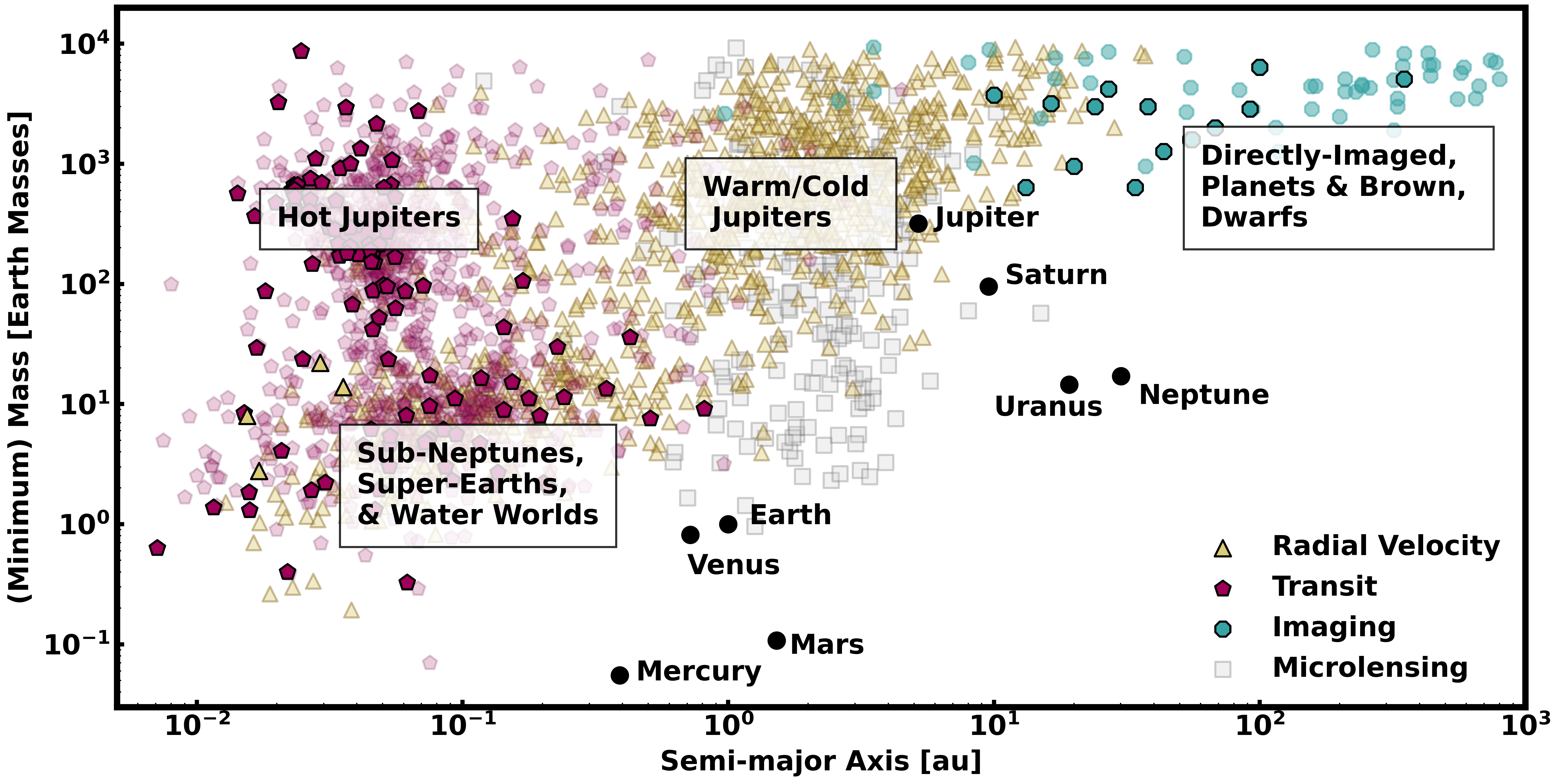} 
\caption{M-a distribution, updated from Figure from \citet{Kempton_Knutson2024}. The solid-colored points have atmospheric spectra as of April 28, 2025 and the transparent points do not. The additional data points come from TESS and DI surveys with JWST and other observatories. Reproduced with permission \textcopyright Mineralogical Society of America.}
\label{fig:keaton_knutson}
\end{figure*}

This review emphasizes the advances that have occurred since the publication of the first edition of Springer's {\it Handbook of Exoplanets} in 2018.  We first discuss the basic characteristics of exoplanets and their host stars, and then protoplanetary disk populations.  We follow this with an extended section on physical processes in PPDs, featuring recent developments in pebble accretion, planet and dust traps, and MHD disk winds in the role of planet formation and migration.  Population synthesis approaches are reviewed (see also the chapter by Burn and Mordasini) culminating in a comparison with observed planetary structure and atmospheres.   The reader may also wish to consult a number of earlier and still highly relevant review articles on the these topics including \cite{Testi2014} for disks and dust evolution; \cite{Turner2014} for disks and angular momentum transport; \cite{KleyNelson2012} for planet migration, and \cite{Raymond2014}, \cite{Helled2014}, and \cite{Benz2014} for population synthesis.  

\section{Exoplanet Populations and their Host Stars}

The basic properties of exoplanet populations can be conveniently summarized in several important diagrams. 

The first and perhaps most fundamental is the mass- semimajor axis (M-a) diagram, shown in Figure \ref{fig:keaton_knutson} and updated from \citet{Kempton_Knutson2024}. It summarizes the demographics of confirmed exoplanets based on their masses and orbital semi-major axes a (or equivalently orbital periods). In this M-a space, exoplanets clearly divide into several planetary populations \citep[see also][]{ChiangLaughlin2013,HP13}.   Based on the measured occurrence rates of exoplanets \citep[for example,][]{Petigura2018}: Hot Jupiters, which orbit within 0.1 AU of their host stars, have the lowest occurrence rates of all, at about 1\% of solar type stars.  The occurrence rate of warm to cold Jupiters increases as one moves away from the host star, peaking at about 14\% at orbital radii of 2-8 AU.  The dominant exoplanet population is the sub-Neptunes to super-Earths, with occurrence rates of 50\% and is arguably the most surprising discovery of the planetary surveys as such planets do not have counterparts in our solar system.    The theory of planet migration that has arisen to account for these planetary orbits rests on how planet-disk gravitational interaction transfers planetary angular momentum to the gaseous disk (see Chapter by Nelson in this volume).  The essential point is that disks exert torques on planets that depend on how disk angular momentum is transported. 

A prominent dynamical feature in the M-a diagram is the significant fraction  of extremely compact systems that are well aligned and having 
short periods \citep{RangMargot2012,HansenMurray2013,ChiangLaughlin2013}.  Although the spacings between orbital pairs seem
to be random, nevertheless, there is an abundance of them that are just wide of major mean motion resonances (MMRs) and a lack of such pairs
just inside these \citep{Lissauer2011,Fabrycky2014}.  One of the explanations for this behaviour is the effect of planet-planetesimal disk interactions
on trapped, resonant pairs of planets (e.g. 2:1) \citep{ChatterjeeFord2015}.  

\begin{figure*}
\centering
\subfloat[M-R diagram from \citet{Otegi2020}. Reproduced with permission \textcopyright ESO.]{
\includegraphics[width=0.5\textwidth]{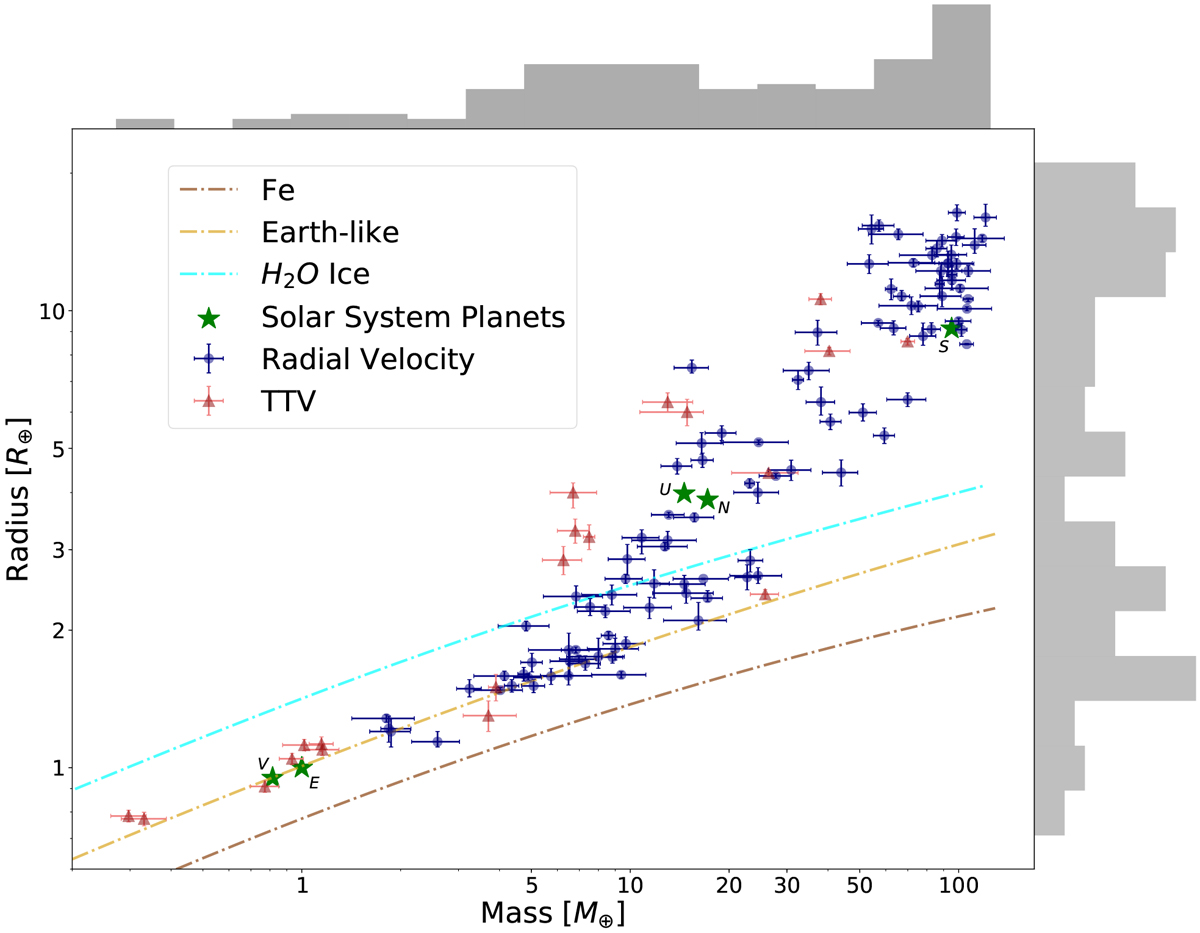}
\label{fig}
}
\subfloat[Radius valley Figure from \citet{Fulton2017}, . Reproduced with permission \textcopyright AAS.]{
\includegraphics[width=0.5\textwidth]{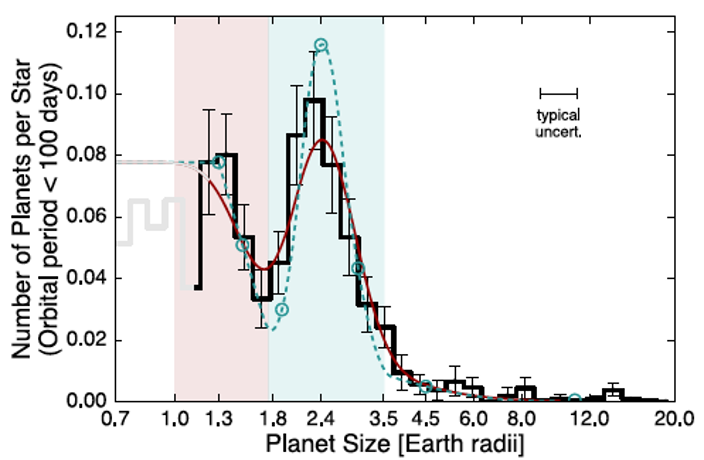}
\label{fig05b}
}
\caption{The M-R diagram for well determined planetary masses and radii (left panelt); and the radius valley separating super-Earths from Mini-Neptunes (right panel).}
\label{fig:otegi}
\end{figure*}

An estimate of the overall composition of planets can be determined if their mass and radius are known.  This requires a combination of radial velocity (RV) or equivalent measurements  to determine a planet's mass, as well as transit observations to deduce its transit radii.  Together, these data allow for the calculation of the mean bulk density of the planet, which, in turn, constrains the planetary and atmospheric structure \citep{Seager2007,Howard2013,Rogers2014,Dorn2017,ChenKipping2017}. 

Figure \ref{fig:otegi} (a) presents the planetary mass - radius (M-R) diagram from a catalogue of exoplanets less than 120 solar masses that have been filtered for the best measurements of planetary radii and masses \citep{Otegi2020}.  Two distinct populations are clearly discernable in this figure; one group follows an Earth-like rocky planet composition while the second group (between $10-25 M_{\odot}$) deviates upward in mass beyond the water-line (blue) curve and is populated with planets that are rich in volatiles.  Rocky Earth-like planets are composed of magnesium silicate rock that is incompressible up to $3R_{E}$.

The sequence of rocky planets in the Figure ends at about $25 M_{\odot}$. \citealt{Otegi2020} note that in the pebble accretion picture of the core accretion theory of planet formation, rapid growth of planetary cores is predicted to occur until a pebble isolation mass of this value is reached, beyond which further pebble flow is cut-off \citep{Lambrechts2014,Johansen2017,Schneider2021}.  The important question becomes how and where in the PPD did these planets acquire accrete these various materials and why is there a division between the rocky planets and the volatiles. 

The breakthrough discovery of the "radius valley" by \cite{Fulton2017} and reproduced in Figure \ref{fig:otegi}(b.), shows that there is a paucity of exoplanets between 1.5-2 Earth radii ($R_E$) at a mass of $1.78 M_E $, for planets around solar type stars, whose orbital periods are less than 100 days.  The smaller mass peak consists of rocky super-Earth planets with radii $1.0-1.7 R_E$, whereas those in the higher mass peak ($1.7-3.5 R_E$) are sub-Neptunes that may be dominated by H/He, or water rich atmospheres. Theoretical ideas for the origin of the valley range from X-ray and EUV driven photoevaporation \citep{Owen2017} and core-heating powered mass loss \citep{Ginzburg2018, Rogers2021}, to the convergence of two distinct planetary populations, the more massive peak representing ice-rich worlds that have migrated in to smaller disk radii \citep{Burn2024}.  Observations have recently identified a radius valley in these planetary populations arounnd M stars, but its character does not seem to be reproduced by photoevaporative models suggesting a different mechanism for its origin in low mass stars \citep{Cloutier2020}. We return to these issues in the section on Population Synthesis. 

The orbital evolution of planetary systems is partially encoded in their eccentricity distribution.  In the M-e digram,  large eccentricities accrue to a significant number of massive exoplanets. The median value of this eccentricity is high; $ \simeq 0.25 $.  The eccentricity of single massive planets can be attributed to planet-planet  scattering interactions after the gas disk has been dispersed  \citep{Chatterjee2008,JuricTremaine2008}.   

Another important result is the observed misalignment between the orbital plane of a traversing planet and the equatorial plane of the rotating star measured via the Rossiter-McLaughlin effect (see chapter by Tibaud). Roughly 1/3 of hot Jupiters show such misalignments. This raises an important question: Did these planets arise through dynamical interactions after migration in the disk had placed them in close-in orbits? Or did they arrive at these innermost orbits by some dynamical process such as the Kozai mechanism coupled with tidal friction?

In the latter case, a distant companion star can cause eccentric motions of a planet whose orbit can shrink and circularize drastically with time due to 
tidal interaction with the star, leading to close-in Jupiters with high eccentricity \citep{FabryckyTremaine2007}.  It may be that the elemental abundances of such planets will ultimately discriminate between planets brought in via disk processes, sampling materials from the inner disk regions, as compared to scattered bodies originally formed in outer disk regions whose compositions reflect the dominance of ices.   

There is a profound link between the metallicity of host stars (defined as the abundance of individual elements heavier than helium relative to hydrogen) and the existence of gas giant or low mass planets.   The planet-metallicity relation \citep{FischerValenti2005,WangFischer2015} shows that massive planets are more likely to be detected around stars only if they have sufficiently high metallicity (solar and above).  These authors found that for a limited range of stellar masses (0.7 - 1.2 M$_{\odot}$) that the probability of a star to host a giant planet scaled as the square of the number
of iron atoms;  $ P_{planet} \propto N_{Fe}^2$.  Later studies, carried out for a wider range of stellar 
masses, found that more massive stars also tend to host Jovian planets, with the scaling $P_{planet} \propto N_{Fe}^{1.2 \pm 0.2}  M^{1.0 \pm 0.5} $ 
\citep{Johnson2010}. The most recent research affirms a strong planet-metallicity relation for Jovian planets while stars of all masses
and metallicities host low mass planets.  These findings suggest that low mass planets can form in all disks but that only a fraction of these in 
high metallicity, or in sufficiently massive disks can grow into massive planets within the disk lifetime  \citep{IdaLin2005,HP14}.

As already noted, the fact that stars and their planetary systems form out of a common disk implies that their chemical abundances are connected. However, many chemical and physical processes can contribute, leading to variations in the relative abundance of elemental species between the host star and its planetary system. The abundance of refractory elements (iron, magnesium, silicon, etc) as an example, prescribes the amount of building material available for the cores and mantles of forming planets.   In a recent review of the star-planet composition connection \citet{Teske2024} concludes that the strongest such connection is still the planet-metallicity correlation noted previously.  Refractory materials make up by far the bulk of the interiors of low mass rocky planets.  Since there is a good match between stellar and planetary abundances of refractory materials, the Fe/Mg ratio measured from stellar spectra provides information about the relative size of a planet's core, while the Mg/Si ratio constrains the mineralogy of the mantle (consisting of minerals such as olivine, pyroxene, etc.).  

\begin{figure*}
\centering
\includegraphics[width=0.75\textwidth]{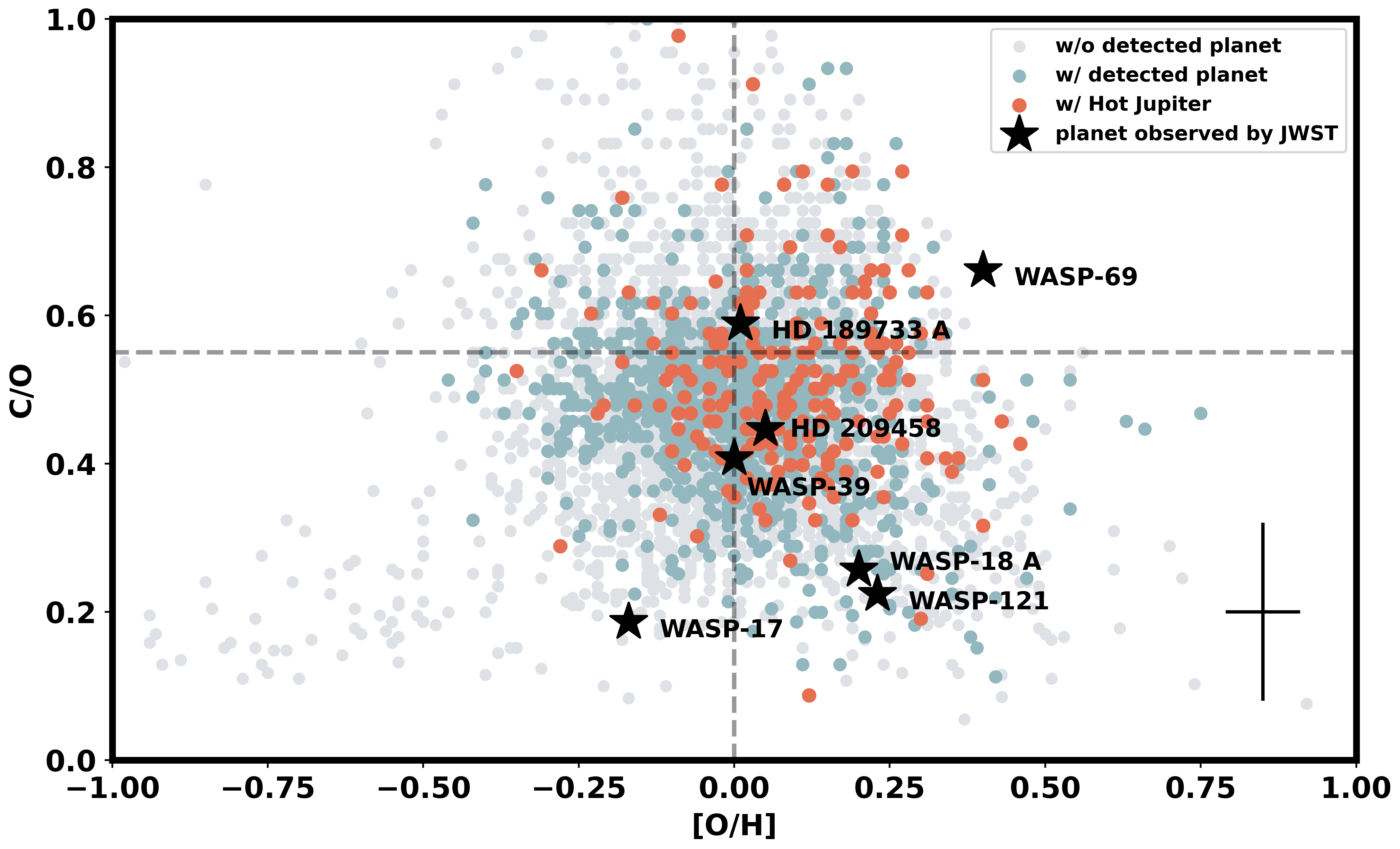} 
\caption{The distribution of C/O and O/H for a wide range of stars with and without planets. The dotted lines denote the solar values (C/O $=0.54$). Clearly there is no preference in C/O in the formation of planets, and particularly for hot Jupiters. There is however, a strong trend in [O/H], where hot Jupiters appear preferentially around metal rich stars. Data for this plot was taken from the Hypatia Catalogue \citep{hinkel2014}. The typical measurement uncertainty is also shown. Reproduced with permisson, \textcopyright AAS}
\label{fig:CtoO_metallicity}
\end{figure*} 

Observations of the element ratios such as C/O and C/N of stellar atmospheres \citep{Brewer16} inform us about the distribution of volatile element abundances in the initial accretion disks out of which both the star and its retinue of planets formed.  These materials were accreted onto the planet as it migrated through the disk.  Recent JWST observations also show that there is a metallicity-planetary mass relation in which the atmospheric metallicity is anticorrelated with the planet's mass (more massive planets have lower atmospheric metallicity) - a relation that solar system giant planets follow very closely \citep{Kempton_Knutson2024, Mansfield2018}. This is also reflected in population synthesis studies of core accretion because the atmospheres of higher mass planets are more diluted by disk gas than low mass planets (see for example \citep{Crid2020b}).

Figure \ref{fig:CtoO_metallicity} shows the C/O and [Fe/H] (normalized to solar values) ratios of 1,617 stars separated into those associated with detected planets (eg. Hot Jupiters), and those that have no known planet. The data is taken from the Hypatia Catalogue \citep{hinkel2014} and overlaid with C/O observations of stars associated with planets observed by JWST. The vertical and horizontal lines mark the solar values, demonstrating that our Sun's C/O ratio is high compared to many planet bearing stars. The figure therefore provides information about the range of compositions of planet hosting stars. As noted above, this plot also shows that many stars hosting hot Jupiters are found to be enriched in metals relative to the sun \citep{FischerValenti2005, WangFischer2015}.

\begin{figure*}
\centering
\includegraphics[width=0.75\textwidth]{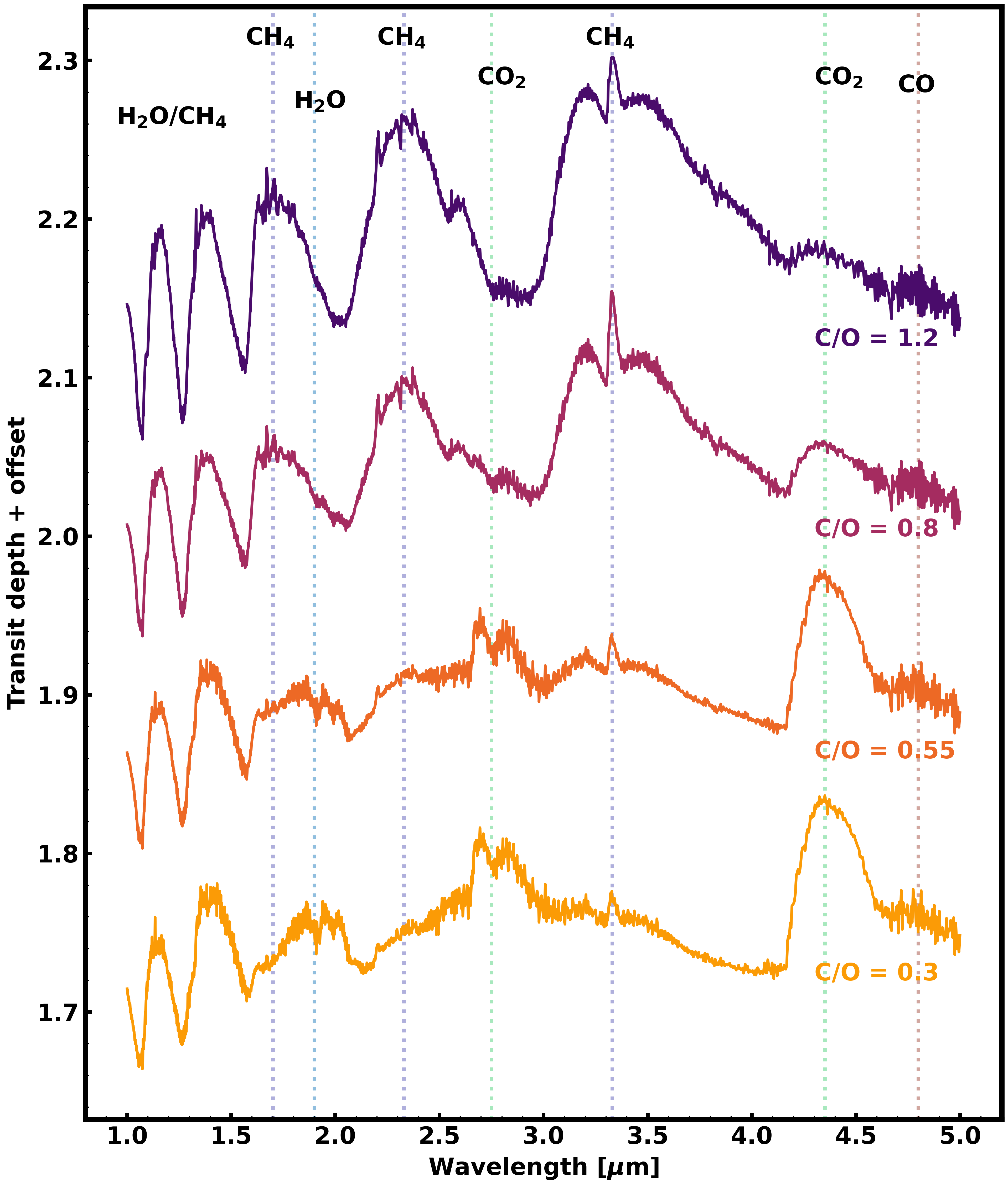} 
\caption{Example cloud-free transmission spectra for a typical hot Jupiter, with properties similar to WASP-39b, demonstrating the variation in dominant carbon- and oxygen-bearing molecules with increasing C/O ratio.}
\label{fig:CtoO_atmosphere}
\end{figure*} 

The stellar [O/H] shown in Figure \ref{fig:CtoO_metallicity} can be thought of as the star's `metallicity' (usually denoted by [Fe/H]) because it is expected that a star's oxygen abundance and iron abundance are (log-)linearly related. This metallicity, however, only roughly relates to the possible metallicities that a companion planet may inherit during its formation. Taking our Solar System as an example, our giant planets each have enhanced `metallicities' (relative to the Sun), as measured by carbon abundance [C/H], due to a lack of oxygen data. Jupiter's metallicity is a factor of a few enhanced relative to the Sun, while Saturn and the ice giants are enhanced by one and two orders of magnitude respectively \citep{Atreya2016}.

The very significant effect that the C/O ratio has on a planetary spectrum is shown in Figure \ref{fig:CtoO_atmosphere}.  Here we have computed a series of cloud-free planetary spectra as a function of the C/O ratio for a typical hot Jupiter type planet. The spectra were generated using the forward modelling package PICASO \citep{batalha_exoplanet_2019,mukherjee_picaso_2023} using the  parameters of the well-studied planet WASP-39b \citep{faedi_wasp-39b_2011}. As the C/O ratio increases and approaches unity, there is a noticeable shift in the carbon- and oxygen-bearing absorbers shaping the transmission spectrum, where carbon dioxide CO$_2$ decreases and methane (CH$_4$) dominates the spectrum. Contributions from water (H$_2$O) also decrease as the relative oxygen abundance decreases. This makes transmission spectroscopy a valuable tool to measure the C/O ratio of a planetary atmosphere.

The difference between the composition of giant planets, and the host star metallicity, is most readily understood as a consequence of where, when, and how planets accreted most of their heavy elements, whether they are in the gas or solid (ice) phase - indicating the possible role of ice lines as places where planets acquired most of their gas \citep[][Bergin and Cleeves review]{Oberg11,Madu2014,Eistrup2018,BoothIlee2019,Crid2019c,Crid2020b}. 

\section{Protoplanetary Disk Populations}

Early surveys of protostellar disks and their lifetimes \citep{Haisch2001,Hernandez2007, Hartmann2008, Andrews2010} suffered from a lack of spatial resolution.  The distribution of disk masses is related to the initial conditions for disk formation and arises from the range of dense core masses and their level of magnetic braking and internal turbulence that will shape their collapse into protostellar disks \citep[eg.][]{Li2014,Seifried2015}.The distribution of disk lifetimes is related to a combination of the processes that carry off disk angular momentum (turbulence, disk winds, or spiral waves)  as well as by disk photoevaporation processes that ultimately dissipate them.

\begin{figure*}
    \centering
    \includegraphics[width=\textwidth]{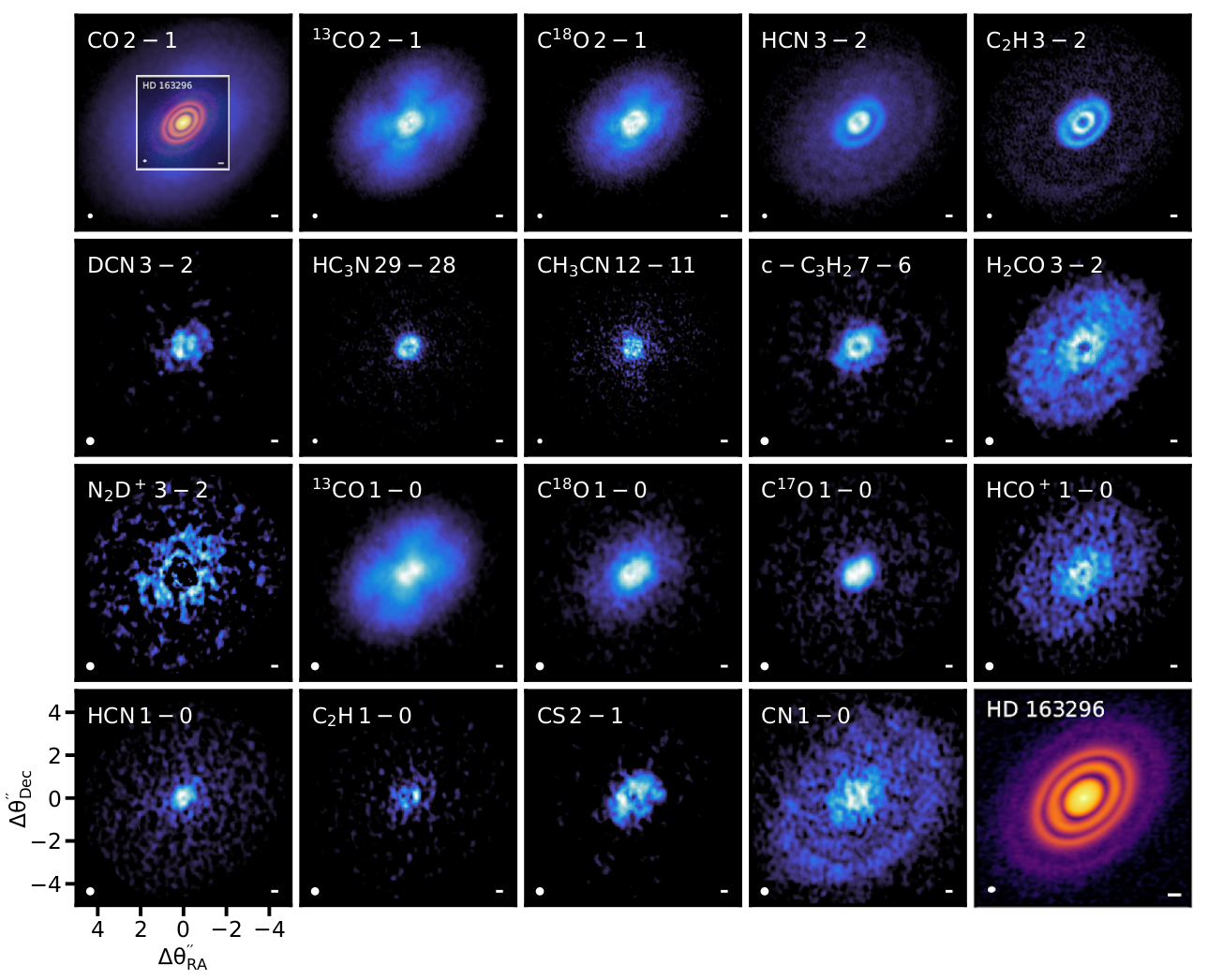}
    \caption{An example of the variety of known molecules and their structures in protoplanetary disks. This particular example is from the MAPS large ALMA program \citep{Oberg2021MAPS} and represents the continuum emission (i.e. dust, in orange, bottom right panel) as well as line emission from a various molecular tracers (blue) for HD 163296. The top left panel shows a size comparison between the dust emission and the emission from the brightest CO line. The dust is much more compact than the gas suggesting that radial drift has played an important role overall. Some gas tracers trace the gas distribution more closely ($^{13}$CO, H$_2$CO, CN) while others more closely tracer the dust distribution (C$_2$H, HC$_3$N). Figure adapted from \citep{Oberg2021MAPS}, reproduced with permission, \textcopyright AAS}.
    \label{fig:e2}
\end{figure*}

This situation has changed completely as a consequence of high resolution observations of disks at millimeter wavelength by ALMA that routinely resolve nearby disks down to 20AU scales.  This has led to the discovery of ring-gap structures in the dust \cite{Andrews2018,Huang2018,Zhang_S2018,Dullemond2018}; as well as at near IR wavelengths by instruments such as SPHERE on the Very Large Telescope (VLT). In a limited number of systems these gaps have been interpreted as containing young, still growing, planets at orbital separations not too dissimilar to the giant planets in our solar system \citep{Zhang_S2018}. The most famous of which, PDS 70 contains two forming giant planets in its ring-gap structure that have been directly observed from their optical and sub-mm emission \citet{Keppler2018,Avenhaus2018,Haffert2019,Garufi2020}.  One of the great observational surprises from ALMA is that disks are far from the smoothly varying structures pictured in the highly idealized theoretical models for accretion disks used for several decades, to study planet formation.  The images show that disks host a  large number of symmetric ring and gap structures, as well asymmetric structure such as spiral waves and lopsided dust distributions revealing that density and temperature inhomogeneities dominate \citep[see for ex. IRS 48,][]{Booth2021}. It is not yet clear whether all of these structures are the consequence of planet-disk interaction \citep{Zhang_S2018}, or whether other processes (eg. MHD) related to planet formation also play a role.

Figure \ref{fig:e2} shows recent results from the Molecules with ALMA at Planet-forming Scales (MAPS) large program for the HD 163296 disk \citep{Oberg2021MAPS}. In the top left panel of the figure an optically think line of CO is compared to the continuum dust emission on the same angular scale. The dust distribution is much more compact suggesting that radial drift has been significant in this disk \citep{Birnstiel2012}.
Figure \ref{fig:e2} also provides a gallery of detected line emission from the HD 163296 disk for a variety of different molecules. The optically thick lines (for eg., CO and most of its isotopologues) trace mainly the upper atmosphere of the disk, while the optically thin tracers like N$_2$D$^+$ and C$^{17}$O (which is the least abundant isotopologue of CO) trace much closer to the disk midplane. These tracers provide observational evidence for the existence of the CO ice line in HD 163296 because N$_2$D$^+$, and its more abundant isotopologue N$_2$H$^+$, are destroyed in the gas phase in the presence of gaseous CO.
These inhomogeneities have important implications for planet formation in that they can give rise to dynamical traps for migrating low mass planets, as well as traps for rapidly moving dust. Finally, observations of debris disks inform the degree to which carbon was frozen out and stored in planetesimals. These can retain imprints of planet formation and disk chemistry processes \citep{Hughes2018}.  

\begin{figure*}
\centering
\includegraphics[width=0.75\textwidth]{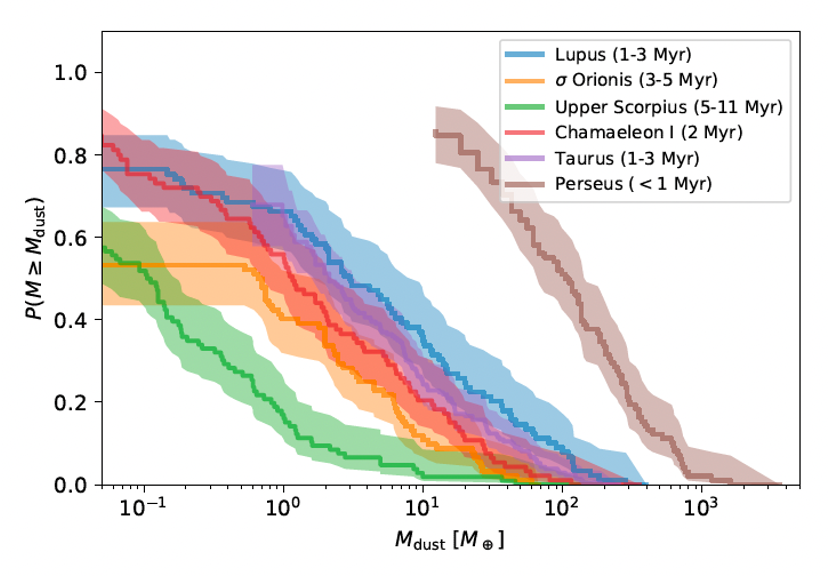} 
\caption{Cumulative distributions of dust mass in PPD populations in star forming regions of different ages.  Adapted from \citet{Miotello2023}, reproduced with permission, \textcopyright ASP }
\label{fig:miotello}
\end{figure*}

Figure \ref{fig:miotello} presents the dust mass distributions of PPDs from star forming regions with a wide range of ages, from less than 1Myr up to 10 Myr.  It is clear that strong evolution of the dust has already occurred by 1 Myr \citep{Miotello2023}.  Noting that the cores of giant planets are likely to have core masses of $10-20 M_E$ in dust, this might suggest that giant planet cores have already formed by this time.  However, other explanations are possible.  Dust could be processed into planetesimals, thereby becoming invisible a mm wavelengths.  It is also possible that rapid radial drift of the dust could have carried most of it into optically thick, inner regions of the disk \citep{Birnstiel2012}.

Stellar irradiation in the form of FUV, EUV, and X-rays plays a central role in driving disk chemistry and limiting disk lifetimes by generating photoevaporative outflows. Specifically FUV irradiation heats the disk via photoelectric heating and $H_2$ pumping \citep{Hollenbach1994,Gorti2016}.  Low energy X-rays may also affect outflow rates especially at low metallicities \citep{Nakatani2018, Ercolano2021}. X-rays effectively determine the ionization state of their protostellar disks and hence their chemistry - topics discussed in the following section on disk chemistry.

\section{Physical Processes in PPDs}

In this section we summarize some of the key physical processes that shape the structure and evolution of PPDs.  These play a major role in how planets form. 

\subsection{Disk formation and initial chemical composition}

 In what are arguably still some of the most comprehensive simulations of star and protostellar disk formation yet performed, \citet{Bate2012} published the highest resolution (down to the opacity limit for fragmentation - a few Jupiter masses), radiation hydrodynamics simulation of a forming star cluster. The initial low
mass, cluster-forming clump (500 $M_{\odot}$ ) had an initial radius of 0.40 pc and temperature of $10^o$ K.  The resulting initial mass function of the stars closely followed the observations \citep{Chabrier2005}.   A comprehensive study of the properties of disks formed in this simulation has also been published \citep{Bate2018}.   The disks show an enormous diversity in types and sizes. Systems can be formed by a wide range of  processes including filament fragmentation, disk fragmentation, dynamical processing, accretion and ram pressure stripping.  Disk morphologies include warped and eccentric disks.   Disk masses increase until $10^4 $ yrs with masses up to 0.5 $M_{\odot}$.   Disk masses range from $M_d/ M_{\odot} ~ 0.1 - 2$ for times $\le 10^4$ yrs,  after which they decline.  Thus, disk masses at these early times are some 30 - 300 times more massive than they are during the Class II phase (when the are $\sim$1 Myr old). The surface density profiles are  $\Sigma_d \propto r^{-1}$, which are flatter than the classic Minimum Mass Solar Nebula (MMSN), $r^{-3/2}$ scaling.

\begin{figure*}
\centering
\subfloat[Streamer feeding a forming PPD \citet{Pineda2023}, reproduced with permission \textcopyright ESO.]{
\includegraphics[width=0.5\textwidth]{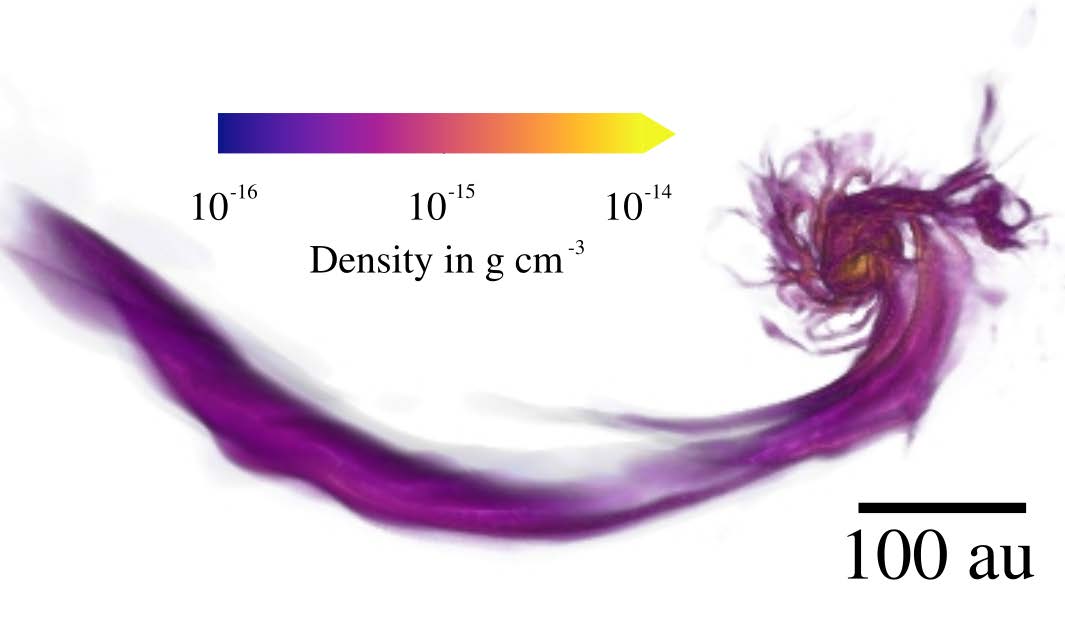}
\label{Kuffmeier_streamer}
}
\subfloat[Three-dimension snapshot from an MHD collapse calculation for the collapse of a 2.6 solar mass core. Black lines are magnetic field lines, blue coloration is of dense filaments bringing gas into the disk forming in the central region. The scale of the box is 1300 AU. Figure reproduced from \citet{Seifried2015}, with permission \textcopyright OUP.]{
\includegraphics[width=0.5\textwidth]{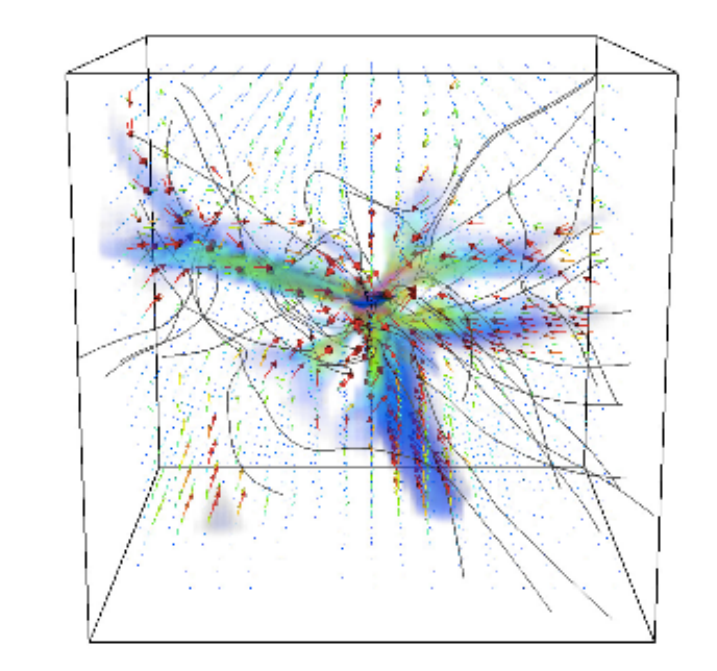}
\label{fig:DiskForm01b}
}
\caption{The late feeding and early formation of PPDs by streamers and filaments.}
\label{fig:streamers}
\end{figure*}

These radiation hydrodynamics simulations did not include magnetic fields.  A reasonably strong magnetic field will strongly brake smooth, rotating clouds so that only very small disks can form - a result known as the ``magnetic braking problem''.   Several solutions have been proposed. Turbulence even at the subsonic level can disorder the magnetic fields connected to disks leading to much reduced magnetic torques leading to larger disks more resembling the hydrodynamic results \citep{Li2014,Seifried2015}.  A number of studies have suggested that non-ideal MHD effects could weaken the magnetic fields in dense regions.  In particular \cite{Zhao2016, Zhao2018} showed that very small grains, being the main charge carriers in these dense cores, can be effectively removed by ambipolar diffusion of the magnetic field.  This process also leads to strong reduction of the threading magnetic field and hence the braking torque.

Figure \ref{fig:streamers} presents 3D simulations of streamers and filaments that play a central role in disk formation. 
Figure \ref{Kuffmeier_streamer} shows a streamer, many hundreds of AU long, flowing onto an embedded disk as a late accretion event \citep{Pineda2023}. Figure \ref{fig:DiskForm01b} shows an image of a forming disk in a turbulent MHD simulation  \citep{Seifried2015} at an earlier stage. The highly filamentary structure arises from the turbulence in the initial cores and the collapse process results in 5 or so filaments that bring material to the forming disk.  MHD torques are inefficient
in magnetic braking  at this earliest phase because the field is disorderd by the turbulence. 
More generally, collapsing, magnetized cores will launch magnetically driven outflows and winds as the disks are forming, and long before the final process of stellar assembly is complete \citep{BanerjeePudritz2006,Li2014}.   Thus, even  in the earliest stages, MHD disk winds will play an important role in the angular momentum evolution of these systems, and this can lead to profound effects on planet formation.

The dust and chemical composition of the protostellar core can, to some degree, be inherited by the disk.  Thus, whereas the largest part of dust growth will occur at the disk mid plane because coagulation is more rapid in high density environments, coagulation helped by ice coated mantles \citep[eg.][]{Ormel2009}, grows grains to several microns at core densities of $10^5$ within $\sim$1 Myr.  Dust can grow up to $\sim$mm sizes within the infalling envelopes \citep[eg.][]{Jorgensen2009}.

Chemical processing also occurs within the dense gas of star forming cores.   
In prestellar cores, the most abundant phase for molecules with elements heavier than hydrogen and helium is a solid.  It has been known for nearly 50 years \citep{GillettForrest1973} that infrared absorption of interstellar ices gives us a glimpse into the chemical composition of star forming material.  Water and CO ice were the first to be discovered, and represent the most abundant molecules after H$_2$. They are followed closely in abundance by CO$_2$ which was not found until the launch of IRAS because of strong absorption in the atmosphere \citep{Oberg11}. With newer space-based studies by ISO and Spitzer, larger, more complex hydrocarbons have been inferred in the infrared absorption of ices towards star forming regions \citep{Oberg2011b}. The formation of these hydrocarbons through the hydrogenation of frozen CO has been studied both theoretically \citep{Walsh2014,Vasyunin2017}, and in laboratory experiments \citep{Butscher2015,Chuang2018}, and represents the first steps towards pre-biotic chemistry. With the latest generation of telescopes these pre-biotic molecules have begun to be found around both young stars \citep{Jorgensen2012}, and in prestellar cores \citep{Ligterink2017,Rivilla2017}.  

As another example, detailed chemical studies of over 39 different molecules, grouped into 4 families of related molecules, have been carried out in the well studied, pre-stellar core  L1544, indicate that significant differentiation of C and N bearing molecules occurs.  Such studies holds great promise for understanding the initial chemical conditions before disks formed  \citep{Spezzano2017}.  Whether these species survive to the protoplanetary disk is still debated. 

In general then, there are two primary pictures for the chemical and elemental abundances of materials in disks based on chemical timescales: `inheritance' and `reset'. In the inheritance regime, most molecular species that were formed in the prestellar core are delivered to the disk intact, whereas in the reset scenario, it is the local thermal equilibrium conditions in the disk that prevail and determine the final chemical products \citep{Pontoppidan2014,Oberg2023}.  In effect, these regimes dominate in different regions of the disk \citep{Oberg2023}.  Inheritance is important in the cold, more diffuse outer regions of the disk where chemical time scales are long \citep{HenningSemenov2013, Eistrup2016, Drozdovskaya2016}, particularly for the volatile content of icy small bodies.  On the other hand, chemical time scales are shorter in the hot, dense inner regions of disks where reset chemistry will dominate. As an example, these high temperature conditions favour the formation of HCN and CH$_4$ and small organics \citep{HenningSemenov2013}. The chemistry in this inner disk region - the abode of forming terrestrial planets - supports chemical processes that are very different than the outer solar system.  Finally, disk chemistry will also vary as a function of height above the midplane: molecules are rapidly depleted onto dust grains at the cold midplane and in the disk's outer regions whereas at higher scale heights the time scale for this process gets longer in the lower density gas. Molecular abundances have their peak values in layers of intermediate scale height below the disk surface \citep{Aikawa_Herbst1999}

Deuterated water could be a good tracer of these different processes because its enrichment is favoured in cold, ionized environments such as prestellar cores (e.g., L1544), and in a protoplanetary disk at large radii \citep{Cleeves2014}.  In the inheritance scenario the deuteration of water would be homogeneous across the disk, while in the reset scenario there would be a deuterium gradient. Of course the true answer may be somewhere between these two extremes.  Carbon deficiency in the solids throughout the solar system could be evidence of reset in the inner solar system, and inheritance in the outer solar system. The number of carbon atoms relative to silicon on Earth is under abundant  by four orders of magnitude relative the ISM while comets like Halley are not similarly underabundant \citep{Bergin2015}. This could be evidence of thermal processing of material because while carbon generally exists in the solid phase in prestellar cores \citep{Bergin2015} if the grains are destroyed upon reaching the disk, the carbon would not re-condense as a solid in the inner solar system \citep{Pignatale2011}.

\subsection{Angular momentum transport and disk evolution} 

 In the earliest stages (within $10^4$ years) of star and disk formation,  the disk mass is  comparable  to that of the protostar \citep{Seifried2015,Klassen2016,Bate2018}.  In this situation, the disk is gravitationally unstable as measured by the Toomre Q parameter ($ Q = c_s \Omega / \pi G \Sigma $ where $c_s$ is the sound speed, $ \Omega $ is the local angular velocity of the disk, and $\Sigma $ is its surface density), is significant \citep[$ Q \simeq 1$][]{Kratter2008}.  The spiral waves that are launched in response are effective in transporting angular momentum. In these early stages,  outflows are ubiquitous for the formation of stars of all masses, and are also the most powerful.

The transport of angular momentum is the essence of disk evolution, star formation, and also, planet formation.  In its most general form, the vertically averaged angular momentum equation that governs a disk undergoing a total stress  $\bf \sigma $  is,\citep{PudritzNorman1986,Turner2014,Bai2016, Suzuki2016}, \begin{equation}
 \dot M_a {d \over dr} (r u_{\phi}) = {d \over dr} ( 2 \pi r^2 < \sigma_{r,\phi} > )+ 2 \pi r^2 \sigma_{z,\phi} \vert_{-H}^{+H} 
\end{equation} 
\noindent where the accretion rate is $ \dot M_a = 2 \pi r  \Sigma v_r$ for a radial inflow speed of the gas $v_r$, and the angle brackets in the first term indicate taking the vertical average of the torque by integrating over z up to a scale height H.  In the second term, the vertical torque is evaluated at the disk surfaces, taken to be at $\pm H$.

The total stress has contributions from both turbulence, and the Maxwell stress of threading magnetic fields.  The first term on the right hand side denotes angular momentum flow in the radial direction, while the second term is angular momentum flow out in the vertical direction due to wind torques.  In the case of shear turbulence, the stress is  the average of the turbulent fluctuations, known as the Reynolds stress:
$ \sigma_{ r, \phi} =  -  \rho  \delta  v_r  \delta v_{\phi}   $.
In the presence of  a toroidal magnetic field $B_{\phi}$ in the disk, a radial field $B_r$ can also contribute to flow in the radial direction through  the Maxwell stress component; 
$\sigma_{r, \phi} =B_r B_{\phi}  $.   This possibility arises in recent models of non-ideal MHD in which the Hall effect can produce an instability leading to a radial field component \citep[eg.][]{Bai2014,Lesur2014,McNally2017}. 

The second term shows that a threading vertical component of the field $B_z$ exerts a torque on the disk; $\sigma_{z, \phi} = B_z  B_{\phi}  $.  This is realized by  
 an MHD disk wind, which is central to the action of the ubiquitous jets and outflows that accompany the formation of all young stars, regardless of their mass \citep{Frank2014,Ray2007,Pudritz2007, pudritz_Ray2019}.  
 There is an important physical difference in the behaviour of these two stresses in how the disk responds.  The radial stress enters as a radial derivative, while the vertical term does not. This implies, as seen below, that the radial term drives a diffusive radial flow that spreads the disk outwards in radius, while the magnetic vertical stress drives and advective inward flow.  

\begin{figure*}
\centering
\includegraphics[width=\textwidth]{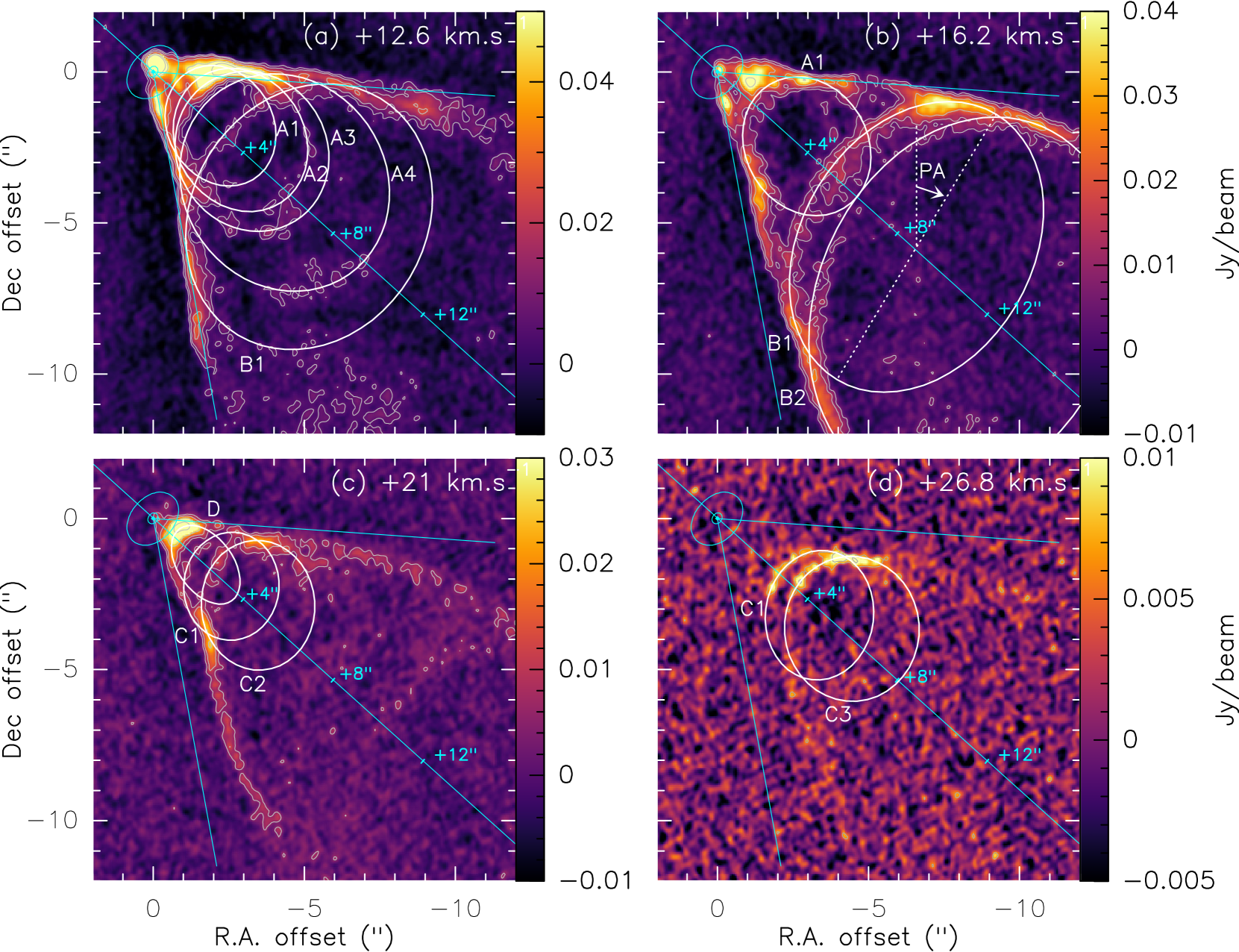} 
\caption{Layered outflow from HL Tau - ALMA observations  Figure from \citet{Bacciotti2025}, reproduced with permission \textcopyright ESO.}
\label{fig:Bacciotti2025}
\end{figure*} 

Figure \ref{fig:Bacciotti2025} is an ALMA image the outflow from the PPD, HL Tau \citep{Bacciotti2025}.  Famous as the first disk to have rings and gaps spatially resolved, these CO maps show that the outflow from HL Tau also has a layered system of arcs that are outflow shells.  These outflow structures can be traced to disk radii at 50-90 AU - the sites of the most prominent rings in HL Tau's disk and therefore capable of removing disk angular momentum even out to the largest disk radii. Their connection with ring-gap structure also suggests that outflows are intimately connected with planet formation itself, as discussed later.  

Models of accretion disks have for decades assumed that angular momentum is transported  by turbulent stress modeled as viscosty, as first addressed in the seminal papers by \cite{SS1973}, \cite{LBP1974}.
In these models, turbulence is assumed to arise from the shearing Keplerian flow and takes the form $ \sigma_{r, \phi} = \nu \Sigma r (d \Omega / dr) $.   For thin disks, the condition of  hydrostatic vertical balance of the disk is $H/r = c_s/v_K$, where the aspect ratio of the disk $h \equiv H/r \simeq 0.05$ for most models. By introducing a parameter that is the ratio of this stress to thermal pressure $\equiv \alpha$ (usually assumed to be a constant), the effective viscosity of the disk $\nu$ can written as $ \nu = \alpha c_s H$.    
Steady state disks then have a radial accretion rate $ \dot M_a $ , which, away from the inner boundary of the disk can be written as

 \begin{equation}
 \dot M_a = 3 \pi \nu \Sigma = const 
 \end{equation}
In order  match the observations of disks,  a value of $\alpha$ must be chosen.  Thus, to drive an accretion flow at the rate observed to fall onto T-Tauris stars, $\alpha \simeq 10^{-2} - 10^{-3} $.  The angular momentum is carried out radially leading to the slow, outward radial spreading of the disk from its initial state.   

A physical origin of the postulated disk turbulence was eventually identified as the magneto-rotational instability (MRI - \citet{BH1991}.   Its derivation has an interesting history. Chandrasekhar in his book ({\it Hydrodynamic and Hydromagnetic Stability}, 1960) derived a striking fact about the stability of magnetized Couette flows (fluid flow between two rotating cylinders).  For the purely hydrodynamic case, the well known Rayleigh criterion for hydrodynamic stability of rotating fluids dictates that the specific angular momentum (i.e., angular momentum per unit mass, $j = v_{\phi}r = \Omega r^2 $ should increase with radius for stable hydrodynamic flows.  However, if one threads a rotating Couette flow with a magnetic field, this criterion is profoundly changed:  stability now requires that $\Omega$ must be an increasing 
 function of radius - even in the limit of vanishing magnetic field strength.   \citet{BH1991},  rediscovered 
 this result in their paper on MHD stability of accretion disks and went on to show that this magneto-rotational instability (MRI) would lead to turbulence.  In the astrophysical context, angular velocity decreases with radius (eg. galactic rotation curves $ \Omega \propto r^{-1}$, Keplerian disks $\Omega \propto r^{-3/2}$ ) and so should be unstable to MRI. The important cautionary note here is that these results pertain only to perfectly conducting magnetized fluids (ideal MHD).   In this regime, growth rates for the most unstable modes in a thin disk are $3/4 \Omega$ \citep{BH1991} with a vertical wavelength $\lambda_z = 2 \pi v_A / \Omega $  where $v_A = B_z / (4 \pi \rho )^{1/2} $ is the Alfven speed in the magnetized gas. A host of other hydrodynamic and thermal instabilities have been discovered in theoretical models and disk simulations, but these generally tend to have low turbulent amplitudes $\alpha \le 10^{-4}$ \citep{Lesur2023}.

The extraction of disk angular momentum by the MHD wind torque can be computed using conservation laws of MHD for steady, axisymmetric flows (see summary in \citep{Konigl_Pudritz2000}).  Imagine a parcel of gas as it accelerates out along a field line, starting from some footpoint radius on the disk $r_o$.   The field enforces the parcel's co-rotation with the disk (at the local Kepler speed $v_K(r_o$) out to that point where it reaches a speed along the field line equal to the Alfvén speed.  Beyond this point, the field is no longer strong enough to enforce corotation and becomes ever more toroidal in character which induces the radial pinching force that collimates the outflow. The radial distance of this point from the axis is the Alfvén radius of the flow, $r_A(r_o)$.  The angular momentum per unit mass at this point in the flow has therefore increased, from the value at the disk by the factor $\lambda \equiv (r_A / r_o)^2$, which is the lever arm of the torque \citep{BlandfordPayne1982, PelletierPudritz1992, Bacciotti2025}. The angular momentum equation for this stress acting on the disk then gives the ratio of the wind mass loss rate to the resulting disk accretion rate as;  
\begin{equation}
\dot M_a / \dot M_w=2(\lambda-1)
\end{equation}
\noindent
\noindent
which amounts to an application of lever arms (Archimedes).

The efficiency of a disk wind is enormous in principle but simulations typically show that $r_A/r_o \simeq 3$ \citep{Pascucci2023}. In a recent review \citet{Pascucci2023} conclude that MHD disk winds provide the best fit to all of the observed properties of outflows on scales $\le 500$ AU. In perhaps the highest resolution observations of a disk wind ever undertaken, \citet{Moscadelli2022}, carried out VLBA observations with a resolution of 0.05 AU on the outflowing water masers in a large scale wind in a protostellar system. The masers followed helical trajectories, well matched to MHD disk wind simulations with a launch point at 6-17 AU on the disk. 

An important fact about PPDs in their planet formation phase is that they have high column densities and therefore, are poorly ionized (see below).  Therefore non-ideal MHD is essential to understand how important turbulence is compared to disk winds.  \citet{BaiStone2013} performed MHD simulations of a vertically stratified shearing box to take into account Ohmic resistivity and ambibolar diffusion. They discovered that both of these mechanisms completely suppressed MRI and that a strong MHD disk wind is launched.  This wind drives a laminar accretion flow through a layer of thickness $\simeq 0.3 H $ that easily accounts for the observed accretion rates onto T-Tauri stars.  The efficiency of the outflow increases as the penetration depths of ionizing FUV radiation increase.  These results have been reproduced in many subsequent publications \citep{Gressel2015,Bai2016, Hasegawa2017, Bethune2017, Lesur2023}.  However, several types of purely hydrodynamic instabilities that are still active remain, such as the vertical shear instability, convective overstability, and the Zombie vortex instability (see Klahr et al., this volume; \citet{Lesur2023}).  All of these drive turbulence at much lower amplitudes, with $\alpha \simeq 10^{-4}$ or less - in other words at levels not consistent with observed accretion rates.  The region in the disk where these non-ideal effects conspire to reduce or eliminate the MRI instability is known as the dead zone \citep{Gammie1996}. 

Recent ALMA observations have made a significant breakthrough in the ability to directly measure non-Keplerian velocity fields in PPDs.   Such disk kinematics studies confirm that turbulence levels in PPDs are generally much smaller than needed to drive observed accretion flows  \citet{Flaherty2017, Flaherty2018, Teague2022}.  A recent survey of these studies indicates that $\alpha \simeq 10^{-4} - 10^{-3}$ \citep{Pinte2023}.  A complementary approach to constraining the observed levels of turbulence in a PPD is by measuring the vertical scale height of dust, which depends on its Stokes number and the strength of turbulence in the disk \citealp{Pinte2016}. A recent compilation of archived ALMA observations of the dust scale height in 23 different ringed systems established that dust settling is prominent \citep{Villenave2025}.  The amplitude of the dust turbulence was $\alpha \le 1.1 \times 10^{-3}$, too small to drive the necessary accretion rates through these systems.  In addition the amplitude of the vertical turbulence was of order the radial turbulence.   This rules out the possibility that the vertical shear instability (VSI) is responsible for the turbulence as it predicts that the ratio of vertical to radial turbulence is much larger than unity.

\begin{figure*}
\centering
\includegraphics[width=.75\textwidth]{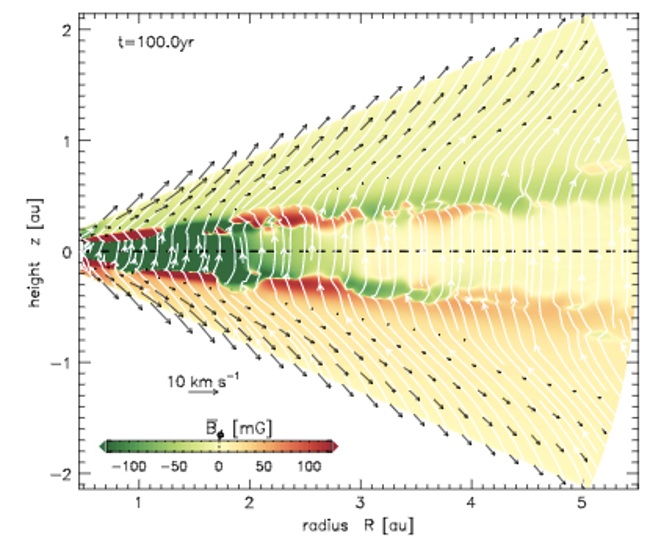} 
\caption{Disk wind from non-ideal PPDs. Figure from \citet{Gressel2015}, reproduced with permission \textcopyright AAS.}
\label{fig:Gressel2015}
\end{figure*} 

Figure \ref{fig:Gressel2015} shows a global simulation of a non-ideal MHD simulation of a disk that includes Ohmic and AD effects under FUV and X-ray ionization \citep{Gressel2015}. This wind drives a laminar accretion flow out to 5AU in the disk, with very little turbulence.  There is also good agreement with the local shearing box simulations of \citet{BaiStone2013}.  The outflows can be driven by rather weak magnetic fields  (magnetic field pressure small compared to gas pressure).

Although 3D MHD simulations play a crucial role in studying winds and turbulence, it is impossible at the present time to follow the full 3D evolution of disks over the Myr time scales of planet formation.  Instead, it is very useful in most models and in particular in population synthesis to use 1D disks whose evolution can be informed by the more detailed 3D simulations.  This is achieved by combining the continuity equation (mass conservation) for the surface density of the disk with the disk angular momentum equation (equation 1) which prescribes the mass accretion rates due to the various torques.  The resulting equation describes the evolution of the disk surface density profile $\Sigma(r,t)$ due to the angular momentum transport  of turbulent and disk wind torques, as well as mass loss,  respectively; \citep{Suzuki2016,Chambers2019,Alessi_Pudritz2022},: 

\begin{equation}
 \frac{\partial \Sigma}{\partial t} = \frac{3}{r}\frac{\partial}{\partial r}\left[r^{1/2}\frac{\partial}{\partial r}\left(r^{1/2} \nu \Sigma\right) \right]\ 
 + \frac{1}{r}\frac{\partial}{\partial r}(rv_w\Sigma)- \dot\Sigma_w 
\label{DiskEvolution} \end{equation} 

The advective flow speed $v_m$ that is driven by the disk wind's angular momentum loss depends on the strength of the magnetic field $v_m = B_{\phi} B_z / 3\pi\Sigma\Omega$; and hence to the wind mass loss rate.  To make these physical effects clearer, note that in the case where the wind mass loss rate from the disk is small compared to the accretion rate (the typical situation) and where viscosity can be neglected, then the steady state solution for $\Sigma$ reduces to solving $\partial / \partial r (rv_m\Sigma) = 0$; thus the wind driven accretion rate is $\dot M_{w, a} = 3 \pi r v_m \Sigma = const$ throughout the disk.   In particular, this relation predicts that variations in the advective speed $v_m$ must lead to inverse variation of the disk column density to maintain this steady state. This may be an important consideration in understanding why the concentration of MHD disk winds occurs in gaps in disks (see subsection below). 

The relation between $v_m$ and $\dot M_w$ has been obtained by numerical simulations. As an example, \citet{Riols_Lesur2019} modeled the mass loss rate efficiency $\zeta = \dot \Sigma_w/\Sigma \Omega^{-1}$ as a power law of the scaled magnetic pressure of the disk due to the threading field $B_z$ as $\zeta \propto B_z^{2p} $ where the index $p=-.6-0.7$ was found by numerical simulations.  Stronger fields produce stronger outflows in this case, and hence drive larger accretion rates. 

It is clear that observations of disk outflows and surface heating mechanisms are needed to inform these MHD disk wind models since disk evolution and accretion depends on what drives mass loss from the disk and on the amount of magnetic flux that threads it \citep{Bai2016}.  In this regard, it is well known that the surface of the disk is heated by FUV and X-ray irradiation from the host star \citep{Glassgold2004}.  In the more tenuous upper regions of the disk where the wind is initiated, magnetic coupling is reasonably good so that ideal MHD conservation laws are applicable.  In particular, mass and magnetic flux conservation can be combined to show that the ratio of the mass to magnetic flux on a field line - known as the mass loading $k = d \dot M_w  / d \Phi $ is conserved along a field line \citep{PudritzNorman1986, PelletierPudritz1992}.  The mass loading of the wind depends on the distribution of magnetic flux along the disk as well as the material that will be launched.  In general, outflow is initiated at the magnetosonic point along a field line where the outflow speed matches the speed of a slow magnetosonic wave (the MHD equivalent of how the solar wind is initiated at a the sonic point in the base of the Sun's corona).  The outflow has been simulated as arising from the base of a disk corona modeled as an adiabatic atmosphere \citep{Ouyed_Pudritz1997} while recent 3D simulations \cite{Aoyama_Bai2023} model an isothermal atmosphere with an assumed atmospheric scale height.

At late times in a disk's evolution, it becomes tenuous enough that  photoevaporative processes, driven by X-ray and UV radiation from the star, truncate the disk and disperse it. The combined effects of accretion and photoevaporation can be combined in a single, time dependent equation for the evolution of the disk's accretion rate \citep{Pascucci2009,Owen2011};  

\begin{equation} 
\dot{M}(t) = \frac{\dot{M}_0}{(1 + t/\tau_{\rm{vis}})^{19/16}} \exp\left(-\frac{t - \tau_{\rm{int}}}{t_{\rm{LT}}}\right)\;,
\label{ViscousAccretion} 
\end{equation}
which includes a viscous evolution term multiplied by an exponentially-decreasing photoevaporation factor. In equation \ref{ViscousAccretion}, $\tau_{\rm{vis}}$ is the disk's viscous timescale, $\dot{M}_0$ is the accretion rate at the initial time $\tau_{\rm{int}} = 10^5$ years, and $t_{\rm{LT}}$ is the disk's lifetime \citep{APC16a}.  In this equation the contribution due to viscous diffusion arises from the analytical model by \cite{Chambers2009} for the evolution of a viscous, irradiated disk.  The exponential factor is due to  rapid photoevaporative truncation of the disk as modelled  by \cite{HP13} who showed that without a sharp cutoff of viscous evolution planets undergo too much migration and accretion to be able to match the distributions in the M-a diagram.  Other authors have used different prescriptions for cut-offs, such as a finite time cutoff to zero \citep{Ruden2004}.  
  
The discussion on non-ideal MHD has so far ignored the Hall effect. One of its most distinct aspects is that the direction of transport of the magnetic flux in disks depends on the polarity of the threading poloidal field component $\bf {B_p} $ with respect to the disk rotation axis.  If its direction is parallel to { \bf $ \Omega$ }, then flux transport 
 is inwards, and if anti-aligned, outwards \citep{BaiStone2017}.  Since the flux distribution affects the strength of the wind torques,
 these Hall effects could be significant for the physics of Type I migration.    

\subsection{Disk heating and ionization: heat transitions and dead zones} 

 The ultimate source of disk energy that is available to drive processes such as turbulence is the gravitational potential energy release across each annulus of the disk due to accretion. Standard models for turbulent viscosity assume that this is ultimately dissipated as heat and radiated away.  Assuming that each annulus of the disk radiates as a black body, one readily derives that viscous heating results in an effective temperature of the disk $\sigma T_{eft}^4= (3/8 \pi) \dot M_a \Omega^2$ and thus the scaling: $T_{eft} (r) \propto \dot M_a^{1/4} r^{-3/4} $.   

The second source of disk heating is external; the radiation field of the 
 central star.   A flaring disk will intercept flux from the star,  and will be absorbed by the dust and re-emitted at IR wavelengths in the disk's surface layer .  Assuming this is a black-body process, the temperature then has a shallower fall off with disk radius $T(r) \propto r^{-1/2} $  \citep{HartmannKenyon1987}.   This can be extended by 
 considering that only the surface layers of the disk are directly heated by the star while the deeper parts of the disk are heated by radiation re-emitted from it, which
 are solved in concert with 
 hydrostatic balance that produces a flaring disk .  The result is a  surface temperature that scales as 
 $T_{surf} \propto r^{-2/5} $ while for the interior $T \propto r^{-3/7} $ for disk radii $ r \le 84 $ AU \citep{ChiangGoldreich1997}. Observations indeed show that the temperature distribution arising from viscosity are too steep to explain the mm and submm observations of disks, having an average temperature exponent $T \propto r^{-q} $, where
 for the dust,  $q_{dust} \simeq 0.5$  \citep{AndrewsWilliams2007}.  The temperature of the gas,  as determined by CO and [CII] line observations,  has a steeper radial decline with  $q_{gas} \simeq 0.85 $. Since the temperature profiles of dust and gas should be similar on the disk mid plane, the difference here suggests a decoupling of gas and dust at high scale heights above the disk \citep{Fedele2013}.   
 
 Unlike viscous heating, radiative heating from the central stars creates a hot surface layer on the disk atmosphere, and a much cooler midplane.  This has several important consequences for disk chemistry and dynamics in that the snow-lines for various species are 2-D surfaces that move outward in radius as one moves away from the disk midplane (see chapter by Bergin and Cleeves).  
 
 The disk radius at which the dominant heating mechanism of the disk transitions from viscous to radiative heating is called  the ``heat transition'' (eg. \cite{Lyra2010, HP11} ), which we will denote $r_{HT}$. 
 Since the temperature of the inner viscously heated part of the disk must decline with time (since $T_{visc} \propto \dot M_a^{1/4}$ ), the heat transition radius $r_{HT}$ moves inwards with time as well. 
 
The ionization of the disk by  stellar X-rays, external cosmic rays, and the decay of radionuclides mixed in with the gas plays a central role in the coupling of the magnetic field - and hence the genesis of MRI turbulence - to the disk.   Disk chemistry is also primarily driven by ionization processes (see Chapter by Bergin and Cleeves). Thus, disk chemistry and angular momentum transport are highly coupled, and as we will see, should therefore be connected to the ultimate element compositions of forming planets.  

Non-ideal MHD effects arise from the finite diffusivity of fields in the background gas.  The ionization fractions are highest at the disk surface and decrease with increasing optical depth as one penetrates down to the disk mid plane.  Thus UV and X-rays are absorbed at column densities of 0.01 and 10 g cm$^{-3}$ respectively.   The greatest penetration can be achieved by cosmic rays (CR) that are attenuated by column densities of 100 g cm$^{-3}$ \citep{UmebayashiNakano2009}.  Unlike X-rays however, CR can be scattered by MHD turbulence.  By decomposing MHD perturbations into their three basic modes (slow, Alfvenic, and fast),  it has been shown
that gyroresonance with the fast modes (sound waves compressing the magnetic field)
is the dominant scattering process \citep{YanLazarian2002}.   
The likely cause of the lack of CR driven chemistry in protostellar disks is their loss via CR scattering as they propagate through protostellar and disk winds \citep{Cleeves2013}.  
 
 Descending from the disk surface through the ever greater densities approaching the disk mid plane, 
 first dust grains, then ions, and finally the electrons decouple from the magnetic field.   The degree of coupling is measured by three different
 magnetic diffusivities \citep{SalmeronWardle2003};  ambipolar diffusion in the surface
low density regions where ions and electrons are well coupled  ($\eta_A$),  the Hall effect at intermediate densities 
where the ions are decoupled from the fields through insufficient collisions with the neutrals ($\eta_ H$), and at the
greatest depths and densities Ohmic diffusion where even the electrons become decouples ($\eta_O$).  Although both 
ambipolar and Ohmic effects behave like diffusive processes, the Hall effect is different in principle.  It drives the field
lines in the direction of the current density with a tendency to twist that can give rise to 
non diffusive dynamical processes, such as the generation of a toroidal field from a radial component. 

The diffusivities depend upon the ionization of the disk, and it is here that models of disk ionization driven chemistry
can play a key role.  As an example, the Ohmic diffusivity depends on the electron fraction $x_e$ and disk temperature as
$ \eta = 234 T^{1/2} / x_e$ cm$^2$ s$^{-1}$.   As one moves towards the disk mid plane, the Ohmic diffusivity grows as 
the X-rays are screened.  Similarly, as the disk evolves, the column density at any radius decreases with time, shifting
the region of Ohmic dominance inwards allowing turbulence to appear.   The temperature at the disk mid plane, where planetary materials are gathering,  is related to the effective temperature of the disk as $ T_{mid} = (3 \tau / 4)^{1/4}T_{eff}$ where $ \tau = \kappa_o \Sigma / 2$ is the optical depth and $\kappa_o$ is the disk's opacity.  Chemistry codes that can follow disk ionization with time are therefore essential \citep[eg.][]{Crid16a,Crid16b}.  

The damping of MRI instabilities can be measured by the ratio of the growth rates to the damping rates predicted by 
these diffusivities.  These are the so-called  Els$\ddot a$sser numbers for each effect: $ A_m =   v_A^2 / (\eta_A \Omega)$, 
$\Lambda_H=  v_A^2 / (\eta_H \Omega)$, and $\Lambda_O =  v_A^2 / (\eta \Omega)$.    Damping of the turbulence will occur if these numbers take a values of typically less unity (see \cite{Turner2014} for a review).   In the case of Ohmic diffusion,  this comparison of damping and growth rates can also be expressed in terms of a comparison of physical scales, 
namely, that the diffusion will erase fluctuations on a scale smaller than $\eta / v_A$, while the fastest growing mode in the disk has a wavelength of $ 2 \pi v_A / \Omega$.  
 
 The presence of dead zones in disks has important implications for planet formation and chemistry.   In order to 
 maintain a constant accretion rate throughout the radial structure of a disk at any time,  the 
 relative roles of turbulence and disk winds in transporting angular momentum must change as one moves from the outer, well ionized regions of the disk, into the region of the dead zone, where MRI turbulence will be damped and the bulk of the angular momentum flow is contingent on wind, and or Hall term transfer.  
 Disk winds do not physically act like turbulent viscosity - the disk does not spread radially outward under the action of a wind but is advected inwards.

\begin{figure*}
\subfloat[Variations in $\Sigma$ and $B$ in disks resulting from disk-wind driven instability. Vertical dashed lines mark the column density peaks.]{
\includegraphics[width=0.5\textwidth]{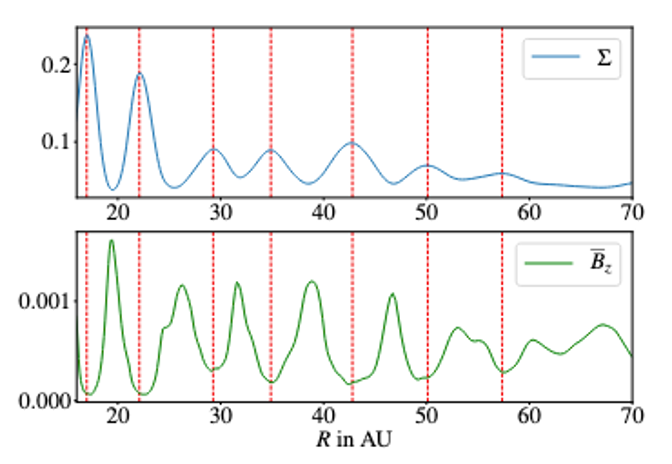}
\label{fig:9a}
}
\subfloat[Zoom in towards disk interior. Colour bar indicates gas density, white lines are poloidal magnetic field lines.]{
\includegraphics[width=0.5\textwidth]{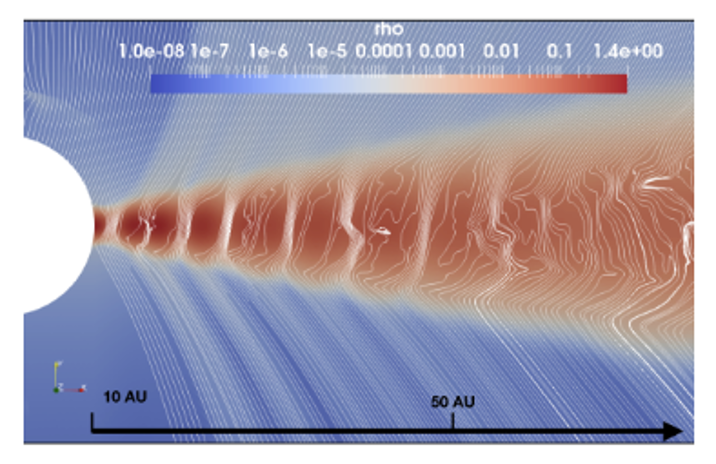}
\label{fig:9b}
}
\caption{ Global 3D simulation of spontaneously produced ring - gap structure by a magnetized disk wind. Adapted from \citet{Riols2020}, reproduced under the creative commons license CC BY 4.0}
\label{fig:Riols2020}
\end{figure*}

\subsection{Disk winds and ring formation}

Theoretical models and simulations have long shown that the observed rings and gaps, and spiral wave structures in PPDs can be produced by a variety of planet-disk tidal interaction models.   As an example, the three main gaps and rings in HL Tau have reproduced by models featuring 3 Saturn-mass planets \citep{Dong2015,Jin2016}, 
or a single Neptune mass planet \citep{Bae2017}.  However, it is still unclear whether all or most ring-gap structures are produced in this way.   Planet formation requires the concentration of dust, pebbles, and planetesimals so other processes must be able to produce this structure to start planet formation. That said,  MHD processes are well known to produce ring-gap structures without the need for planets, either through non-ideal MHD processes involving MRI turbulence \citep{Flock2015} or disk winds \citep{Suriano2019,Riols2020, Aoyama_Bai2023}.  We examine these processes in more detail, below.  

Figure \ref{fig:Riols2020} presents the results of a 2.5D (axisymmetric) global simulation of a non-ideal MHD (ambipolar diffusion) disk that includes dust \citep{Riols2020}.  The simulations are run with the PLUTO code \citep{Mignone2007}.  The disk develops self-organized rings and gaps, the most prominent of which are in the inner disk regions.  These structures are long lasting.  The magnetic flux threading the disk (the $B_z$ field) is concentrated into very thin shells at minima (gaps) in the column density.  This effect has been observed in other simulations \citep{Bai2014,Suriano2019,Riols2019} of both ideal and non-ideal MHD processes in disks.  

The MHD wind associated is also non-uniform and is concentrated in plumes that are connected with these magnetic flux concentrations.   Ambipolar diffusion of the field prevents these structures from being swept along with the accretion flow and remains steady throughout the simulation. Dust also gets concentrated into the pressure maxima of the rings whose widths in the dust  are of the order $0.3-0.4 H$.  The accretion flow is mainly concentrated towards the disk surface. 

The concentration of magnetic fields into gaps is also seen in fully 3D non-ideal MHD simulations.  \citet{Suriano2019}.  In this case, one observes that the field is concentrated into gaps where significant magnetic reconnection occurs.  The accretion flow in this situation is concentrated at the disk midplane.  In these simulations the MHD structures become less prominent in disks with weaker initial fields.  As ambipolar diffusion increases, both the structures and the wind become more laminar.  This favours the settling and more rapid grwoth of dust grains.  

These simulations of wind-produced rings may play a significant role in planet formation.  First, these MHD produced structures prevent the rapid loss of dust grains to the inner regions of the disk (see next section).  The concentration of pebbles in these rings, if sufficient, may be the sites where the seeds of the first planets are born.  

\section{ Chemistry of evolving disks }

\subsubsection{Gas}

As the disk evolves, its changing physical structure is imprinted on its evolving chemistry. Of principle importance is the reduction of gas temperature and the increasing ionization as the disk accretion rate decreases. Because reactions between ions and neutrals lack an activation barrier, the ionization rate plays an important role in dictating the rate of reaction for many gas phase reactions \citep{Eistrup2016}. As the disk surface density drops, and ionizing radiation can more easily penetrate to deeper regions of the disk, driving the chemical system to a (mathematically) steady state - where molecular abundances no longer change with time - more quickly. This steady state differs from the thermodynamic equilibrium solution for a set of reactions, whose final molecular abundances are dictated by Gibbs free energy minimization in that steady states are not necessarily global minima for the Gibbs free energy.  

Generally speaking, the gas changes its chemical structure through only a few chemical pathways. They are: freeze out onto and sublimation off of grain surfaces, neutral-ion gas phase reactions, neutral-neutral reactions on grain surfaces, and neutral-neutral gas phase reactions. The rates of each of these reaction pathways sensitively depend on the temperature, density, and ionizing flux of the disk's gas.

\begin{figure*}
\centering
\includegraphics[width=0.5\textwidth]{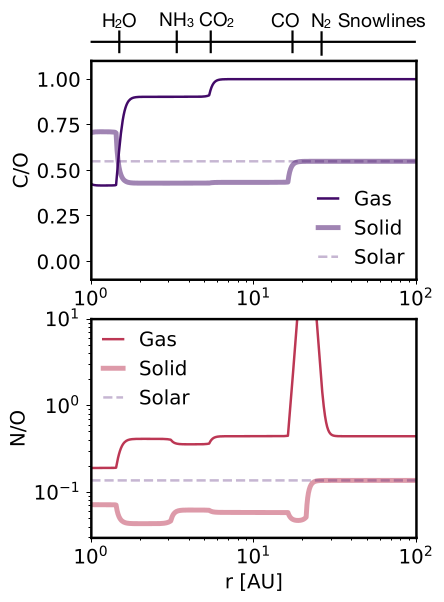}
\caption{The abundance of carbon (top) and nitrogen (bottom) relatively oxygen as a function of radius in a protoplaneary disk. The major jumps in the C/O and N/O ratios occur at the particular ice lines of abundant volatiles throughout the disk. Adapted from \citet{ObergBergin2021}, repoduced with permission \textcopyright ELSEVIER.}
\label{fig:04}
\end{figure*}

Figure \ref{fig:04} \citep[from][]{ObergBergin2021} illustrates a well known, elegantly simple model of the elemental distribution through a disk. In the top panel, it shows the ratio of the total carbon and total oxygen elements (counting the most abundant molecules), known as the `carbon-to-oxygen ratio' (C/O), for gases and solids. At the ice lines of H$_2$O, CO$_2$ and CO, C/O changes as particular volatiles freezes onto dust grains. This process depends on the local gas temperature, so as the temperature of the gas cools the location of the ice lines (and their jumps in C/O) will move inward.

In the bottom panel of this figure, the nitrogen-to-oxygen ratio is shown, including the NH$_3$ and N$_2$ ice line. Since most of the nitrogen in disks is retained in its molecular form, the largest change in N/O occurs at the coldest temperatures when N$_2$ freezes out. In this model, the only chemical process that is taken into account is the freeze out of volatiles onto grains, which is balanced by their sublimation. In reality once a gas species has frozen onto a grain, it can be chemically processed while in the ice phase. This can be particularly important for the production of molecules like methanol which is produced through the hydrogenation of frozen CO \citep{Walsh2014}. Similarly, the gas phase CO can be depleted through grain surface reactions that convert it to CO$_2$ ice between the CO$_2$ and CO ice lines \citep[][see below]{Eistrup2018}. Such processes have been used to explain the unexpectedly low CO abundance and inferred disk gas masses from molecular line emission surveys \citep{Ansdell2016,Miotello2016,Krijt2020}, and could have an important impact on the overall C/O or N/O inherited by growing giant planets.

\begin{figure*}
\centering
\subfloat[The evolution of the C/O in a standard protoplanetary disk model due to gas and surface chemistry. The dashed line shows the gas ratio while the solid lines shows the ice. The disk starts with a similar step function as in the simple \"Oberg et al. model, but after 1 Myr the gas has lost some of its carbon to the ices. And by 7 Myr the C/O distribution is nearly flattened. Some chemical reactions occur rapidly, such that the C/O has changed by 0.5 AU while other, slower reactions are responsible for the flattening of the C/O over longer timescales. Originally from \cite{Eistrup2018}, repoduced with permission \textcopyright ESO.]{
\includegraphics[width=0.5\textwidth,valign=c]{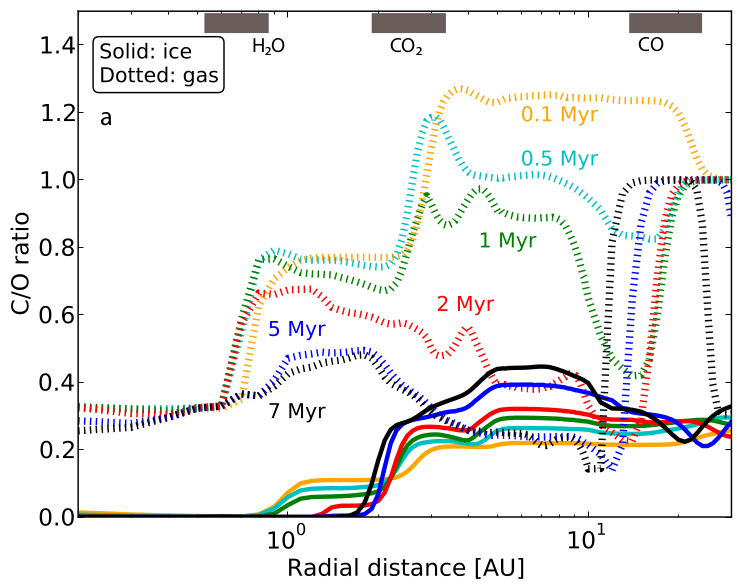}
\label{fig:05a}
\vphantom{\includegraphics[width=0.5\textwidth,valign=c]{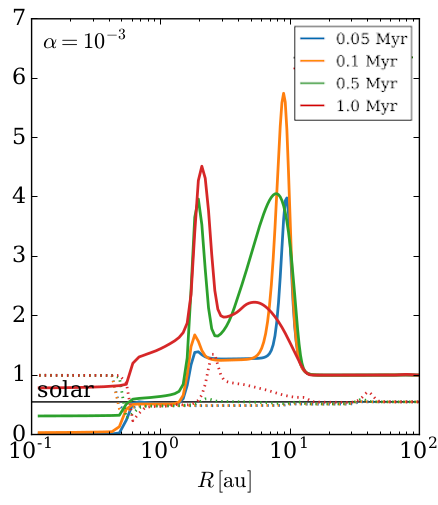}}%
}
\subfloat[The evolution of C/O due to the radial drift of volatiles through a protoplanetary disk. As the dust crosses particular ice lines, the dust loses that ice to the gas phase, enhancing the C/O in the gas (solid line). This occurs very quickly for the outer ice lines, like for CO, while it takes much longer to occur for closer ice lines, like H$_2$O. Over time, the enhancement of C/O caused by the release of the gas is gradually suppressed as the source of drifting dust is exhausted. Originally from \cite{BoothIlee2019}, repoduced with permission \textcopyright OUP.]{
\includegraphics[width=0.5\textwidth,valign=c]{IMAGES_edition_2/fig5b_e2_booth_ilee_2019.png}
\label{fig:05b}
}
\caption{Two examples of how chemical and/or physical evolution can shift the C/O away from the simple step function. Radial drift can greatly impact C/O very near to the location of the ice lines (note the change in y-axis between the two panels).}
\label{fig:05}
\end{figure*}

While the distribution of carbon and oxygen of figure \ref{fig:04} is a good starting point, it ignores chemical and physical processes that can influence the partitioning of carbon and oxygen between the gas and ice phases. In figure \ref{fig:05} we show two such processes and their impact on the distribution of carbon and oxygen in a typical protoplanetary disks. 

Figure \ref{fig:05a} shows the chemical evolution of the C/O for an evolving disk presented by \cite{Eistrup2018}. They find significant partitioning of the carbon and oxygen between the gas (dashed line) and ice (solid line) phases caused primarily by reactions on the surface of dust grains. Of particular importance is the conversion of CO into CO$_2$ ice between their respective ice lines (noted at the top of the figure). In this region of the disk CO is in its gas phase, however there is a non-zero probability that a CO molecule will collide with a dust grain, temporarily sticking to the grain. Because the temperature is higher than its freeze out temperature the CO molecule does not spend long chemically bonded to the dust grain, however it is long enough that a mobile OH molecule on the grain surface has enough time to collide and react with the CO molecule. The net effect is the production of CO$_2$ (and a free $H$ atom), and since the reaction occurs beyond the CO$_2$ ice line, this newly formed molecule remains frozen on the dust grain. By between 1-2 Myr much of the CO in the outer part of the disk has been converted to CO$_2$ ice in this way. In their model, the disk cools and loses mass with time, but there is no radial transport of the gas or ice.

The effect of radial transport is shown in figure \ref{fig:05b} and explored by \cite{BoothIlee2019}. There the solid line shows the C/O in the gas phase while the dashed line show the ice phase C/O. The primary physical mechanism at play is the `radial drift' of dust grains: the transport of dust mass through the protoplanetary disk towards the host star caused by a head wind felt by the grains which drain their angular momentum (we discuss this in more detail below). The ice frozen onto dust grains in the outer disk is transported inward through this radial drift and is then released from the dust once the grains crosses important ice lines. For the most abundant volatiles (H$_2$O, CO$_2$, CO) when the icy dust grains cross their ice lines, they are released into the gas phase, drastically changing the local gas C/O. 

For an initially constant dust-to-gas ratio and a standard disk model, the bulk of the dust mass is found in the outer disk. As such the C/O enhancement occurs first at the outer most ice lines and proceeds inwards with time. Over time the dust mass reservoir in the outer disk is exhausted which cuts off the source of ice, and the enhanced gaseous C/O gradually decreased as it gets mixed with the rest of the gas through viscous spreading. The enhanced C/O is localised to just inside the ice lines because the sublimation rate are faster than the transport rate through the disk. \cite{BoothIlee2019} find that the sublimation and dust transport rates were generally faster than any other chemical timescales they analysed and thus concluded that radial transport dominated over regular chemical evolution on the evolution of the disk C/O.

Both of these are one dimensional models that parameterize the vertical properties of the disk in a simple way. Full three dimensional gaseous models that include dust growth/drift and complex chemistry is largely beyond our computing abilities at the moment. Some models, like \cite{Krijt2020}, have expanded on the models of \cite{Eistrup2018} and \cite{BoothIlee2019} by including a simple chemical network and used detailed dust physical models in two dimensions. This type of model can include not only the (inward) radial evolution of volatiles due to radial drift but can also model the settling of large dust grains which is thought to transport volatile species from the upper atmosphere of disks down to their midplane \citep[see for ex.][]{Salyk2008,Krijt2016,Xu2017}, where they would be available for accretion onto forming planets.

The chemistry of the inner disk is of particular importance for the formation and chemical composition of terrestrial planets and their atmospheres.  The availability of water for such planets is a pre-requisite for life as we know it.  Recent JWST observations in the MINDS program have detected water in the terrestrial planet formation region of the PPD in the famous source PDS70 \citep{Perotti2023} - known to host forming Jupiters in its outer regions. The most common stars in the galaxy have masses $\le 0.2$ M$_{\odot}$ and are therefore likely to host the bulk of terrestrial planets, making them desirable targets for planet formation studies.  MINDS observations on the JWST have detected rich hydrocarbon chemistry (C$_2$H$_2$,  C$_4$H$_2$) as well as a high C/O ratio, likely exceeding unity, in the inner 0.1 AU of the disk around the dwarf star 16053 (0.14 M$_{\odot}$.  The evidence suggests that oxygen is strongly depleted in the region leading to the question as to whether terrestrial planets formed in such a region would be carbon rich \citep{Tabone2023} - unlike Earth which is carbon poor.

\subsubsection{Solids and Ice Lines}
 
Left out of the above discussion is the chemical composition of the grains (often called the refractory material) on which grain chemistry occurs. This is because it is generally assumed that the grains condense very early in the disk lifetime. This assumption is supported by the fast ($\sim$ hour) condensation rate timescales that have been observed in lab experiments \citep{Toppani2006}.  Because these  reaction times are shorter than the disk's viscous evolution timescale, equilibrium chemistry methods can be used to compute the chemical abundance of the refractory material.

Equilibrium chemistry utilizes the thermodynamic result that a chemical system in equilibrium will have its total Gibbs free energy minimized. The Gibbs free energy of the system can be expressed as,

\begin{equation}
G_T = \sum_i^N X_i(G_i^0 + RT\log X_i),
\label{eq:Sol01}
\end{equation}

\noindent for a set of $N$ molecules, each with mole fraction $X_i$, Gibbs energy of formation $G_i^0$ at a temperature $T$ \citep{Pignatale2011,APC16a}. A second restriction is that the total number of elements:

\begin{equation}
\sum_i^N a_{ij} X_i = b_j \quad (j=1,2,...,m),
\label{eq:Sol02}
\end{equation}

\noindent where $m$ is the total number of elements in the system and $b_j$ are their initial mole fraction. $a_{ij}$ is the number of the $j^{th}$ element present in the $i^{th}$ molecule \citep{Pignatale2011,APC16a}.

Solving the equations \ref{eq:Sol01} and \ref{eq:Sol02} is generally done with a commercially available software package {\it HSC}, which
has been used in several astrophysical settings \citep[eg.][]{Pasek2005, Bond2010, Pignatale2011, Elser2012, Moriarty14, APC16a}.

\cite{APC16a} split the majority of refractories into two primary families: mantle, and core. Core materials are iron and nickel refractories that would settle to the core of a differentiated planet. While mantle materials are silicate, aluminium, and magnesium refractories that would end up in the mantle of a differentiated planet. They report that the primary core materials are iron (Fe), troilite (FeS), fayalite (Fe$_2$SiO$_4$), and ferrosilite (FeSiO$_3$). The primary mantle materials are enstatite (MgSiO$_3$), forsterite (Mg$_2$SiO$_4$), diopside (CaMgSi$_2$O$_6$), gehlenite (Ca$_2$Al$_2$SiO$_7$), and hibonite (CaAl$_{12}$O$_{19}$).

A third solid that can be tracked with equilibrium models are ices, however thermochemical data in equilibrium software is often incomplete for astrophysically relevant ices like CO$_2$ and CO. As a result, the only ice that is generally discussed in equilibrium contexts is water. 

\begin{figure*}
\centering
\includegraphics[width=0.75\textwidth]{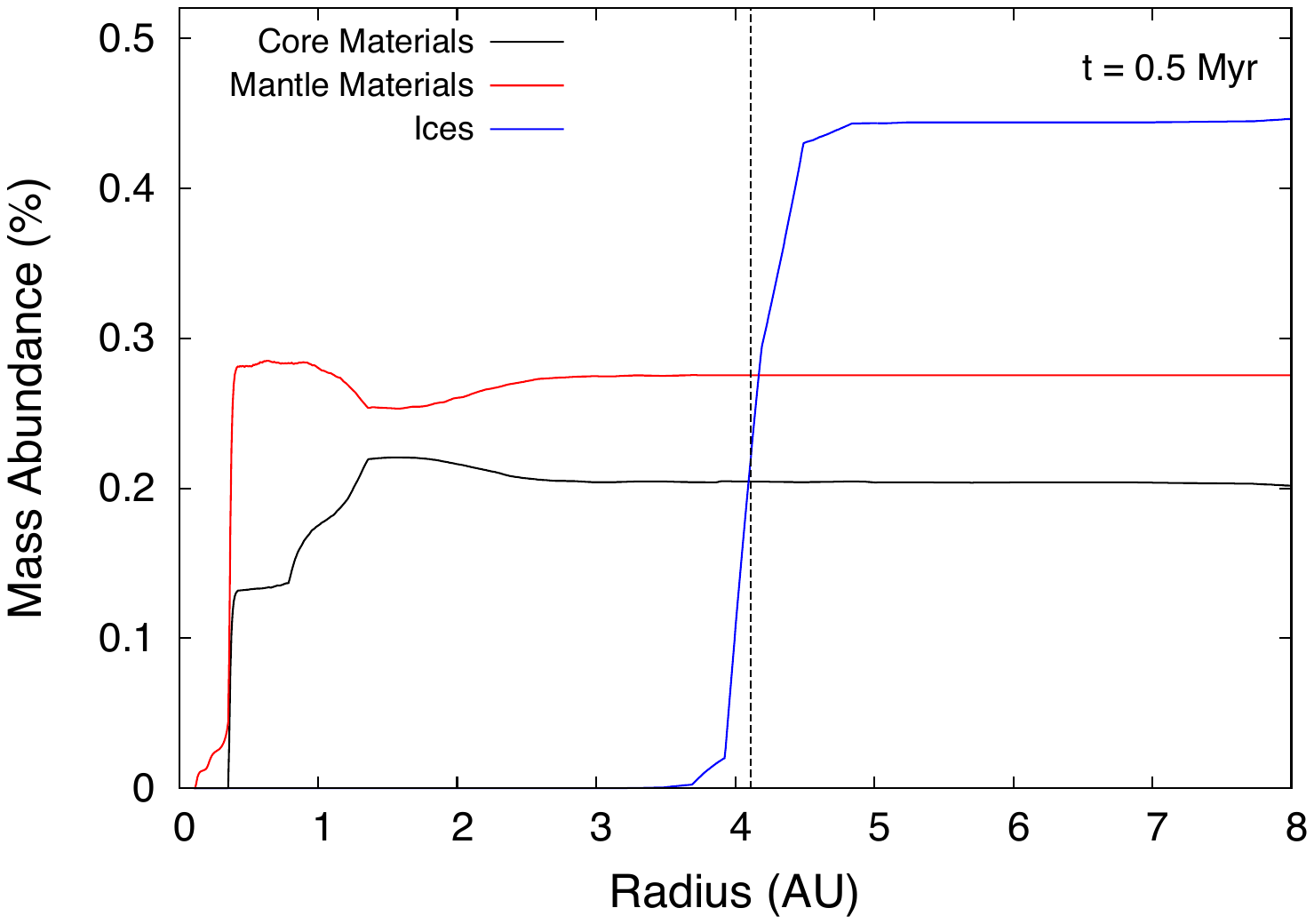}
\caption{A snapshot at t=0.5 Myr of the radial distribution of cumulative solid materials derived from a thermochemical equilibrium calculation in an evolving disk. Individual silicate and iron based refractories are combined into the more general `core' and `mantle' materials. Where the former will typically end up in the core of a differentiated planet, while the latter would end up in the mantle of a differentiated planet. The two phases of water, solid and gas, are found predominately outward and inward of the water ice line (dotted line) respectively. Figure reproduced from \citet{APC16a}, with permission \textcopyright OUP.}
\label{fig:6}
\end{figure*}

In Figure \ref{fig:6} \cite[from][]{APC16a} we show the radial dependence of the core, mantle, and ice material at the disk mid plane, early on in the disk lifetime. The mid-plane solid abundances are quantitatively similar
to those found by  \cite{Bond2010} and \cite{Elser2012}, who also performed equilibrium chemistry calculations on a disc of solar abundance.  There is little variation between the abundance of core and mantle material over most of the disk, apart from the inner ($< 1$ AU) regions of the disk. In these high temperature regions of the disk, more complex core material like fayalite are less energetically favourable, hence the extra silicon is available to produce more mantle material like enstatite and forsterite \citep{APC16a}. As the disk evolves these features move inward with the inward motion of the gas and dust. The abundance of ices show the largest variation, due exclusively to the location of the water ice line (vertical dotted line in Figure \ref{fig:6}). This feature likewise moves inward as the disk ages. The large variation in available ice abundance during the formation of planetesimals implies that the resulting composition (and hence structure, see below) will depend on where the initial planet core accretes.

As has already been noted, ice lines play an important role in disk chemistry and planet formation.  The freeze out of volatiles onto grain surfaces changes the chemistry of both the gas and solid phases.  As an example, ALMA observations of CO isotopologues in the TW Hya PPD locate the CO snowline on the disk surface at a radial distance of 30 AU, which constrains its disk mid-plane ice line to 17-23 AU \citep{Schwarz2016}.  This freeze out also explains the low oxygen abundance in TW Hya.

\subsubsection{Pebble radial drift}

The dynamical behaviour of dust particles, or more generally, pebbles is not defined by their physical size, although this is a factor, but by their aerodynamic size.  Given the drag forces that small particles undergo in PPDs, the critical factor in their dynamics is their stopping time $t_{stop}$, that is, on what time scale do drag forces slow the motion of particles.  The relevant time scale to compare this with is the local dynamical time scale in the PPD, which is measured by the local angular velocity $\Omega$ of the gas.  The dimensionless stopping time, or Stokes number of a pebble is then defined as $St = t_{stop}\Omega$. 

As outlined in the chapter by Ormel, the solid component of the disk (ie. the dust) evolves radially either due to hydrodynamic flows (eg. by turbulence or laminar flow) when the Stokes number $St << 1$; or by radial drift when $St \sim 1$. For very small Stokes numbers, the particles are tightly coupled to the gas dynamics.  Vertical settling of various dust species occurs until hydrostatic balance is achieved, given by the vertical scale height $H_d$. This depends on both the amplitude of the turbulence that keeps grains aloft, and their Stokes number:   $H_d = H \sqrt{\alpha_{turb} / (\alpha_{turb} + St) } $ where recall that $H$ is the scale height of the gas \citep{Birnstiel2016}.

Radial drift on the other hand, occurs when a grain's Stokes number is large enough that its dynamics decouples from the gas. When this occurs it is no longer affected by gas pressure, and begins to orbit faster than the gas (in Keplarian orbits). This velocity difference produces a head wind on the grain that drains angular momentum, moving the grain to a smaller orbit. The radial drift of the particle speed is related to $St$ through \citep{Weidenschilling1977, Birnstiel2024}:
\begin{equation}
v_{drift} = -\frac{2 \eta v_K}{St + (1 + Z_{mid})^2St^{-1}},
\end{equation}
where $\eta $ is the power-law index of the local radial density gradient of the gas such that the gas orbital velocity in the disk is $v_{\phi,gas} = (1 - \eta)v_K$, and $Z_{mid}$ is the local, mid-plane dust to gas ratio. The solution shows that the drift is the largest for pebbles with St $=1$.  Pebbles with $St > 1$ are defined as large and can exceed meters in size.  They can drift over significant portions of the disk within the disk's lifetime. 

This drift has important implications for the chemical structure of the disk because dust plays an important role in setting both the opacity of higher energy radiation (hence impacting the ionization state of the gas), and the rate of gas freeze out (by providing the sites necessary to freeze). Radial drift of dust can also act as a transport mechanism for ices, as discussed above.

It has long been known that ice lines could play an important role as the sites for rapid particle growth by  condensation.  The original suggestion by \cite{StevensonLunine1988}
was that there could be a considerable enhancement of material at an ice line from vapour that diffuses from the inner part of the disk.  The enhancement of vapour by the evaporation of materials moving inwards across the ice lines was addressed by \cite{CuzziZahnle2004}.  
The transport properties of dust grains have been investigated in the context of volatile transport across ice lines \citep{Stammler2017,Booth2017,Bosman2017b}. These works have demonstrated that radial drift is sufficiently fast to transport ices across their ice line before sublimation returns the volatile back to the gaseous state. \cite{Bosman2017b} report that grains with a Stokes number of unity (the most susceptible to radial drift) have drift timescales of approximately 100 yr. The rate of sublimation (per volume) is given by \citep{Bosman2017b}: 
\begin{equation}
f_{sub} = p_x\sigma_{dust}n_{grain}N_{act}\exp\left[-\frac{E_{bind}}{kT}\right],
\label{sq:sub01}
\end{equation}
where $p_x$ is a prefactor, $\sigma_{dust}$ is the surface are of the dust, $n_{grain}$ is the number density of grains, $N_{act}$ is the number of ice layers available for sublimation (usually 2), $E_{bind}$ is the volatile's binding energy, and $T$ is the temperature of the dust and gas (assumed to be the same). Using the values from \cite{Bosman2017b}, and assuming a Stokes number of unity we estimate a reaction time ($n_{H2O,ice}/f_{sub}$) of a few 100 yr for grains crossing the water ice line. This implies that as the grain crosses the ice line, it does not immediately lose its ice layer. However, because the sublimation rate scales as $\exp(-1/T)$, the grains do not travel far inward of the ice line before losing all of its ice. In fact an increase of only 30 K \citep[radial change of about 0.1 AU for the disk model of][]{Bosman2017b} results in a reduction in the sublimation time of 2 orders of magnitude! Once in the gas these enhancements spread out through diffusion, delivering some of the volatile outward of its ice line where it refreezes onto grains - continuing a cycle of freeze out, transport, sublimation, and diffusion \citep[see also][]{RosJohansen2013}.  The end result is change in the C/O within ice lines in the planet forming regions of the disk \citep[recall figure \ref{fig:05b},][]{Booth2017}, possibly being imprinted into the atmosphere of a forming planet.

Radial drift also plays a role in dictating the ionization state of the gas, because the dust (which contributes highly to the disk opacity) is rapidly cleared from the outer regions of the disk. \cite{Crid16b} showed that  when this happens,  the region of the disk with low ionization rapidly shrinks, moving the outer edge of the dead zone inward (to $\sim 1$ AU) quickly (within 1 Myr). Because of its tendency to drive chemistry on short time scales, a rapidly changing ionization structure will have an important impact on the chemical structure of the disk.

A fully self consistent treatment of the dust physics and photochemistry has not yet been undertaken because of the technical challenge of incorporating the radial movement of material in a chemical evolving system of equations. However, some of these complexities have begun to be incorporated by \cite{Bosman2017b} for a limited chemical network dedicated to the formation and destruction of CO$_2$ in disks.  Without a full chemical network as seen in \cite{Walsh2014}, \cite{Helling2014}, \cite{Eistrup2016}, or \cite{Crid17} the detailed evolution of the C/O remains elusive.

\section{ Planet formation: pebbles, planetesimals and gas }

Although the core accretion picture of planet formation is supported by several types of evidence, gravitational instability of massive disks has long been proposed as a mechanism for giant planet formation.  In this scenario, a massive planet forms by the gravitational fragmentation of spiral arms induced in a gravitationally unstable disk \citep{Boss1997, Gammie2001, Rice2003,Rogers2012}.  Simulations tend to show that very massive Jovian planets, or brown dwarfs are the likely fragments to form in this way.  For the first time, high resolution ALMA observations a disk's velocity structure around the young star AU Aurigae have revealed such a spiral wave \citep{Speedie2024}.  In this system, the ratio of the disk to stellar mass is surprisingly high; $\sim$ 1/3. Another intriguing aspect of this result is that this process is taking place in a disk that is in a late stage of disk evolution (Class II). 

Planet formation in the core accretion picture involves growth from interstellar dust with sizes starting at $\mu$m, up to Jupiter-sized planets at  $10^5$ km, a range of 14 orders of magnitude.  Rocky planets or planetary cores can grow by accreting either pebbles or planetesimals, and current theory suggests that the former dominates the latter in the outer regions of PPDs beyond 5-10 AU, while planetesimal accretion is more important in the inner regions \citep{Drazkowska2023}.   These issues are addressed in detail by chapters by Ormel, and Klahr et al. (this volume), as well as in the overview by Armitage. 

 The distinguishing feature of pebble accretion is its aerodynamic rather than ballistic character.  Pebbles are captured within that region of the disk dominated by a planet's gravity - the Hill radius ($R_H = r(M_p/3M_*)^{1/3}$) - and settle to the planetary surface.  The Hill radius grows with orbital radius so pebble accretion will be more important in the outer disk. The condition for pebble settlement is that the interaction time between the pebbles and the planet pebbles and the planet exceed the stopping time; $t_{enc} > t_{stop}$. Equating these time scales, in the limit of thin disks, accretion rates reach their highest values (shear or Hill limit), with growth times approaching $2 \times 10^4$ yrs for Stokes numbers $St = 10^{-2}$ (see Ormel's article for details).  This is an unrealistically short time because of the depletion of the entire Hill sphere. 

Pebble accretion starts to become efficient for bodies with initiation masses of the order of Ceres and the Moon \citep{Lambrechts2014,Johansen2017}.  As a planet grows by pebble accretion, it starts to clear a gap in the disk.  This creates a pressure bump to the exterior.  Noting that the sign of the drift speed depends on the sign of the pressure gradient index $\eta$, a positive drift speed develops exterior to the planet and pebbles pile up where this cancels the (negative) disk value of $\eta$.  The relevant physical scale for gap opening is when the Hill radius equals the disk scale height $H$.  The pebble isolation mass is the planetary mass at which the inflow of pebbles from larger disk radii is halted at the edge of the gap,  cutting off the pebble flow onto the planet and starving interior regions of the disk \citep{Bitsch2018, Bitsch2019}: 

\begin{equation}
M_{iso}= 42M
_E (h/0.05)^3f_{fit}
\end{equation}

 \noindent 
 where $f_{fit}$ is a numerically computed correction factor that depends on $\eta$ and $\alpha$.  This isolation mass should also mark the maximum limit of accreted solids that could be detected in an atmosphere retrieval study as an example - thereby constraining the theoretical model.  

 The pebble isolation mass has been invoked as a mechanism for creating planetary systems with low mass planets in the interior disk regions and giants at larger orbital radii \citep{Bitsch2019}, as is the case for the Solar System.  It may explain the relatively low abundance of inner super-Earths in systems that contain outer giants \citep{Mulders2020}, although recent observations indicate that there is a positive correlation between them \citep{Bryan_Lee2024}. However, recent global, multifluid 2D and 3D simulations by \citet{Huang2025} show that such dust traps are quite leaky and that the planet isolation mass does not present an impassable barrier to dust flow to the disk interior.  Most of this dust flows to the disk interior and is not accreted onto the planet.  Such traps act as a filtration mechanism that allows the small dust to pass through \citep{Zhu_Nelson2012}.

 These results may explain why water vapour is detected in the inner disks of some PPDs \citep{Bryan_Lee2024}.  As an example, despite the presence of giant planets in the outer regions of the large PPD in the PDS 70 system, water is detected in the inner disk, as we have already noted \citep{Perotti2023, Henning_MINDS2024, Schwarz2025}, possibly carried there by a pebble flow through a leaky dust trap, from the outer ice-rich regions of the disk.  ALMA observations of compact  (as opposed to the large) disks, where one expects that the pebble flow has proceeded without too much hindrance, have detected cool water emission.  This is interpreted as arising from 1-10 AU in the disk and may be evidence of the sublimation and release of water vapour from the flow of icy pebbles through the snow line \citep{Banzatti2023}.

The maximum size that pebbles achieve through agglomeration is set by the fragmentation scale wherein dust grains can shatter due to grain-grain collisions.  A typical value for the fragmentation speed is $v_{frag} \simeq 1$ m s$^{-1}$ \citep{Stammler2022}.  Collision speeds depend upon the turbulent amplitude $\alpha$; the smaller its value the larger dust grains can become, and thus the faster their radial drift speed \citep{Birnstiel2012}.  The evaporation of pebbles as they drift inside the CO, CO$_2$, and possibly the CH$_4$ icelines enriches these innermost regions.  This combination of drift and evaporation can raise the C/O ratio of disk gas to superstellar values, which is inherited by the atmospheres of accreting planets in such regions \citep{Schneider2021}.

In the planetesimal accretion picture, 10-100 km size planetesimals collide to build rocky planets.  Growth from interstellar dust to planetesimals requires crossing the `meter-barrier' - a size where fragmentation and radial drift destroys the object faster than it can grow (see chapter by Andrews and Birnstiel) . A method for circumventing the meter-barrier is through the streaming instability \citep{YoudinShu2002,YoudinGoodman2005,Raettig2015,Schafer2017}. Planetesimals form in the filamentary structures produced by this instability when enough mass has been accumulated to become mildly self gravitating.  Simulations
by  \cite{Simon2016} found that the resulting planetesimal mass distribution scales as $ dN/dM_p \propto M_p^{-1/6 \pm 0.1}$ up to objects
on the scale of the large asteroids or Kuiper Belt objects.  Hydrodynamic simulations of the streaming instability can rapidly build up a population of planetesimals available for further growth \citep{Johansen2007,Schafer2017}.

In the classical planetesimal picture, runaway growth of planets by planetesimal collisions results from dynamical friction as well as the gravitational focusing of planetesimal trajectories by the growing planet. This process ceases when the gravitational stirring of the planetesimal cloud by the planet exceeds that of the planetesimal cloud itself.  These objects are known as oligarchs of the order of $10^{-2}M_E$ and the inner region of a PDD may have 100 - 1000 of these objects. The eccentricity of their orbits grows as the planetesimal disk is depleted. A well known problem with purely planetesimal planet formation is that the gravitational scattering process by the growing planet limits the planet's mass 

Both of these mechanisms grow a planetary core with a mass of between 5-10 M$_\oplus$, at which point the planetary core begins to accrete its gaseous atmosphere (see the chapter by Youdin). Gas accretion occurs in two phases - a slow phase where the planetary envelope remains connected to the surrounding protoplanetary disk gas, and a fast phase where the planetary envelope decouples from the disk. 
Beginning in this slow phase of accretion, the gas is first accreted into an envelope that remains connected with the surrounding disk, but slowly contracts. This contraction is limited by the rate that the envelope can radiate its energy away, and hence contracts on the Kelvin-Helmholtz timescale (\cite{IdaLin2004}). In this phase the contraction occurs at a slower timescale than the (trapped) migration timescale ($\sim 10^6$ yr), and the planet evolves nearly horizontally across the diagram. 

For masses exceeding a critical mass $M_p > M_{\rm{c,crit}}$, gravitational instability ensues and the planet's gas envelope grows by accretion from 
the disk on the Kelvin-Helmholtz timescale \citep{Ikoma2000},
\begin{equation} 
\tau_{KH} \simeq 10^c \, \textrm{yr}\left(\frac{M_p}{M_\oplus}\right)^{-d}\;.\label{Gas_Accretion}
\end{equation}
The values of parameters $c$ and $d$ in the Kelvin-Helmholtz timescale are physically linked to the opacity of the 
accreting planet's atmosphere, $\kappa_{\rm{env}}$.   This is included in the model by using the fits shown in \citet{Mordasini2014}, that relate results of a numerical model of gas accretion to the Kelvin-Helmholtz parameters for a range of envelope opacities of $10^{-3}-10^{-1}$ cm$^2$ g$^{-1}$. The fit given for the Kelvin-Helmholtz $c$ parameter is,
$ c = 10.7 + \log_{10}\left(\frac{\kappa_{\rm{env}}}{1\,\rm{cm}^2\,\rm{g}^{-1}}\right)\;.\label{KHc}$
The Kelvin-Helmholtz $d$ parameter has a more complicated dependence on envelope opacity, ranging from $\approx$1.8-2.4 over the range of $\kappa_{\rm{env}}$ considered,
the details being given by a  piecewise-linear function shown in \citet{Mordasini2014}.

Once the planet becomes sufficiently massive the contraction becomes rapid and the envelope decouples from the surrounding disk. In this rapid phase the gas accretion timescale 
($\leq 10^5$ yr) is much shorter than the migration timescale.  Alternatively, the gas accretion can be limited by other accretion rates once a gap is opened in the disk. In these models, the gas accretion is limited by the Bondi accretion rate, or the global disk mass accretion rate (see above). Because the majority of the gas mass is accreted during this fast accretion phase, changing the gas accretion prescription could result in changes in the chemical composition of the gas as it accretes onto the planet. 

\begin{figure*}
\centering
\includegraphics[width=\textwidth]{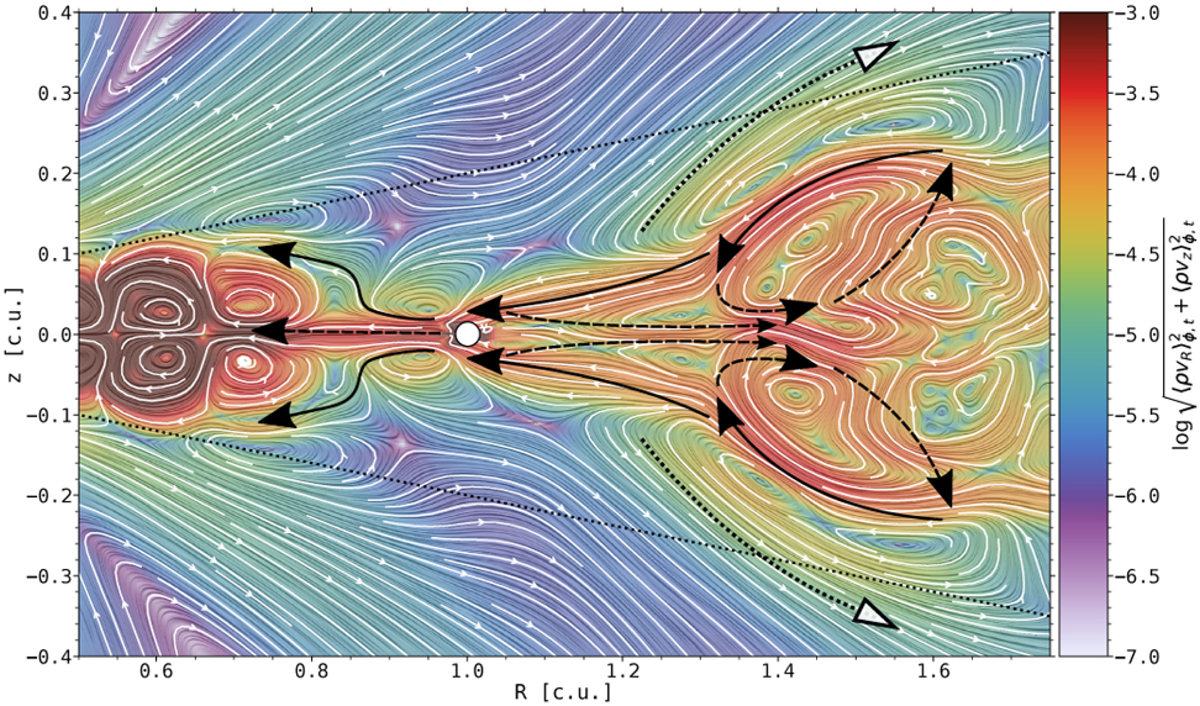}
\caption{ 3D MHD simulation of planet-disk interaction in presences of an MHD disk wind.  Background colour is log of poloidal mass flux; white stream lines and the texture in the line integral convolution (LIC) correspond to the radial and vertical components of this flux.  Planet - white circle. Black dotted lines - upper and lower disk surfaces at $\pm 4 H(R)$ Dashed black arrow - planet driven flow.  Solid black arrow - upper layer accretion flow.  Dotted black and white arrows - wind driven flows.  Figure from \citet{Wafflard_Lesur2023}, reproduced under the CC Attribution 4.0 license.}
\label{fig:Wafflard}
\end{figure*}

3D simulations of gas-disk interaction and gas accretion onto a growing gas giant play a critical role in understanding gas accretion in magnetized 3D disks.    Figure  \ref{fig:Wafflard} presents the results of the process of gap opening in a disk undergoing and MHD disk wind \citep{Wafflard_Lesur2023}.  As in other MHD simulations, magnetic field is concentrated in the gap with the often observed anti-correlation of field strength with gap density. Strong field in the gap leads to efficient angular momentum removal by the MHD wind. Although there is a complex interplay between angular momentum removal by the wind and the effects of the planet on the gas, the accretion rate through the gap region is nearly constant and near the same value as that in the disk outside of the gap.  The wind torque plays a role in planet migration, and this leads us into the next section. 

\section{Planet migration, planet traps, and gap opening}

\subsection{Planetary torques - Type I migration}

The idea that disk-planet interaction gives rise to planet migration was proposed decades ago. Planets induce the launch of waves at Lindbland resonances in the disk where the forcing frequency of the planet equals the epicyclic frequency of fluids motions oscillating around their guiding Kepler orbits (eg. \cite{GoldreichTremaine1979,Ward1986,LinPapaloizou1986}.  For torques exerted by the outer and inner Lindblad resonances the corresponding resonances are at epicyclic frequencies of $ \kappa (r) = [ (m / m + 1) ; (m / m-1) ] \Omega_p $  respectively, for waves with azimuthal wave numbers $m$.  The collection of these waves manifest themselves as leading and training sprial waves that exert torques in opposite directions, with the outer resonance (outward angular momentum transport) being the larger as it is closer to the planet due to the gas' sub-Keplerian orbital velocity - a consequence of the local gas pressure gradient. Thus, the net Lindblad torque generally leads to the planet losing angular momentum.  

The resulting migration rate is high in these models.  If the planet mass is small enough, the disk response is linear. The migration rate is then proportional to the planet and disc masses, independent of the viscosity and weakly dependent on the disk surface density and temperature profiles. This is the so-called Type I migration \citep{Ward1997}.  If unopposed, these torques would push an Earth mass orbiting at 1AU into the central star within $\sim 10^5$ years, a timescale that gets shorter as the mass of the planet increases.   A detailed analysis  of the magnitude of the Lindblad torque that fits the results of 2D numerical simulations to disk models and that includes the effects of smooth power-law behaviour of the disk surface density $\Sigma \propto r^{- s} $ as well as the disk temperature $ T \propto r^{- \beta}$ gives the following expression ( \citet{Paardekooper2011}, equations 46, 47);

\begin{equation}
\gamma \Gamma_L / \Gamma_o = -2.5 - 1.7 \beta + 0.1 s 
\end{equation}  

\noindent where $\gamma$ is an effective adiabatic index of the gas and the torque $\Gamma_o = (q / h)^2 \Sigma_p r_p^4 \Omega_p^2$ can be derived by calculating the change in the angular momentum of a fluid element perturbed in passing a planet
of mass $M_p$ for quantities evaluated at the position of the planet.   

What saves the day is the counteracting co-rotation torque. In the frame co-rotating with the planet, gas parcels close in to the co-rotation region  near the planet orbital radius,  undergo nonlinear perturbations and move along "horseshoe" orbits wherein cooler, higher  angular momentum fluid at the planet's exterior undergoes a sharp U- turn in front of the planet and swapped into an inner orbit where gas is hotter and has less angular momentum.  
The opposite occurs for inner fluid moving outwards in the U.  Since entropy must be conserved during these motions,  a density enhancement near the planet develops resulting in an outward net co-rotation torque (\cite{Ward1997},  Nelson's chapter).  Similar effects arise from the conservation of vortensity during these motions.  Finally, heating of the co-rotation region by the accretion luminosity of the forming planet can also give rise to a thermal torque that is relevant for sub-Earth mass planets \citep{Masset2017}.

The key feature of the co-rotation torque is its dependence on how angular momentum is transported in disks.  A density fluctuation in the co-rotation region will be sheared apart on a libration time scale $t_{lib}= 8 \pi/3 x_s\Omega_p$ for horseshoe orbits, where $x_s $ is the dimensionless radial half width of the horseshoe orbits at the planetary orbital radius $r_p$. This can only be replenished by the inflow of fresh material into the region.  

If viscosity is the major mechanism of angular momentum transport, then this will be the viscous timescale across this band of orbits, or $t_{vis} = (x_sr_p)^2/\nu$. The amplitude $F(p)$ of the co-rotation torque peaks when these time scales are comparable \citep{Masset2006, Paardekooper2010,Paardekooper2011}. This is measured by the saturation parameter of the torque:
\begin{equation}
    p = \frac{4}{3 \sqrt{3}}\sqrt{\frac{t_{\nu}}{t_{lib}}}
\end{equation}
\noindent
In the limit that the viscous time scale greatly exceeds the libration time, which occurs as the mass of the planet increases sufficiently ($p >> 1$), the co-rotation torque vanishes or "saturates". At these higher masses, the Lindblad torque dominates once again and will be arrested once gap opening occurs.

The co-rotation torque in the unsaturated regime \cite{Paardekooper2010} takes the form ( (their equations 45, 46):

\begin{equation}
\gamma \Gamma_{HS} /  \Gamma_o = 1.1 ({3 \over 2} - s) + 7.9 (\zeta / \gamma) 
\end{equation}

\noindent where $\zeta = \beta - (\gamma - 1) s$ is the power law exponent of the entropy profile of the gas, and 
the first and second terms address the effects of entropy and vortensity.  

The co-rotation mass $M_{corot}$ is the planetary mass at which the co-rotation torque attains its highest value and corresponds to  the mass at which the amplitude of $F(p)$ peaks.   An analytic expression for this mass in the case of viscous disks, which is a good match to these numerical simulations, is given by \citet{Speedie2022}:

\begin{equation}
M_{p, \nu,corot}=4.43 \left (\frac{\alpha}{10^{-3}} \right)^{2/3} \left (\frac{h}{0.05} \right)^{7/3} M_E 
\end{equation}

In the case that the disk angular momentum is carried off by disk winds, the replenishment of material into the co-rotation region takes place not on a diffusion time scale, but on the advective time scale through the region based on the disk wind driven flow speed: $t_{w}= x_s r_p/ v_m$, which replaces the diffusion time scale in equation 13. \citet{Kimmig2020} simulated the migration of planets in disks undergoing MHD disk winds, although MHD was not explicitly included in the simulation.  They showed that inward or outward motion of planets could be controlled by the strength of the disk winds.  Specifically, when the parameter $K=t_w/t_{lib} \ge 10$, then the co-rotation torque was strong enough to drive the planet outward in the disk (see also \citep{Speedie2022}). 

In MHD wind driven co-rotation regions where flows are essentially inviscid, \citet{McNally2017}, review in Nelson's chapter) show that the shape of the horseshoe orbit region near the planet can be modified by the winds, leading to a more complex,  "history-dependent" evolution of the horseshoe torque.

\subsection{Planet traps}

One of the main results of the early population synthesis studies of planet formation  arose when the effects of Type I planetary 
migration  \citep{IdaLin2008} were considered.  Synthesis studies of planets migrating in evolving,  standard, Shakura-Sunyaev smooth disks showed that rapid loss of such bodies occurs within $10^5$ yrs.   The model introduced a parameter - a slowing down factor - needed in order  to match predicted and observed populations in the M-a diagram.  The result was that standard migration in smooth disks needed to by slowed down by a factor of 30-300 (see Nelson's review) - i.e. - slowed to speeds more reflecting viscous evolution of the disk.  The theory of planet traps, sketched below, provides a physical solution for this problem. 

The total torque that a low mass planet undergoes dying Type I migration is  the sum of these two the Lindblad and horseshoe co-rotation torques; $\Gamma = \Gamma_L + \Gamma_{HS}$.  Thus low mass planets that migrate without opening a gap may move inwards, outwards, or be caught in planet traps where the net torque vanishes $\Gamma = 0$ \citep{Masset2006, HP11, HP12}.  These traps would still move inwards in disks at the viscous diffusion or advective speed rates.  A planetary core trapped there moves with the trap, and accretes materials at the trap position as the latter moves through the disk.  There is a range of masses that can be trapped in regions of zero net torque.  Detailed simulations by \cite{ColemaneNelson2016} examined the growth of planets that grow in traps caused by radial variations in the 
disk.  Their results showed that null points for the torque can trap planets up to 10s of Earth masses.  For the more massive planetary cores, being released from the trap will, with rather little additional mass accretion, result in gap opening and the transition to slow  Type II migration.

We examine several of the major traps for Type I migration in disks:

(i) Heat transition trap:  The values of the power law indices appearing in the combined torque formula  have been worked out for \cite{Chambers2009} disk
models \citep[see][]{Crid16a} for an adiabatic index of $\gamma = 1.4$. As has been observed by several authors the net torque is outward in the inner viscous part of the disk and inward in the outer  radiatively heated regime.  This implies that there is a radius of net zero torque $\Gamma = 0$  at the heat transition radius $r_{HT}$ discussed above  \citep{Lyra2010,HP11,Dittkrist2014}.   Thus a low mass planet can be trapped at the heat transition, and this cuts off its rapid inward migration.     The evolution of $r_{HT}$ with time is governed by the disk evolution equation discussed above, and in particular depends on the accretion rate, which falls as a function of time as the column density falls, and later as the disk undergoes photo evaporation.   These motions are much slower functions of time than the Type I migration time scale because they reflect the slower viscous evolution of the disk which is responsible for the falling accretion rate $\dot M_a $.    

(ii) Dead zone trap:   Noting that the dead zone must have very low levels of MRI turbulence, while the active regions at larger radii can support active MRI, there is a discontinuity in the dust scale height as one proceeds through the outer dead zone radius .  For disk radii   $ r \le r_{DZ} $,  dust will rapidly settle into the disk mid plane whereas at $ r \ge r_{DZ}$, turbulence will keep the dust stirred up to higher scale heights.   This means that radiation from the star will see a "wall of dust" at $r_{DZ} $ which will reflect and, as for a garden bed in front of a sunlit wall, will be heated by the back-scattered radiation.   This alters the 
temperature profile of the gas at smaller radii, in such a way as to create a planet trap \citep{HP11,Crid16a}.

The position of the dead zone, as we have seen, depends on the ionization of the disk and hence is directly connected to ionization driven disk chemistry. A planet trapped at the dead zone radius migrate with the trap's evolution, acquiring a composition reflecting the disk materials encountered as the trap moves inwards through the disk. 

(iii) Ice line traps:  The entire class of ice lines that result from the freezing out of various chemical species on grains.  As has already been noted, three of the potentially most important ice lines are those of water, CO, and CO$_2$.  In its most general form,  astrochemical models for  the disk predict the distribution of ices (eg. water).  These results can then be used to compute the change in dust opacity in the disk.  This opacity change has a direct effect on the temperature profile across the ice line, which in turn sets the direction of the torque by the temperature dependence of equation 9. In order to have a sufficient large change in the opacity across an ice line, one can anticipate that the relevant volatile must be abundant - as is the case for water \citep{Miyake1993}.   Water ice lines have accordingly been suggested as trapping points for planets \citep{IdaLin2008b,HP11}.

Why does an ice line act as a potential trap?  At the ice line, the opacity $\kappa$ is reduced as the dust
grains are coated with ice, and the associated cooling rate of the gas is increased \citep[since the cooling rate $\propto 1/ \kappa $][]{Bitsch2013}. Coupled to the cooling rates, the local temperature and thus the disk scale height, is reduced ($H = c_s \Omega  \propto T^{1/2} $) . Since the disk accretion rate is  constant  across the disk at any instant, and because the viscosity is dependent on the local gas temperature ($\nu \propto c_sH \propto T  $), a reduced temperature results in an enhancement in the local surface density at the ice line to maintain a constant mass accretion. Both the modified temperature and density gradients impact the net torque, resulting in a trap \citep{Crid2019a}.

Further analysis of opacity effects at ice lines 
indicates that water is sufficiently abundant $ 1.5 \times  10^{-4} $ 
molecules per H) to trap planets at its ice line due to an opacity transition. Volatiles that have mass abundances lower than a factor of $\sim$40 with respect to water do not result in a sufficiently strong opacity transition to trap planets in a disk that is viscously heated as shown in these numerical simulations.   These results suggest that CO, while sufficiently abundant, also does not trap planets at its ice line.  Like water, the CO$_2$ ice line occurs in the viscously heated part of the disk and hence could act as a trap. However it is not easily produced in the gas phase, and hence its availability as a trap depends on the choice of initial conditions (dark cloud vs. diffuse ISM chemistry) and chemical network (gas only vs. gas-grain chemistry).

We reach an interesting and important conclusion about planetary migration that has direct consequences for both planet
formation and composition.   Low mass planetary embryos move along with traps that move inwards through the disk
as it evolves on a viscous time scale.  Planets eventually break away from their traps when they become sufficiently massive as shown in numerical simulations of \citep{ColemanNelson2014} - typically up to 10 Earth masses.    
The solid materials accreted during this time reflect the composition of the evolving disk visited by the relevant trap.
The heat transition, being typically the furthest out in the disk, is beyond the ice line and so planetary cores can be
expected to have a strong contributions of ices \citep{APC16a}.  A core building at the ice line would be expected
to have less ice, and a dead zone, which is often inside the ice lines, would be expected to have a very small ice content.
Detailed simulations bear out these general results (see subsection on solids).  Although the dynamics of embryos as they approach planet traps has not yet been investigated in any detail, it is likely that a chain of embryos can come into mean motion resonances with a trapped planet.

\begin{figure*}
\centering
\includegraphics[width=.75\textwidth]{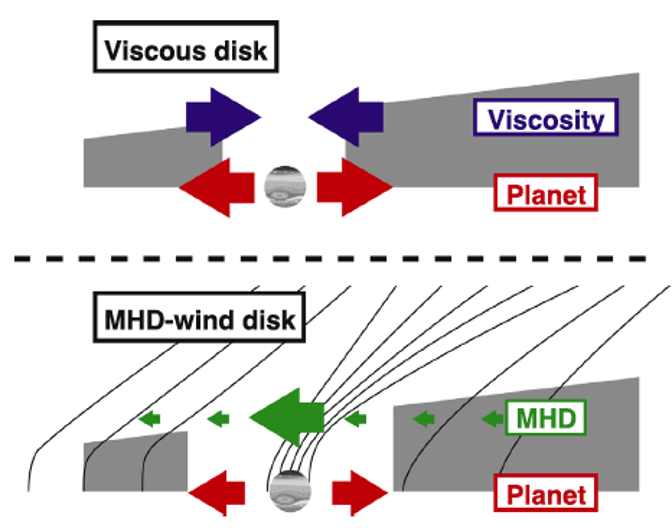}
\caption{Comparison of planet-induced gap opening in viscous vs MHD disk wind dominated disks.  Figure from \citet{Aoyama_Bai2023}, reproduced under the CC Attribution 4.0 license.}
\label{fig:Aoyama_Bai2023}
\end{figure*}  

\subsection{Type II migration and gap opening}

As the mass of a planet increases, its angular momentum exchange with the disk can lead to the 
opening of a gap, a process called Type II migration \citep{Ward1997}. The gap opening mass and gap properties depend on how angular momentum is transported.  The most thoroughly analyzed case in the literature is for viscous PPDs.  The contrast between the forces that open and close a gap in the presence of a planet are illustrated in Figure \ref{fig:Aoyama_Bai2023}.

 The `gap opening' mass requires the planet's Hill radius to be larger than the pressure scale height at the planet's location, otherwise the gap will be closed by gas pressure. A second requirement is that the torques from the planet on the disk equal or exceed the torques caused by viscous stress which fill in the gap. These requirements are summarized by, \citep{LinPapaloizou1993,HP11},
 
 \begin{equation}
{M_p \over M_* } =  min[ 3 h_p^3, \sqrt{ 40 \alpha_{turb} h_p^5} ]
\end{equation} 

\noindent where $h_p = H / r_p$ is the disk's aspect ratio at the orbit of the planet. The depth and width of the gap depend on this balance \citep{PapaloizouLin1984}.  The 
Lindblad resonances that drive the disk-planet angular momentum exchange fall within the opening gap and therefore the migration rates are drastically reduced compared to Type I.  The planet then becomes locked to disk migration at the radial velocity of $u_r = \nu / r $, and the disk is essentially split into an inner and outer region.   This result is based on the assumption that gas doesn't enter the gap once formed.  This is in fact too simplistic a view since horseshoe orbits can readily facilitate a flow through the gap.   Numerical studies carried
out by \cite{CridaMorbidelli2007} and \cite{Edgar2008}, have been generalized by \cite{Duffell2014} who showed that Type II migration can, on this
basis, be faster or slower than the viscous rate depending on disk parameters such as the turbulent Mach number.   

A more refined picture of gap opening in viscous disks takes into account the fact that as material piles up on the edge of the gap, the steep pressure gradient causes a departure from Keplerian rotation of the gas.  Also, waves that carry the angular momentum from the planet are not instantaneously damped, but carry angular momentum to larger distances thereby widening the gap.  Taking these effects into account, \citet{Kanagawa2015a, Kanagawa2015b} derived an expression for the depth of the ratio of the column density of gas in the gap to that of the undisturbed disk;
\begin{equation}
    \frac{\Sigma_{gap}}{\Sigma_o} = \frac{1}{1 + 0.04K}
\end{equation}
\noindent
where $K=q^2h_p^{-5}\alpha^{-1}$ is a non-dimensional parameter and $q\equiv M_p/M_*$. ALMA observations can detect gaps at the level of $\Sigma_{gap}/\Sigma_o \simeq 0.5$, which when applied to this formula recovers the opening mass scaling; $q_{crit}\propto h_p^{5/2}\alpha^{1/2}$ \citep{Armitage2020}. This formula can be used to predict the masses of planets that may produce observed gaps: a planet mass of 0.4 M$_{Jup}$ could explain the gap with a dept of 1/3 at 30 AU in HL Tau \citep{Kanagawa2015b}.

Unsurprisingly, MHD disk winds operate differently than viscous torques in opening a gap.  First of all, the inflow in such inviscid (low viscosity) disks is nearly laminar, requiring special care in analyzing the time dependence of the specific angular momentum as the gap is opened  \citep{Cordwell_Rafikov2024}.  An important planetary mass scale in gap opening is the thermal mass $M_{therm} \equiv h_p^3 M_* = c_s^3 / G \Omega_K $ where the second identity follows from the condition of hydrostatic balance of the gas; $H = c_s/ \Omega_K $.  The waves excited by planets below the thermal mass fail to shock, damp, and deposit their angular momentum in this region.  The results show that gas that is normally trapped in a ring in the co-orbital region is gradually evacuated, while the shape of the gap has a universal profile that grows linearly in time.  In full 3D MHD simulations of \citet{Aoyama_Bai2023}, a planet is introduced into magnetized MHD PPD.  The flux concentration in the gap, already noted in the absence of a planet, is a factor of 2-4 greater than that in the surrounding disk.  Strong angular momentum extraction from the disk occurs, that drives a greater advective flow speed in the gas, as shown with the larger green arrow in Figure \ref{fig:Aoyama_Bai2023}. The gap is much deeper than in the purely viscous or advective cases.  At the same time, the outflowing wind is driven from the surface of the disk just exterior to the gap region.  Finally, the migration (co-rotation) torque is primarily negative (inward) in these MHD simulations - the reason perhaps being due to short wind time scales in the region.  Explicit formulae for MHD gap opening masses have also been derived from simulations by \citealt{Wafflard_Lesur2023}.

\subsection{Visualizing planetary migration:  torque maps}

\begin{figure*}
\centering
\includegraphics[width=\textwidth]{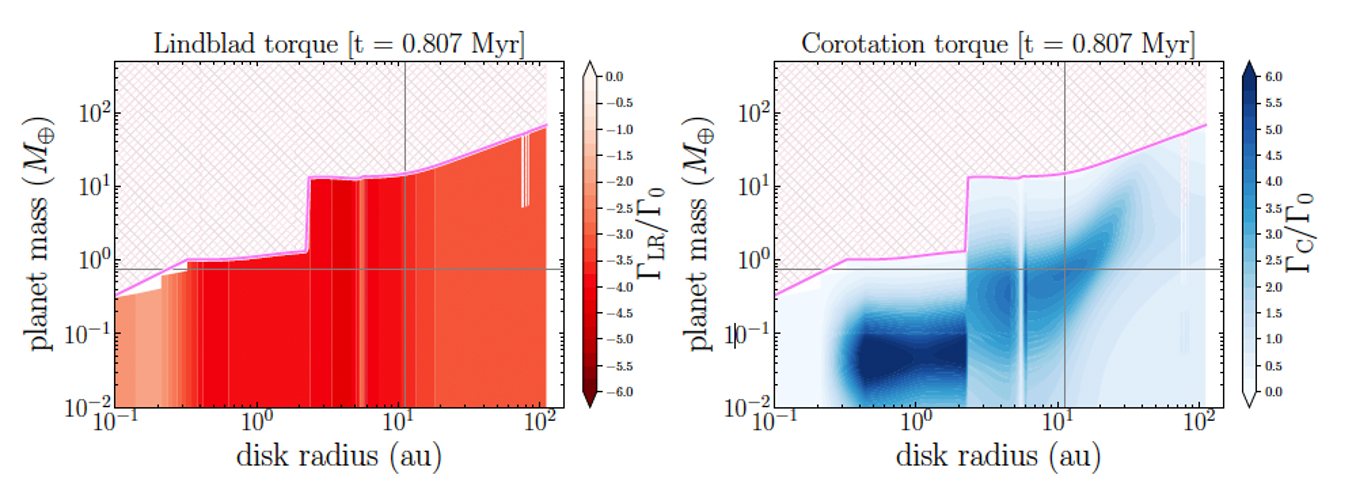}
\caption{Snapshot of  Lindblad (left) and co-rotation (right) torque maps at time t=0.807 Myr in an evolving PPD. Red is inward directed motion and blue is outwards.  The top pink line demarcates the domain of Type II migration (above) from Type I (below). Figure from \citet{Speedie2022}, reproduced with permission \textcopyright OUP}
\label{fig:Speedie_2torques}
\end{figure*}

\begin{figure*}
\centering
\includegraphics[width=\textwidth]{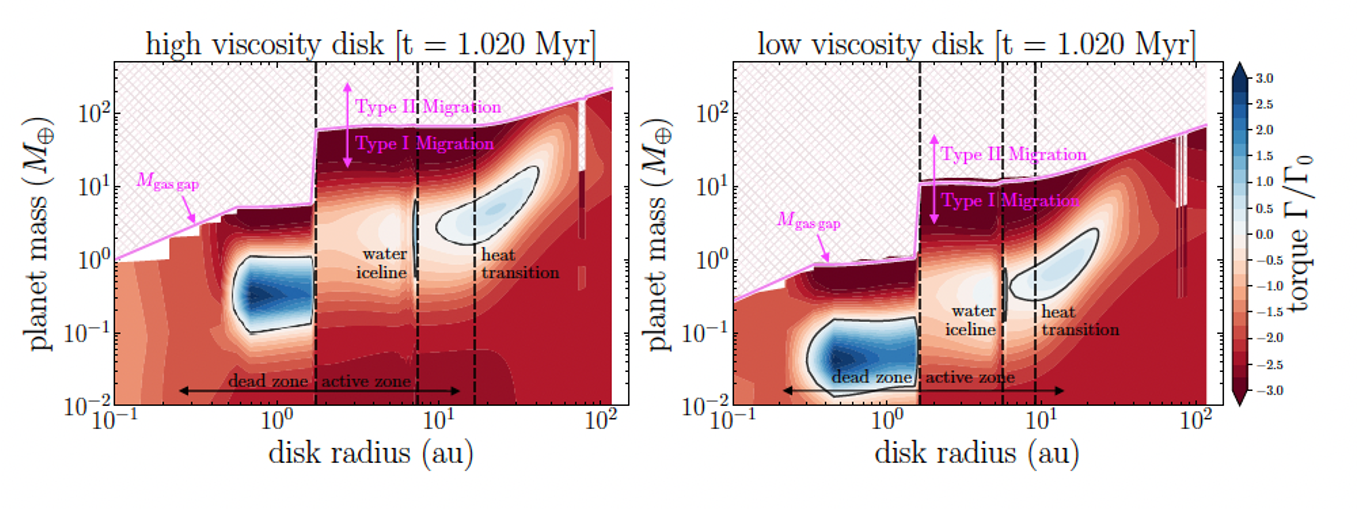}
\caption{Snapshot of the total torque for two different viscosities; $\alpha= 10^{-3}$ (left), and$\alpha= 10^{-4}$ (right) torque maps at time t=1.02 Myr in an evolving PPD. Red is inward directed motion and blue is outwards.  White indicates a regime of net zero torque $\Gamma = 0$ that corresponds to a planet trap. Three different traps can be seen; the innermost dead zone, followed by the water ice line at a larger disk radius, with the outermost heat transition trap seen as a more extended region in disk radius. The top pink line demarcates the domain of Type II migration (above) from Type I (below). Figure from \citet{Speedie2022}, reproduced with permission \textcopyright OUP.}
\label{fig:Speedie_nettorque}
\end{figure*}

A most useful approach to visualizing and quantifying the migration of accreting planets in any disk model and for any type of angular momentum transport mechanism, is to create a "heat map" of the net torque that the disk exerts on a planet at any disk radius for each value of the planetary mass \citep{ColemaneNelson2016}. As we have seen, the inward directed Lindblad torque can be offset for some range of planetary masses by an outward co-rotation torque.  As the planet grows in mass, it experiences Type I migration, and becomes trapped in one or more of the various traps previously discussed.  When it reaches a gap opening mass, the torque is sufficient to open a gap.  Forming planets in an evolving PPD execute distinctive tracks in this $M_p - r_p$ space, where the direction and magnitude of the net torque are shown as background colours - red for inward directed motion, and blue for outward.

Figure \ref{fig:Speedie_2torques} presents such a torque map from the study of planet formation and migration in evolving PPDs of \citet{Speedie2022}.  This paper tracks planet formation in a background evolving analytical model for PPDs by \citep{Chambers2009}.  It includes evolving disk chemistry \citep{Crid16a, Crid2019a} which allows for the self-consistent computation of ice lines, dead zones, and heat transitions.  The left panel shows snap shots of only the inward directed Lindblad torque, and the right panel only the outward directed co-rotation torque.  Note the peak in the co-rotation torque that matches the analytic prescription for the co-rotation mass $M_{p,\nu,corot}$ in \citet{Speedie2022}.

From that same paper, Figure \ref{fig:Speedie_nettorque} shows the net torque that is the sum of the panels in the previous Figure \ref{fig:Speedie_2torques}. The colour scale gives the magnitude and direction (red vs blue dominant colour) of the net torque.  The various planet traps show up as regions of white colour; specifically the outer dead zone, water ice line, and heat transition traps.  The two panels show the torque maps for high ($\alpha=10^{-3}$) and low ($\alpha=10^{-4}$) disk viscosities. Since the co-rotation mass depends upon the disk viscosity, we see that the positions of peak co-rotation mass and the corresponding planet trap position are offset with respect to one another. 

\begin{figure*}
\centering
\includegraphics[width=\textwidth]{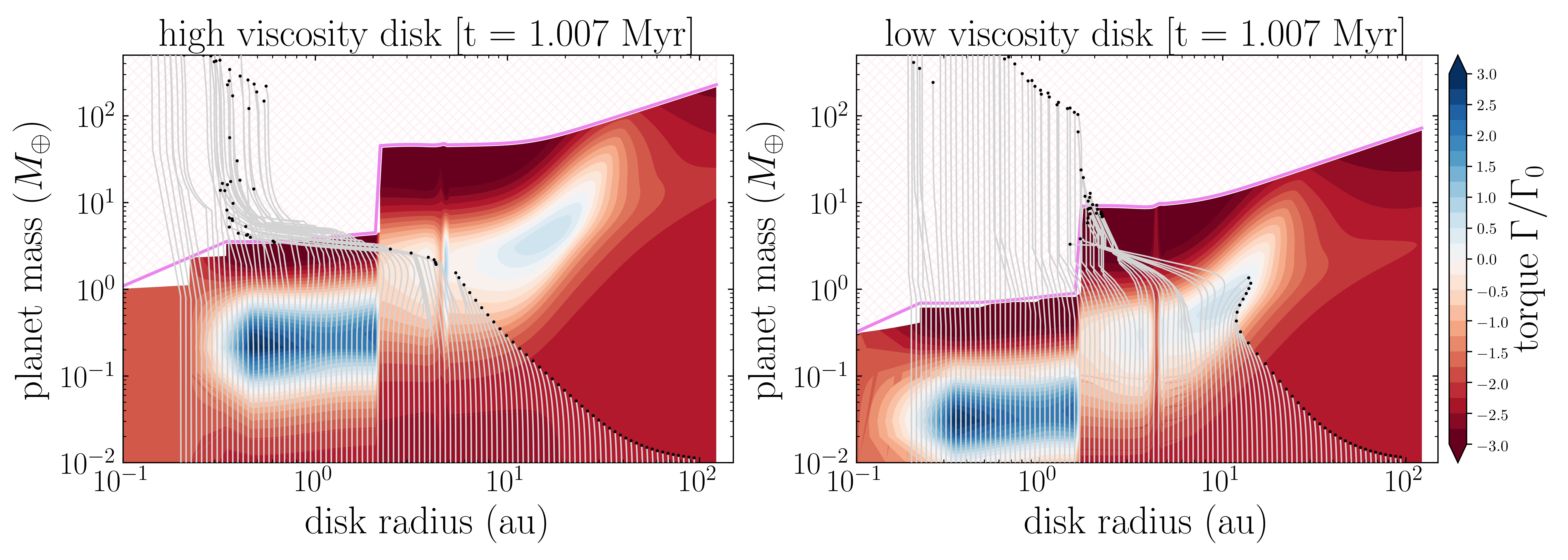}
\caption{Snapshots of planetary evolution tracks superposed on torque maps for high (left) and low viscosity PPDs (right) at time t=1.007 Myr in an evolving PPD. Tracks in the outer region of the low viscosity PPD move outward to 20 AU or more as they are caught in the heat transition trap, whereas this does not happen in the high viscosity case  Figure from \citet{Speedie2022}, reproduced with permissio \textcopyright OUP.}
\label{fig:Speedie_tracks}
\end{figure*}

The evolution tracks of forming planets in these PPDs are superimposed in the net torque map shown in Figure \ref{fig:Speedie_tracks}.  In the low viscosity case, tracks of forming planets can be pushed outward in disk radius as they get caught in the co-rotation region associated with the heat transition trap.  Tracks upon entering the TypeII migration regime tend vertically upwards as planets undergo rapid gas accretion and gap opening. 

Of particular interest in these results is the bifurcation of planet tracks that occurs in the low viscosity disk - that is the splitting of tracks into inwards and outwards directed migration.  The latter occurs because accreting planets get caught in an outward directed co-rotation torque, which is greater in more massive disks.  Eventually, these outward directed planets move inwards as they become massive enough to leave the trap and experience the inward directed Lindblad torque once again. This behaviour may help explain why higher mass Jovian planets at large disk radii occur in more massive disks \citep{vanderMarel2021}.

\section{Roadmap for a new paradigm: MHD disk winds and planet formation}

 A brief summary of the specific ways that disk winds affect planet formation, as discussed in the review, is given below:

$\bullet$ Disk formation - truncation of disk radius by magnetic braking and tower flow

$\bullet$ Disk angular momentum transport – via rotating jet/wind.

$\bullet$ Accretion rate -  driven by disk wind torque
 
$\bullet$ Disk flow dynamics -  advective laminar flow, not viscous spreading;  

$\bullet$ Dust/pebble growth -  enhanced in low turbulence, and nearly laminar flow

$\bullet$ Ring-gap structure - before the appearance of planets, induced by MHD disk winds, which may be sites of first planet formation

$\bullet$ Type I migration:  –  co-rotation torque depends on wind outflow rate.

$\bullet$ Type II migration: – planetary gap opening against advective flow rather than viscosity.

$\bullet$ Planetary populations – disk winds create observable effects in planetary populations, as can be seen in M-a, and M-R diagrams.  This will be addressed in the next section. 

Taken together, disk wind driven disk evolution impacts all aspects of planet formation and composition through its central role in disk evolution and planetary migration. We now turn to addressing how these issues translate into building models of and characterizing planetary populations.

\section{Population synthesis: chemical evolution, disk winds, accretion, and migration}

The pioneering work of \citet{IdaLin2004} used a statistical Monte Carlo approach to create planetary populations in the M-a diagram.  This was achieved by incorporating the best available physical models for planet formation into the computation of the evolution tracks for thousands of planets under different initial PPD conditions.  This was followed for stars of different metallicities and masses \citep{IdaLin2004b,IdaLin2005}, more comprehensive treatments of various migration processes \citep{IdaLin2008},  and an examination of the ability of the ice line to act as a potential migration trap \citep{IdaLin2008b}. Investigations of the effects of stellar masses on planet populations were carried out by \citet{Alibert2011}.  The differences between the predicted and observed populations offered insight into how theory needs to be further developed \citep{Benz2014}.  The reader is encouraged to read the comprehensive review by Burn and Mordasini (this Volume) for a comparison of recent population synthesis models.    

The basic components of an end-to-end theory of planet formation that also includes the chemical composition and angular momentum transport by disk winds are briefly summarized below.  Given that entire populations must be computed over the 3-10 Myr lifetime of disks only 1D models, informed by more detailed 3D MHD simulations for details of local planet formation processes,  are feasible.  This is a major constraining factor in designing the population synthesis simulations, which are constructed as follows:

-  (i) Adopt a model for the structure and evolution of protostellar disks, from the initial conditions (reflecting their formation), through disk evolution due to the proposed mechanism of angular momentum transport, to the end phases in which photo evaporation of disks leads to their rather quick demise 3 - 10 Myr after their formation.  Equations 4 and 5 as an example, treat gas evolution under combined viscous and MHD wind torques.  These must be supplements by equations that consider the back reaction on the magnetic fields.  

-  (ii) Prescribe the evolution of solids within such disks.   Dust may arrive in the disk during 
disk formation by the collapse of an initial  protostellar core, or later introduction via accretion streamer.  In any case, it is subject to a condensation sequence wherein minerals appear at different disk radii depending on their condensation temperatures in the PPD.   Subsequent dust settling into the PPD midplane leads to rapid coagulation while at the same time undergoing radial drift due to drag forces. Radial drift changes the dust to gas ratio throughout the disk and will help dictate where planets may form (see chapter by Andrews and Birnstiel).  Dust evolution in such disks has been incorporated in various recent codes, such as DustPy \citep{Stammler2022}.

- (iii) Prescribe the solid accretion process.  These processes most likely take place within dust traps where the column density of dust can increase to values exceeding the gas column density as it is trapped in pressure bumps.  In the absence of pre-existing planets, these traps could arise in MHD wind-induced pressure bumps.  Accretion is mediated by collisions of planetesimals to build oligarchs, pebble accretion onto rapidly growing bodies, or a combination of these which most likely depends upon the orbital radius of the dust trap (eg. \citet{Lau2022}).  The composition of these materials will play a basic role in determining the composition of the rocky planets and )super-Earths.

- (iv) Prescribe the migration of embryos and forming planets. Theories of migration for bodies with masses that are too small to open gaps (Type I migration) have mainly focused on Linblad  and co-rotation torques prescribed for viscous disks. This needs to be generalized to disk wind torques, a topic of current interest in 3D MHD simulations.  Real disks also have inhomogeneities in temperature and densities, and these prove to be crucial in providing zones of "zero net torque" or planet traps.  Dust and planet  traps are often co-spatial (eg. the water ice line, and dead zone boundary).

- (v)  Prescribe how gas accretion takes  place onto sufficiently massive cores.  This involves accretion runaway that quickly build Jovian planets.  The composition of gas accreted during this phase will determine a great deal about the properties of the Jovian atmospheres.   The latter is best followed using time dependent gas chemistry codes \citep{Fogel2011,Helling2014,Crid16a,Eistrup2016}. Accretion of solids into the atmosphere will also strongly affect the final atmospheric C/O ratio of the planet.

- (vi)  Prescribe the mechanism for gap opening.  In disk-wind dominated disks, gaps are opened in opposition to advective flow driven by the wind.  This is followed by Type II migration and the end of accretion from the disk.  The late accretion from planetesimals may affect the chemical composition of the atmospheres.  Gap opening in MHD wind dominated disks provides a new criterion for gap opening in which the local disk wind strength, not disk viscosity, controls the process.  

- (vii)  The end of planet- gas disk interaction arises as photo evaporation of the tenuous disk sets in.  This does not yet mark end of planetary chemical enrichment of atmospheres. 

- (viii) The dynamical evolution of the gas-free planetary system in which planet-planet interactions will rapidly lead to high eccentricities, the loss of mean motion resonances, and probably the loss of some planets (see chapter by Petit et al., this Volume). The scattering of planetesimals onto colliding trajectories with gas giants may lead to considerable metal enrichment in Jovian atmospheres.  

- (ix) The structure of a planetary atmosphere depends on its pressure-temperature (P-T) profile as well as its chemical composition.  For massive planets, this is usually computed using chemical equilibrium models based on the elemental abundances of gas and solid materials delivered to the forming atmosphere during its formation.  For super-Earths, the composition of secondary atmosphere that forms due to outgassing will depend on the accretion history as well, as volatiles undergo outgassing from the newly formed planet.

In order to connect the end-products of this population synthesis program with the observed composition of the resulting populations,  the chemical composition of the material that is accreted during the planet formation process can be tracked and computed with the kind of machinery that has been discussed here. Such models are called either `end-to-end' or `chain' models because they link physical and chemical models together in succession. In general, the construction of such end-to-end models follows a scheme like this: {\it disk model} $\rightarrow$ {\it chemical model} $\rightarrow$ {\it planet formation \& migration}. Variations and extensions of this general chain have been made, including a dust evolution model \citep{Crid16b} and a planetary atmosphere model which includes the generation of synthetic spectra \citep{Mordasini16}. The chemical model can be constructed empirically \citep{Mordasini16}, through chemical equilibrium models \citep{APC16a}, or from time-dependent chemical models \citep{Crid17}, depending on the focus of the chain.

Several important differences arise in the resulting populations depending on the physics of planet formation. First, there differences in the populations produced by planetesimal and pebble accretion.   \citet{Brügger2020} addressed this issue and found that population differences depended  upon the accretion model adopted. In the pebble accretion picture, a lot depends on timing - specifically when the first embryo appears in the disk. The earlier it appears, the more massive will be the resulting planet up to 1000 Earth masses. Planetesimal accretion by contrast, is less sensitive to start time and depends strongly on its radial distance from the star in the disk - with very low rates of growth occurring at the largest distances. By following both accretion and migration, these authors found that pebble accretion produced mainly super-Earths because Type I migration of planets was extremely efficient.  This latter effect is due to the very rapid increase in core mass up to the isolation mass.   Giant planets tended to appear more frequently in populations built by planetesimal accretion.  

Recent complete population synthesis models have also included different treatments of solid accretion; by planetemals \citep{Emsenhuber2021, Kimura_Ikoma2022} - the former representative of the Bern population synthesis model; and by pebbles \citep{Brügger2020, Bitsch2015}.  These are summarized by Burn and Mordasini (this volume).  All four have included Type I and II migration, disk and stellar irradiation heating, and similar values for disk viscosity $\alpha \simeq 10^{-3}$.  In general, these four models do not fit the observed distributions in the M-a diagram although the separation in planet masses between 10-100 $M_E$ is generally recovered. 

\begin{figure*}
\subfloat[Representative planet evolution tracks assiciated with 3 different planet traps, in the M-a diagram for disk wind evolved planetary populations: the case of high tubulent amplitude $\alpha=10^{-3}$.]{
\includegraphics[width=0.5\textwidth]{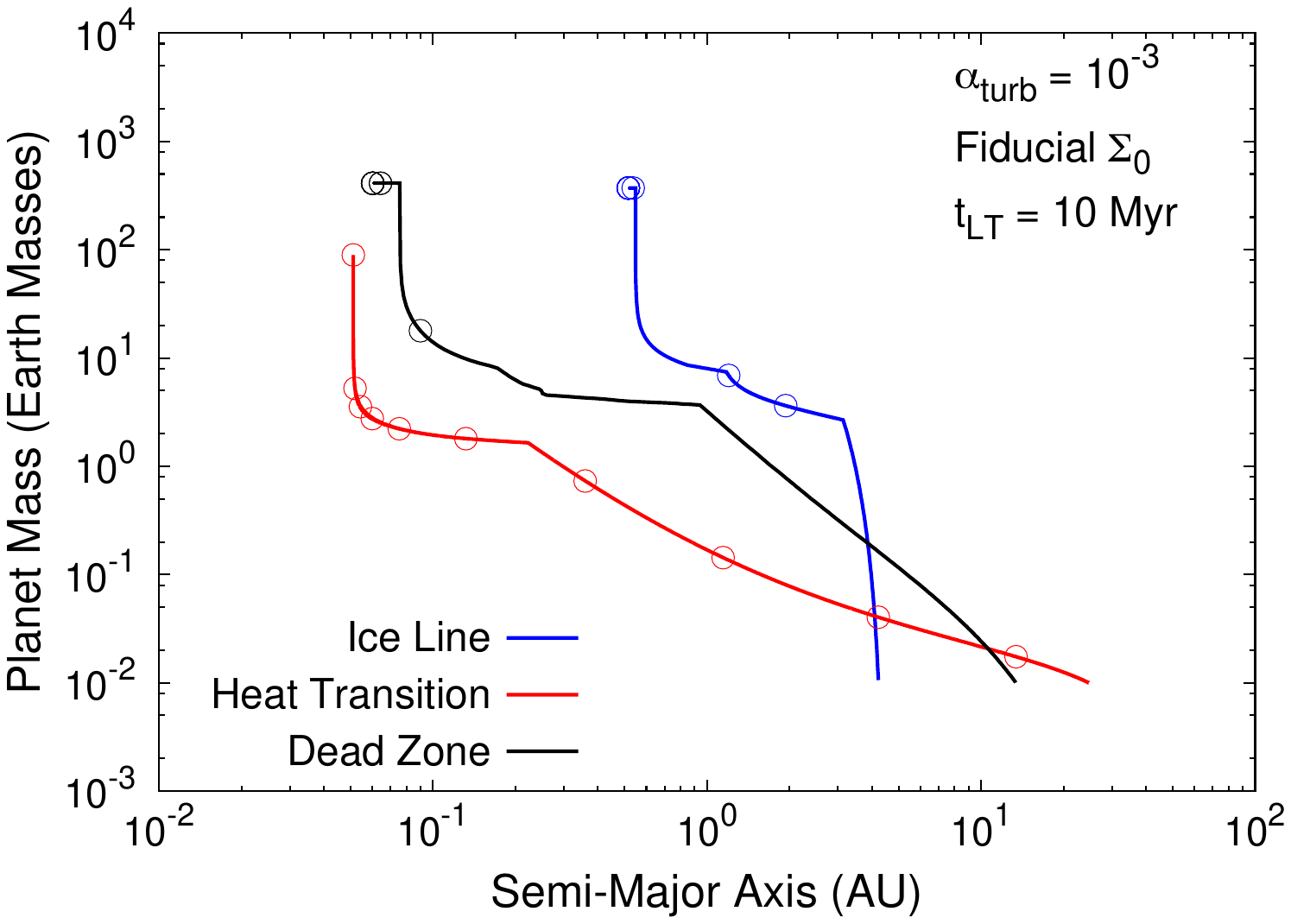}
\label{fig:AP_tracks_highalpha}
}
\subfloat[Representative planet evolution tracks associated with 3 different planet traps, in the M-a diagram for disk wind evolved planetary populations: the case of low tubulent amplitude $\alpha=10^{-4}$]{
\includegraphics[width=0.5\textwidth]{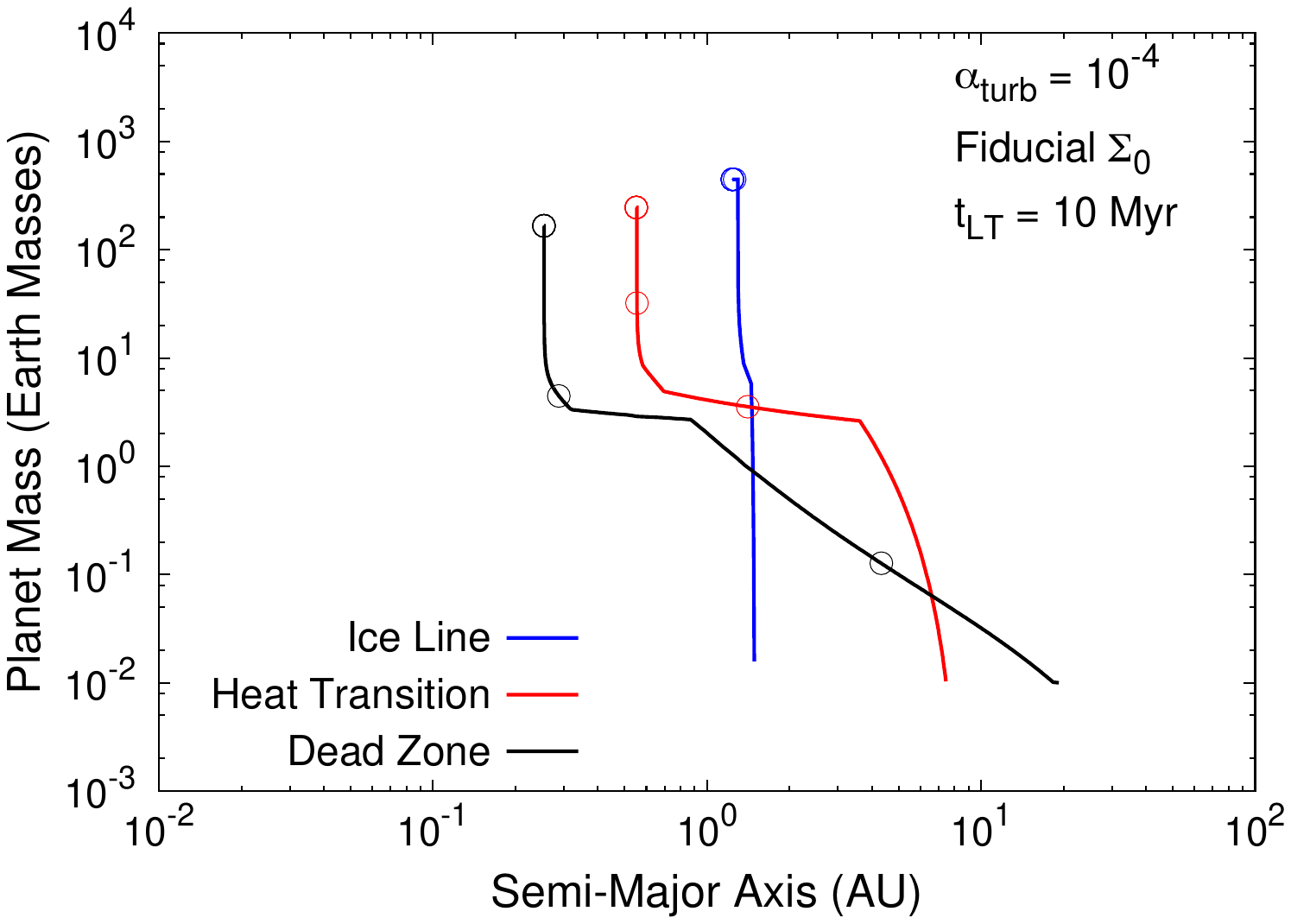}
\label{fig:AP_tracks_lowalpha}.
}
\caption{Planet evolution tracks - for giant planet formation - in the M-a diagram, for MHD disk winds and two different turbulent viscosities.  Figures from \citet{Alessi_Pudritz2022}, reproduced with permission \textcopyright OUP.}
\label{fig:AP_tracks_2022}
\end{figure*}

\subsection{The effects of MHD disk winds on planetary populations}

Another important difference that arises between planet formation theories and the populations they predict is in how angular momentum is transported in the PPD phase.  Although most publications focus on viscous transport, the \textit{McMaster population synthesis model} (\citep{Crid16a, APC16a, Alessi_Pudritz2022} is distinguished  by the incorporation of detailed disk astrochemistry, planet traps, and angular momentum transport by MHD disk wind. In this subsection we summarize the results of the population synthesis of planet formation in disks governed by both turbulence and MHD disk winds \citep{Alessi_Pudritz2022} whose column density evolution is described in equation (4) \citep{Chambers2019}. In this work, planetesimal accretion is assumed for the growth of embryos, and planetary migration is followed in an evolving astrochemical environment. Population synthesis calculations are performed for thousands of individual planets evolving in the M-a diagram, in which key disk parameters are varied through Monte Carlo simulations.  In the final step, the final planetary radii and atmospheres are computed given the composition of the accreted materials and gas, the final results being plotted in the M-R diagram.  Disks are evolved for 10 Myr.  The wind mass loss rate that needs to be fixed in the disk wind model is assumed to be $\dot M_w/\dot M_a \simeq 10^{-1}$ as is observed for most disk winds \citep{Watson2016, Pascucci2023}.

Figure \ref{fig:AP_tracks_2022} shows individual planetary evolution tracks in the M-a diagram, typical growing planet in three
traps indicated by the colour code.  Tracks are shown for models of high (left panel) and low viscosity (right panel) disks, while most of the disk angular momentum is being carried off by the disk wind.  The size and evolution of dead zones in disks depend on the type of 
disk ionization that dominates.  In this model, a constant dust to gas ratio is assumed for the gas.
 Shown in the figure are tracks for planets in dead zone traps,  for disks ionized primarily by X-rays.  
In general, planets generally evolve from right to left, as their traps move inward in the evolving disk. This migration time is typically of the order of Myr.  Planetary growth by solid or gas accretion moves the planet from the bottom to top of the diagram.   Planet evolution tracks start in planet traps because the embryos  migrate very quickly until
they encounter a trap.   They then  begin their growth in the Oligarchic stage \citep[see][]{KokiboIda2002,IdaLin2004} by accreting solids until the majority of solids have been cleared out of the planet's feeding zone. The timescale associated with this accretion ($\sim 10^5$ yr) is much shorter than the migration timescale ($\sim 10^6$ yr) if the planet begins close ($< 10$ AU) to the host star.  (It is reasonable to assume that planetesimal rather than  pebble accretion dominates solid accretion within this 10 AU inner disk region).  Thus,  the planet evolves nearly vertically on the diagram (see the blue and black curves). Farther out in the disk, the solid accretion rate is comparable to the migration rate, and the planets evolve more diagonally (red and black dashed line). Once the rate of solid accretion drops, the core can cool to begin accreting gas, and enters into a phase of slow gas accretion.  Gas accretion requires a low envelope opacity ( $\kappa_{env} \simeq 0.001$ cm$^2$ g$^{-1}$  )
that is three orders of magnitude smaller than the opacity of materials in the disk, as has also been found in the  Bern models.

\begin{figure*}
\subfloat[M-a diagram for disk wind evolved planetary populations.]{
\includegraphics[width=0.5\textwidth]{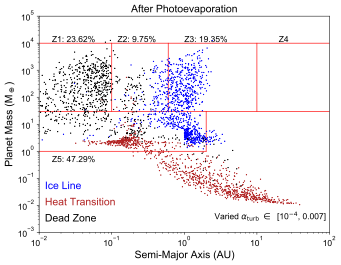}
\label{fig:AP_M-a_2022}
}
\subfloat[M-R diagram for disk wind evolved planetary populations ]{
\includegraphics[width=0.5\textwidth]{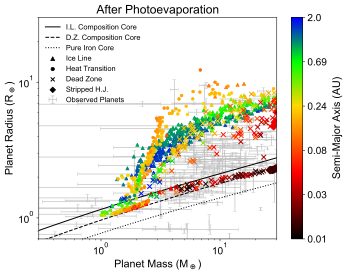}
\label{fig:AP_M-R_2022}
}
\caption{Population synthesis results for disk wind evolved planetary populations. Figures reproduced from \citep{Alessi_Pudritz2022} with permission \textcopyright OUP.}
\label{fig:AP_2022}
\end{figure*}

In Figure \ref{fig:AP_2022} (a) we show results from a population synthesis of our formation model \citep{Alessi_Pudritz2022}. Each planet has evolved through different regions of the disk at different times, because they are the result of different disk initial conditions and parameters, and hence have accreted gas with potentially different chemical histories.   It is assumed that the values of the turbulent $\alpha$ for these tracks are drawn from a lognormal distribution - the physical reason being that disks of different column densities are likely to have variations in the strength of turbulence within them.  These models assume that 80\% of the angular momentum is carried off by the wind.   

A striking result is that the super-Earth population is a mixture of planets with contributions from each of the three traps. Some of these planets have migrated inward to small periods that originated far beyond the ice line and are ice rich.  Warm Jupiters are primarily formed in the ice line trap - a conclusion also reached by \cite{HP13} and
first suggested in \cite{IdaLin2008}.  Ice line planets also populate the sub-Neptune sector in the 0.5 - 2 AU range.  Planets trapped at the heat transition, being the furthest out in disk radius, form  the most slowly and contribute substantially to super-Earths between 0.1 and 0.6 AU. 
Overall there are too many Hot Jupiters compared to observations and a smaller deficiency in the super-Earths and mini-Neptunes.

Figure  \ref{fig:AP_2022} (b) gives the M-R diagram for the populations.  The background light gray points and error bars are from earlier planetary M-R data, and not the more recent filtered set from \citet{Otegi2020}.  Note the population on the magnesium-silicate line that defined Earth-like rocky planets, and the branch that marks the mini-Neptunes, are in good agreement with the M-R observational data in Figure \ref{fig:otegi} (b).

\begin{figure*}
\centering
\includegraphics[width=0.75\textwidth]{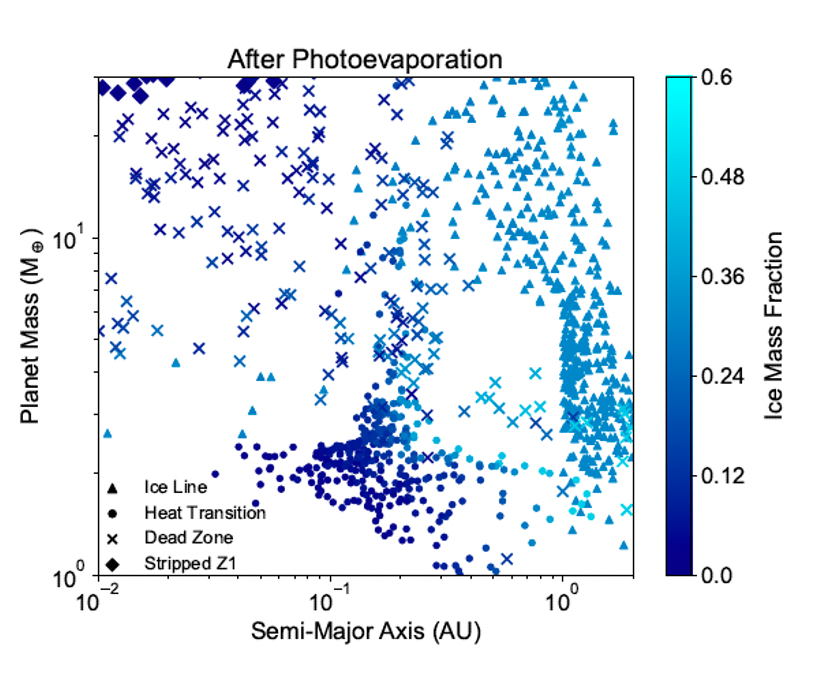}
\caption{Ice fraction of simulated planets, as situated in the M-a diagram.  Note concentration of very dry, super-Earths at short periods in bottom left of Figure (dark blue symbols), with a very icy population just above it in mass (light blue symbols).  These populations have undergone substantial migration, characterized by the three different planet traps.  Figure from \citet{Alessi_Pudritz2022}, reproduced with permission \textcopyright OUP}
\label{fig:ice_fraction}
\end{figure*}

Figure \ref{fig:ice_fraction} shows the ice fractions of these populations, focusing on the short-period planets in Figure \ref{fig:AP_M-R_2022} \citep{Alessi_Pudritz2022}.  There is a very large range of values ranging from very dry planets to ice fractions of 60\% or more.  There is a clear separation in mass between the dry and ice  populations; the super-Earth mass range (dark blue symbols) gives way to ice rich planets (light blue symbols) in mass range  6-8 $M_E$.   This has the hallmark of a radius valley that separates two different populations, which is under current study (Skinner, Pudritz, and Cloutier 2025, in preparation).  

This result supports the finding of \citet{Burn2024} that the radius valley can be created by the superposition of two planetary populations, dry super-Earths and a population of ice-rich more massive planets that arrive at short period orbits via migration from beyond the ice line.  It depends on an updated equation of state (EOS) for water ice, which we address in the next section.

\section{Panetary structure}

\subsection{Planet Cores} 

Planets acquire their rocky materials while migrating through the disk as opposed to in situ formation scenarios.  Depending on how much of this migration occurs while in a planet trap, the accreting materials will be characteristic
of those traps (eg. ice lines). Thus, it is the movement of 
the trap through the evolving disk that can set, to a significant degree, the cumulative inventory of all of the different kinds
of solid materials that are accreted along the way.  
The materials delivered to a planet during the accretion phase are modified as a consequence of their equations of state within the pressure-temperature structure of
the planet's interior and atmosphere.  
Terrestrial planets are modelled as having a crust, mantle, liquid core, and a solid core.  The most massive of these are the mantle and 
the core, comprised of silicates and iron alloys.  These materials differentiated from one another because iron is denser than silicates.  Just four elements
(oxygen, iron, magnesium and silicon) account for 95 \% of the total mass of the Earth  \citep{Javoy1995} . 

The M-R relation 
for rocky material at a constant density is $ R \propto M^{1/3} $.  
The dependence of the M-R relation on the types of materials in the planet can be  
simplified by including
3 basic types of materials;  ice,  mantle materials, and iron (the latter two being categories 
containing many minerals).   Following the early work on zero-temperature models  for single
compositions by \cite{ZapolskySalpeter1969}, \cite{Valencia2007} built models that include all
three of these basic materials where the structure  was computed using EOS used for modelling 
the Earth.  The results of such calculations are often illustrated in terms of ternary diagrams often used in Earth sciences. 
For super-Earths, pressures will be much higher and so different EOS are required.
Thus, the resulting M-R power law exponent is actually less than $ 1/3$ (0.274) because, as 
the density increases with increasing pressure, the temperature remains roughly constant \citep{Grasset2009}. 

The detailed internal structure of planetary interiors and cores depends on the planet's mass; the equations of state (EOS) of the primary
materials such H, He, and materials made of water, silicates, and iron; as well as the overall chemical composition of the planet \citep[see the review][]{Baraffe2014}. The insolation of the planet by the host star gives the equilibrium temperature of the planet at the outer boundary. The atmosphere is included in solving the complete structure of the planet, and also influences the structure of the core.  Knowledge of the EOS relies on experimental studies of the properties of materials under high pressure, as well as by satellite probes of the density structures of the giant
planets such as Jupiter, recently achieved by the Juno mission \citep{Bolton2017}. 

A planet's internal structure is assumed to consist of a number of homogenous layers, which are solved by what is in essence a suitably modified set of stellar structure equations (eg. \citep{Haldemann2024}.  The outermost layer is assumed to consist of water/ice and the most recent compilation of EOS for ice that ranges continuously from low to high pressures, as in planetary interiors, is tabulated in the AQUA EOS  in \citet{Haldemann2020}.  A detailed discussion of EOS and improved structure calculations for planets ranging over $0.5 - 30 M_E$ may be found in the BICEPS paper of \citet{Haldemann2024}.  Although the change in the radii for rocky planets is just a few percent, changes for volatile rich planets can be 10\% or more.  In the latter case, the EOS of water plays an important role

\begin{figure*}
\centering
\includegraphics[width=\textwidth]{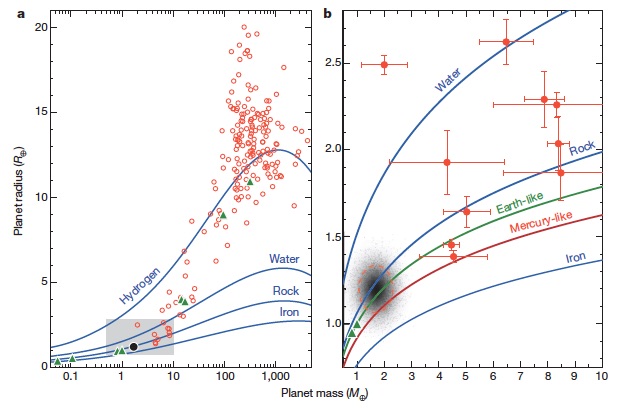}
\caption{The mass-radius (M-R) diagram for a set of Hot Jupiters and sub-Earths. In the left panel, the M-R curves denote planets that are made from pure hydrogen, water, rock and iron. The right panel is a blow up of the greyed region in the left panel, and additionally shows the M-R curves for Earth-like, and Mercury-like planets. Clearly there is a diverse set of compositions in the super-Earth's (right panel) as no internal model uniquely describes every planet. Likewise, Hot Jupiters (left panel, top right) can be highly irradiated, and hence are `puffier' than the hypothetical `pure hydrogen' planet. Figure from \citet{Howard2013}, reproduced with permission \textcopyright Springer Nature.}
\label{fig:8}
\end{figure*}

The observations show that exoplanets are surprisingly diverse.  In Figure \ref{fig:8}, left panel \citep{Howard2013}, note that planets with a given mass can have a wide range of sizes.  Thus Jovian planets (a few hundred Earth masses) range over a factor of 2 in planetary radius for a given mass.    There are also  "inflated"  Jovian planets whose radii greatly exceed those predicted for a composition of pure hydrogen. 
 A variety of models have been proposed to explain such planets including stellar heating  \citep{ShowmanGuillot2002,Weiss2013} and heavy element gradients in the planetary interior that would decrease the rate of heat transport thereby slowing down the cooling and contraction of the planet \citep{ChabrierBaraffe2007}. Similarly, for super-Earth masses, there is a wide range of radii for any given mass, indicative of planets that could be dominated by heavy H/He atmospheres, or are water rich. The right panel of Figure \ref{fig:8} is a blow-up of the super-Earth mass regime (1 - 10 Earth masses), and clearly shows the existence of planets with densities corresponding to rock-iron mixtures. 

The effects of heavy metal enrichment on planetary structure were first carried out by \citet{ZapolskySalpeter1969}, for planets made of individual elements at $ T = 0 $.   If convective energy transport dominates in the planetary interior, the temperature gradient will be nearly adiabatic.  Increasing the heavy element content in giant planets leads to a decrease in their radii. Composition gradients within their interiors suppress their interior cooling, leading to models that are hotter than the adiabatic models \citep{LeconteChabrier2013}.  

The core accretion picture of Jovian planet formation predicts the existence of rocky, 10 Earth mass cores  (\cite{Mizuno1978,BodenheimerPollack1986,Pollack1996} - and chapter by Youdin and Zhu).   What evidence is
there for the existence of such a core?  If Jovian planets form by gravitational instability, as an example, then there would be no initial core.  Even if the core accretion picture is correct, cores could erode away over the billions of years since their formation as they slowly dissolve in liquid metallic hydrogen \citep{Stevenson1985,Gonzalez2014}.  This process would enrich the envelope above it.  

These fundamental questions were addressed  by one of the most significant experiments in planetary science over the last decade - the Juno spacecraft.  The mission goal  is to improve our understanding of the origin and evolution of Jupiter, the history of the solar system, and planetary system formation in general.   On August 27, 2016, this probe flew less than 5000 km over the equatorial cloud tops of Jupiter acquiring a wealth of measurements of the state of the planet's atmosphere, magnetic field, and interior structure \citep{Bolton2017}.  The gravity measurements, found by determining small deviations of the spacecraft trajectory due to low order harmonics ( $J_4, J_6$ in particular ) of Jupiter's gravitational field, were an order of magnitude more sensitive than any before it.   

The modelling of the planetary interior and comparison with the data \citep{Wahl2017} considered interior density profiles that are in hydrostatic equilibrium;  

\begin{equation}
\bigtriangledown P = \rho \bigtriangledown  \Phi
\end{equation}

\noindent where a barotropic pressure $P(\rho) $ corresponding to isentropic profiles constructed from various EOS is used and $\Phi$ is the gravitational potential.
Numerical simulations  of H-He mixtures from (\cite{MilitzerHubbard2013}, MH13) were employed.  Density functional theory molecular dynamics simulations are the best technique  for determining densities of hydrogen-helium mixtures in most high pressure conditions in a giant planet interior ( $ P > 5 $ GPA).   

The results show that Jupiter has a core in the range of 6-25 Earth masses.   The results are generally consistent with the core accretion - collapse model \citep{Pollack1996}. The larger masses correspond to having a more dilute density profile in the core, equivalent to extending about 10 Earth masses of material out to 0.3- 0.5 Jovian radii,  $R_J $.
This agrees with models that account for the dissolution of planetesimals \citep{Lozovsky2017} as the reason for 
a dilute core structure.   It is not known whether there is enough convective energy available to lift so much material.   The overall results clearly depend on exactly how the planet formed, and how mixing occurred during these early stages \citep{LeconteChabrier2012}.   The mass of the heavy elements in the envelope depends strongly on the EOS, with MH13 predicting 5-6 times solar heavy metal fraction in Jupiter.  

In the upcoming years, we may see similar measurements made for the interiors of exoplanets. Ultra hot Jupiters such as WASP-12 b orbit close enough to their host stars to experience significant tidal deformation \citep{akinsanmi_tidal_2024}. The extent of this deformation depends on the interior structure of the planet and in particular the density profile \citet{wahl_tidal_2021,akinsanmi_effect_2023}. Extremely precise transmission and phase curve observations with JWST are capable of measuring the distortion of these planets and constraining the core mass fraction with within $20\%$ \citep{akinsanmi_effect_2023}. Upcoming programs with JWST have been approved to attempt these measurements and could provide more insight into the interiors of giant planets.

\begin{figure*}
\centering
\subfloat[A breakdown of the refractories that are accreted into a super-Earth with a final mass of 5.4 Earth masses, formed
at the heat transition trap $r_{HT}$, in a calculation with a disk life time of 3 Myr. Figure reproduced from \citet{APC16a}, with permission \textcopyright OUP.]{
\includegraphics[width=0.55\textwidth]{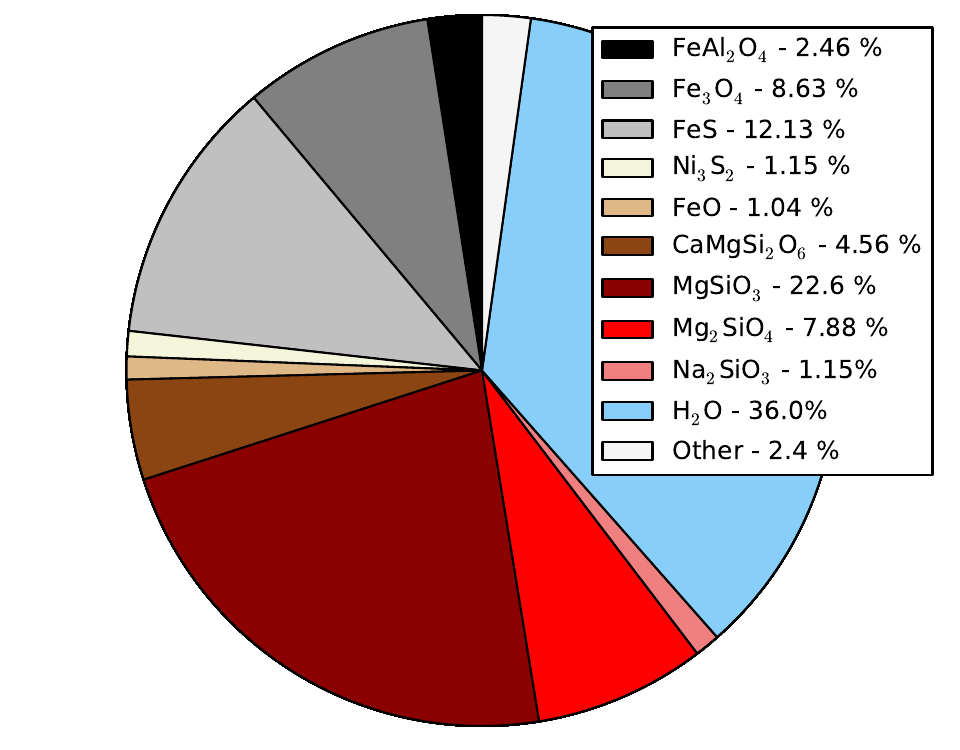}
\label{fig:9a}
}
\subfloat[Models of the interiors of planets with varying degrees of water sequestration in iron and silicate materials. Figure reproduced from \citet{Luo2024}, with permission \textcopyright Springer Nature.]{
\includegraphics[width=0.45\textwidth]{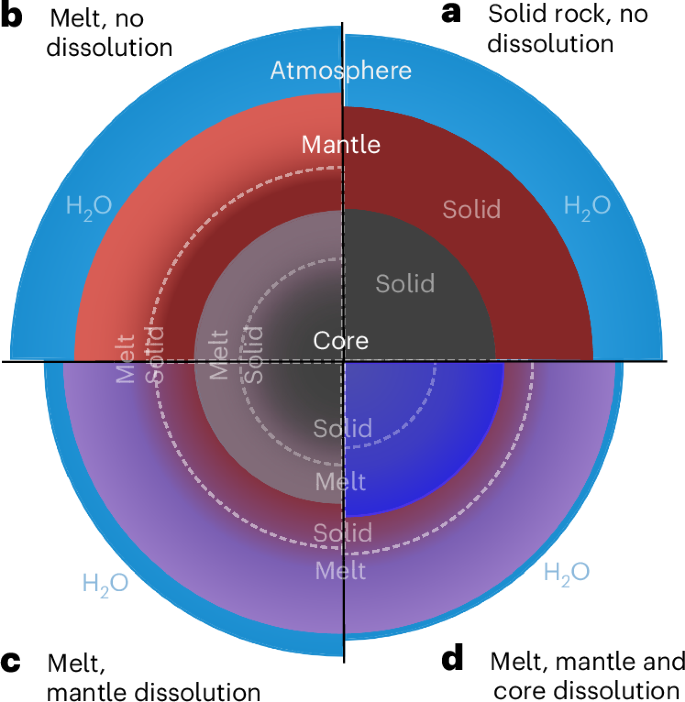}
\label{fig:9b}
}
\caption{The link between a super-Earth's size and its internal composition is complicated, but generally dominated by the abundance of ice that it accretes and retains during its formation. The left panel shows an example of the fractional abundance of refractory materials produced in a particular model, giving the fractional abundances of mantle and core materials as well as a water fraction. The right panel shows the effect of varying degrees of water dissolution in the core and mantle on a planet's interior structure and radius}
\label{fig:9}
\end{figure*}

Returning to the internal structure of super-Earths and mini-Neptunes, 
Figure \ref{fig:9a} gives an example of the detailed breakdown of materials accreted from the
disk by a forming super-Earth planet moving with the heat transition trap \citep{APC16a}.   The final mass in this model reaches 5.4 Earth masses.
In this calculation, the ice fraction is 36 \%,  the mantle 36 \%, and core materials 27 \%.  Since 
cores trapped at
the heat transition will spend most of their life beyond the water ice line, these will generally have the largest ice mass fractions.  
Planets trapped at the water ice line will have a smaller proportion of ice, while dead zone planets will have the least since the dead zone radius
occurs typically inside the ice line.   As an example of the sensitivity of the compositions of super-Earth compositions to disk parameters, planets formed in a disk with a 2 Myr life time
in the heat transition planet had an ice content reached 48 \%, whereas for the dead zone planets only reached 6 \%.   

 Given the prediction of large ice fractions by various models of super-Earths and mini-Neptunes (eg. \citet{Alessi_Pudritz2022, Burn2024}), it is essential to address the water mass fraction and how water is distributed within the planet's interior structure.  Most of the literature assumes that water is layered near the surface of the planet's interior.  However, recent detailed molecular dynamics simulations have shown that water can be sequestered in the iron core where it will be locked up and not outgassed \citep{Dorn2017}. The partitioning of water between iron and silicates is sensitive to pressure and hence to the planet's mass \citet{Luo2024}. Iron can sequester up to 70 times more water than silicates, depending on the pressure. Water not stored in either of these components stays on the surface.

Figure \ref{fig:9b} shows the results of four different models for planetary interiors \citep{Luo2024}.  These are: Model A - rocky interior with no melting,  water as a separate layer; Model B - rocky interior with (dry) melting in mantle and core; Model C - same as B but with water dissolution in the mantle; and Model D - same as C except water dissolution in the iron core. Planetary radii are the smallest for Model D, as seen in the Figure. 

\begin{figure*}
\centering
\includegraphics[width=0.6\textwidth]{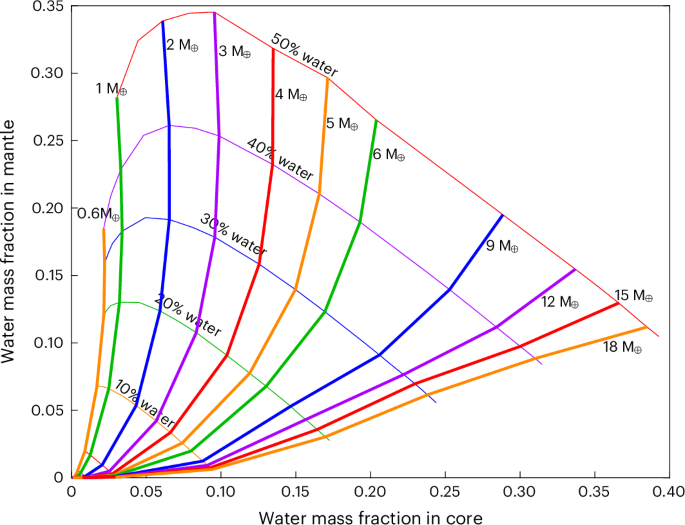}
\caption{Water mass fractions stored in the mantle and core depends on pressure, and hence the mass of the planet; dependence on planet mass (thick lines) and bulk water mass fraction (thin lines). Figure from \citet{Luo2024}, reproduced with permission\textcopyright Springer Nature.}
\label{fig:Luo_watermassfract}
\end{figure*}

Figure \ref{fig:Luo_watermassfract} presents a comparison of the water mass faction in a planet's iron core versus that stored in its mantle \citep{Luo2024}.  The results show that water is stored mainly in the mantle for planets $\le 6 M_E$,  and predominantly in the iron core for planets above this mass.  In these more massive planets, most of the water will be stored in the core which results in a change in a planet's radius by up to 25 \% depending on its mass. 

\subsection{Atmospheres} 

As for solids, gas abundances of planet atmospheres reflect the temperature  of the regions in the disk where they accreted material. Abundances of CO, N$_2$ and SiO result from gas accretion in hot regions
of the disc, while accretion from colder regions of the disc results
in higher abundances of H$_2$, O, CH$_4$, and NH$_3$. 
These results clearly illustrate
how bulk densities of planets can reflect where they form in disks.  

A key observable in studying the chemical composition of exoplanetary atmospheres is the carbon-to-oxygen ratio (C/O). This is because (as noted earlier) it can be naturally linked to chemical processes in the PPD, and also because the elemental ratio is less sensitive to chemical processes within the atmosphere.  In particular, the C/O ratio determines the abundances of spectrally active species  made up of C and O in the disks - molecules such as $CO$, $CO_2$, $HCN$, $H_2O$, $CH_4$, etc.  in the mid-infrared band \citep{molliere2022}.

Nitrogen bearing species can also be important diagnostics for where a planet formed.  Nitrogen in the form of $N_2$ is highly volatile and only freezes out at large disk radii, beyond where O and C species freeze out \citep{Oberg2019,Crid2020b}.  To give an important example - the high nitrogen content of Jupiter's atmosphere implies, therefore, that it may have formed beyond the nitrogen iceline, which is out at 30 AU. 

\begin{figure*}
\centering
\subfloat[Planet formation tracks for three planets formed in the water ice line (blue), dead zone (black) and heat transition (orange). The points on each track denote the position of the growing planet a 1,2,3, and 4 Myr. The dotted lines denote the location of the water ice line at each of the labels times. Reproduced from \citet{Crid17} with permission \textcopyright OUP.]{
\includegraphics[width=0.5\textwidth]{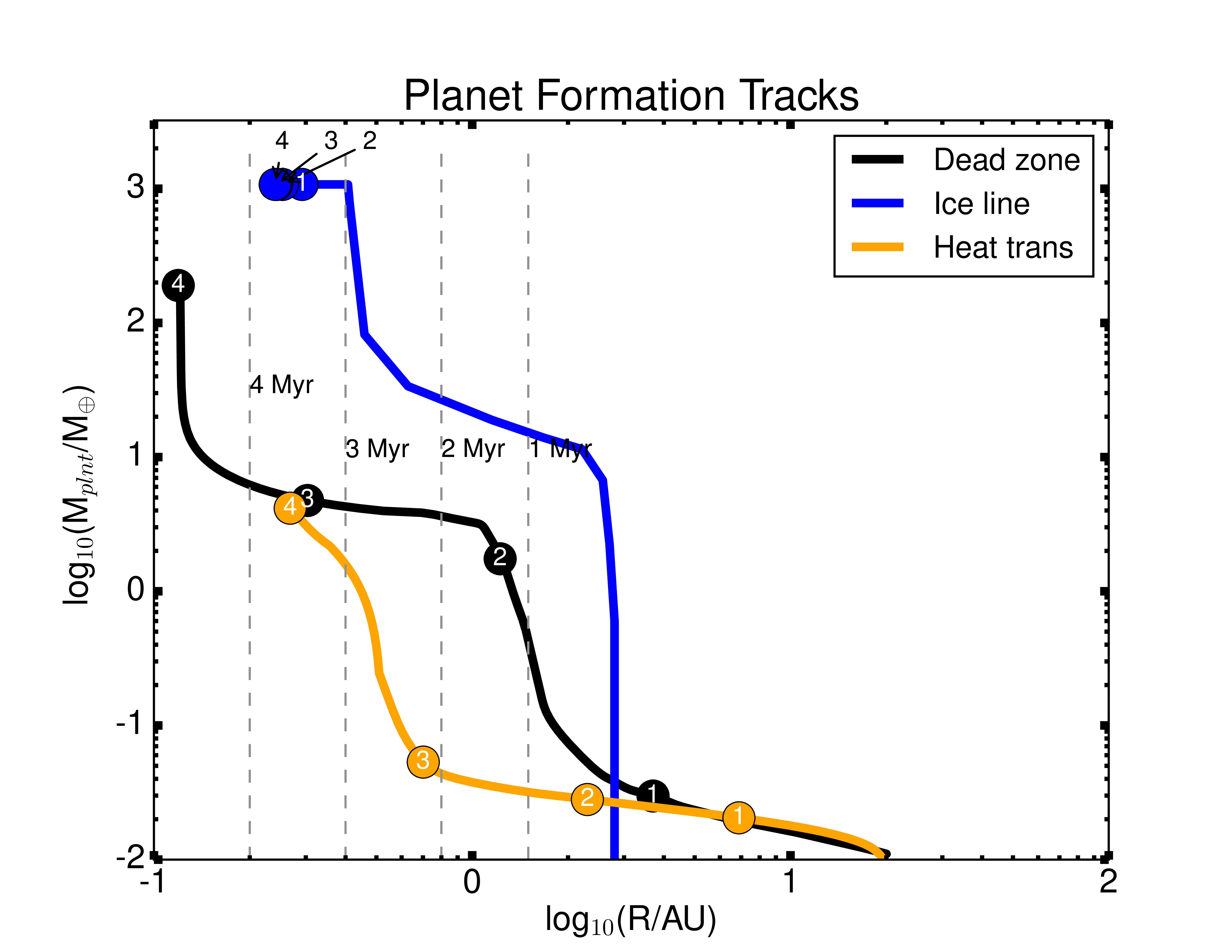}
\label{fig:10a}
}
\subfloat[The resulting atmospheric C/O for hot- and warm-Jupiters from \cite{Crid2019c} and \cite{Crid2020b} respectively. The coloured points denote planets forming in particular planet traps. Higher mass planets tend to have higher C/O, while at a given mass low C/O planets formed from the dead zone trap and accrete most of their gas inward of the water ice line.]{
\includegraphics[width=0.5\textwidth]{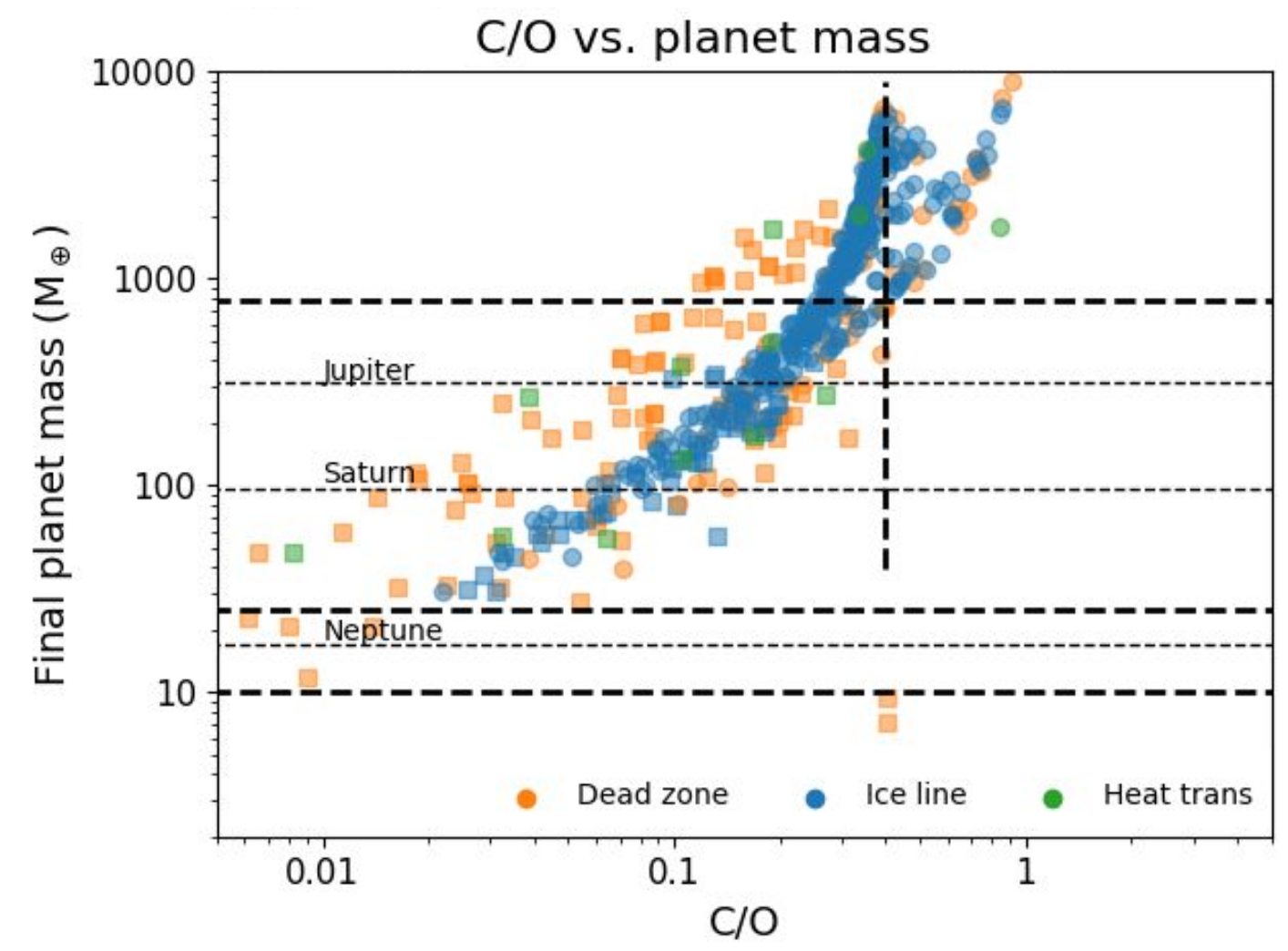}
\label{fig:10b}
}
\caption{An example of planet formation tracks (left) and the result of a chemical population synthesis (right). Figure \ref{fig:10a} is reproduced from Cridland et al. (2017), MNRAS, 469, 3910. While figure \ref{fig:10b} is reproduced from \cite{Crid2019c,Crid2020b}.}
\label{fig:10}
\end{figure*}

In Figure \ref{fig:10} we provide a concrete example of how the C/O ratio in the atmospheres of  planetary populations is connected to a specific  model of planet formation (as computed in \citet{Crid17}). The left panel (Figure \ref{fig:10a}) shows a set of planet formation tracks used in \cite{Crid17}. Each track denotes the evolution of a planet in each individual planet trap, and we label the time (in Myr) when the planet appears at a given position.   We denote the location of the water ice line with vertical dotted lines at 1-4 Myr. The main difference between each of the planet traps is how quickly the trap moves inward and how far out in the disk the trap begins.

Traps that form at larger disk radii grow their planetary embryos more slowly because planetesimal accretion is less efficient there. This can be seen in figure \ref{fig:10a} where the heat transition (yellow) and the dead zone (black) begin their evolution outward of 10 AU. It takes them nearly the whole 4 Myr to fully form their planets. The ice line (blue) on the other hand, evolved from a closer disk radii and reached its final mass in less than 1 Myr.

The dead zone moves inward more quickly than the heat transition trap and thus its planetary embryo is fully formed by 2 Myr. Over the next 2 Myr it proceeds to migrate in where it accretes its atmosphere inward of the water ice line. The heat transition trap evolves in slowly and only produces a super-Earth of a few Earth masses. The variety of planet properties involves a variety in the migration rates for the different planetary embryos.

The results of a full population synthesis model are shown in figure \ref{fig:10b}. There are two types of planets in the population, each with planetary masses larger than 10 M$_{\oplus}$: hot-planets that end their migration inward of 0.1 AU, and warm-planets that orbit between 0.5-4 AU \citep{Crid2020b}. There is a correlation between the final planet mass and the atmospheric C/O that we will discuss further below. For a given mass, planets forming from the dead zone planet trap (orange points) tend to have lower C/O because they migrated closer to their host star than the water ice line before accreting most of their gas (see tracks in figure \ref{fig:10a}).

The chemical calculation of \cite{Crid2019c,Crid2020b} assumed as an initial condition that the global C/O of the protoplanetary disk 0.4 (vertical line in figure \ref{fig:10b}). Almost all of the formed planets had atmospheres that were more oxygen-rich (relative to carbon) than their natal disk. This is a consequence of solid accretion on the final C/O in the atmosphere. While the carbon abundance in the gas is approximately one carbon atom per 10000 hydrogen atoms, it makes up approximately a third of the solid mass \citep{Mordasini16}. This implies that only a small quantity of solid accretion is needed to dominate the carbon and oxygen abundance over gas accretion.

\begin{figure*}
    \centering
    \includegraphics[width=0.8\textwidth]{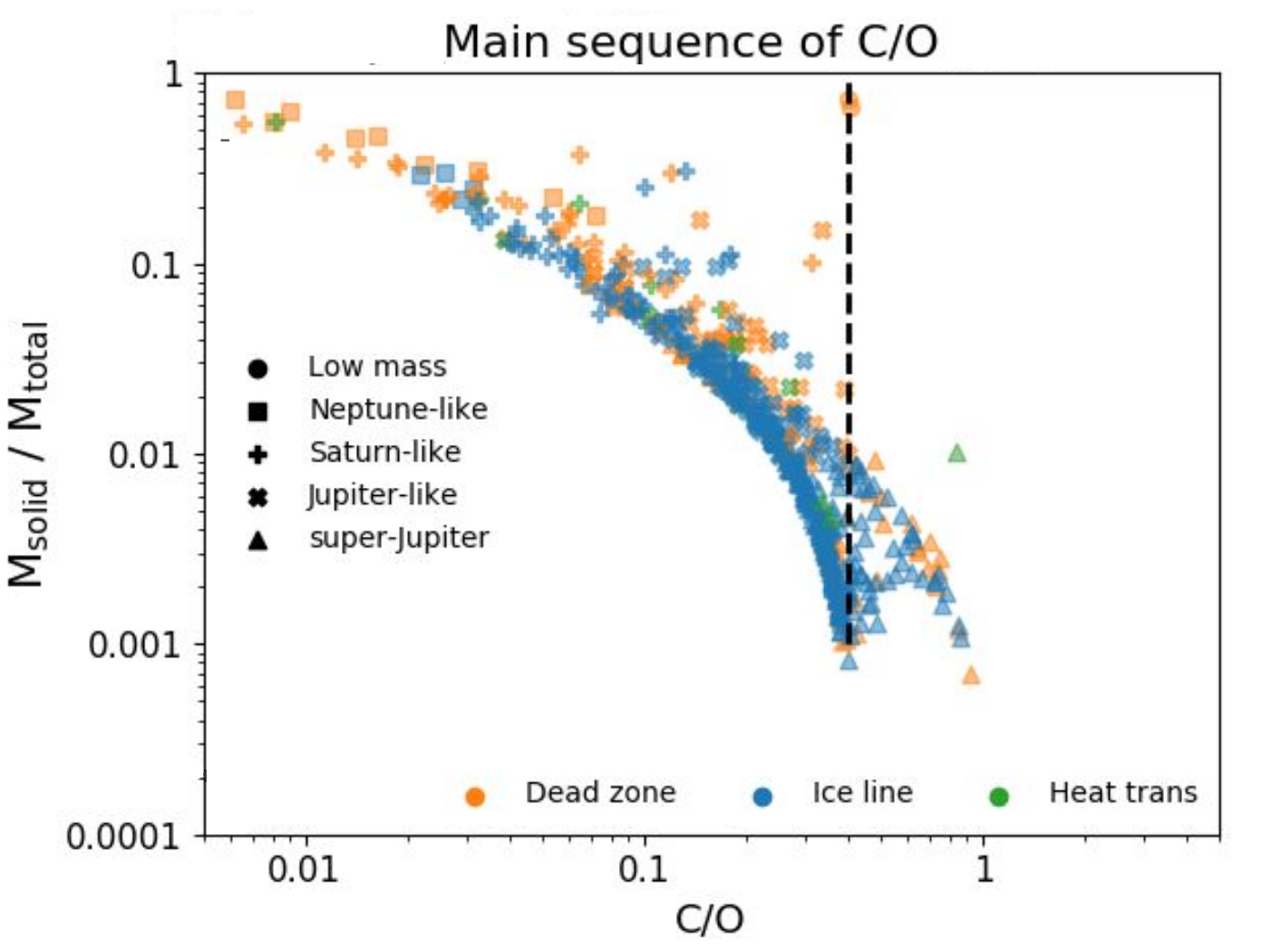}
    \caption{The C/O main sequence for hot- and warm-Jupiters based on the population of \cite{APC2020} and presented in \cite{Crid2019c,Crid2020b}. The colour points denote the particular planet trap from which the planet originated. The main sequence is a close correlation between the atmospheric C/O and the quantity of mass accreted into the atmosphere as solids (planetesimals) relative to the total mass of the planet. The vertical dashed line shows the initial C/O of the volatiles in the protoplanetary disk.}
    \label{fig:11}
\end{figure*}

Furthermore, \cite{Crid2019c,Crid2020b} discovered the C/O main sequence: a tight correlation between the atmospheric C/O and the fraction of mass accreted into the atmosphere as solids, relative to the total mass of the planet. We combine their figures into Figure \ref{fig:11}. The shape of each symbol in the Figure denotes one of 5 different mass bins from both populations, while their colour denotes their natal planet trap. We see that the low mass planets (Saturn-like, Neptune-like, Low mass) are nearly completely dominated by solid accretion in setting their carbon and oxygen abundance - and resulting in low C/O.

On the high mass end, the solid accretion makes up a tiny fraction of the overall contribution to the atmospheric mass, and thus their C/O is dominated mainly by the gas that they accrete. As discussed previously, the gas tends to be more carbon rich than the solids, which produces the observed relation. There are two branches at the high mass (high C/O) end of the figure that are related to how quickly the planetary embryos grow. In the population of disks there is a subset that are of high metallicity ($[Fe/H] > 0$) and these produce embryos faster than lower metallicity disks. High metallicity disks (and their host star) should thus produce a larger variety of chemistries in the atmospheres of their planets.

The importance of the chemical contributions of solid materials has been explored primarily in the context of accretion directly into the atmosphere. However, the core can be eroded by the extreme pressures of the gaseous envelope and atmosphere sitting on top of it. This was discussed in \citet{Madhusudhan2017}, who suggested that convective motions throughout the giant planet can mix the eroded core through the envelope, enriching the atmosphere and changing its composition. These effects suggest that the accretion rates and compositions of both gases and solids need to be considered to fully understand the compositions of planet atmospheres throughout the formation phase.

Observations of ultra-hot Jupiters at optical and UV wavelengths can provide insight into the relative contribution of solid and gas accretion to final atmospheric chemistry \citep{chachan_breaking_2023}. As species like Fe, Si, and Mg condense at relatively high temperatures, they exist in the solid phase throughout most of the disk, and therefore their atmospheric abundances is directly dependent on the volume of solids accreted. Ultra-hot Jupiters have high enough temperatures for atomic species to remain in the gas phase. HST observations of WASP-121 b found a rock/ice ratio of accreted material of greater than 2/3 for this close in planet, indicating significant enrichment by rocky material \citep{lothringer2021}. High-resolution optical spectrographs have also already demonstrated the ability to detect atomic species and constrain their abundances \citep[e.g.][]{gibson_relative_2022,damasceno_atmospheric_2024}  These observations provide necessary context for interpretation of the atmospheric C/O ratio.

Once accreted, the atmosphere will evolve in time both physically and chemically. These processes are complicated, and their study involves three-dimensional hydrodynamic simulations \citep[eg.][]{Cooper2006}, time-dependent photochemical networks \citep[eg.][]{Agundez2014},  and synthetic spectra (eg. \citep{Molliere2017}). The formation of oxygen-rich silicate clouds such as enstatite (MgSiO$_3$), quartz (SiO$_2$) or forsterite (Mg$_2$SiO$_4$), can act as an oxygen sink, removing up to 20$\%$ of atmospheric oxygen according to models of brown dwarfs \citep{burrows_chemical_1999,calamari_predicting_2024}. This increases the apparent C/O ratio of the remaining gas, as has been noted in studies of brown dwarf atmospheres \citep{line_uniform_2015,calamari_atmospheric_2022}. These clouds have recently been observed in the atmospheres of transiting planets with JWST \citep{grant_jwst-tst_2023,inglis_quartz_2024}, suggesting that they will similarly affect these atmospheres. Additionally, the formation of water clouds in the solar system has been invoked to explain the apparent oxygen deficiencies in the upper atmospheres of the giant planets, Jupiter and Saturn \citep{visscher_chemical_2005,visscher_deep_2010}.

As of this writing, although substantial progress has been made over the last decade, no complete model of exoplanetary atmospheres has been developed, and observations of chemical complexity in atmospheres are in their early stages of development \citep{molliere2022}. That said, the JWST exoplanet atmosphere observations that are now in full swing \citep{Rustamkulov2023,Alderson2023}, and the ground based observations with the extremely large telescope (ELT) due to begin  in the early 2030s,  stronger observational constraints on atmospheric abundances of many species will soon transform the field.

\section{Conclusions}

Having reviewed the basic multiscale processes involved in planet formation, we return to our original question: are there clearly discernible links between planetary composition and structure and their formation history?  The wealth of recent ALMA, JWST, TESS observations and the developments in theory and complex MHD simulations of PPDs and accretion processes lead to several new and important insights. 

\begin{itemize}
\item Observations, theory, and simulations strongly implicate MHD disk winds as the main agent of angular momentum loss from PPDs.  This is the basis of an emerging new paradigm for planet formation that is distinct from turbulent models.
\item A range of simulations show that MHD processes, such as disk winds, concentrate magnetic flux to create gaps and rings.  The resulting layered disk winds predicted by this picture are supported by recent JWST observations of layered outflows.  
\item Rings formed as a consequence of MHD disk winds and gap formation could be the sites of first planet formation if enough pebbles can be accumulated.
\item The low observed turbulent amplitudes in disks together with laminar, disk driven accretion are processes that promore more rapid grain growth and migration.  This may speed up planet formation, transport, and mixing of materials from a greater range of disk radii by pebble flows. 
\item Disk winds may play a central role  in planet migration. This is most clearly evident in their effects on   co-rotation torques, and could lead to outward migration. The magnitude of such torques depends on the strength of the winds.   
\item The core accretion picture has ever more empirical support - in particular with the detection of a core in Jupiter discovered by the Juno probe. Additionally, upcoming high-precision phase curves and transit observations from JWST have the potential to measure tidal deformation of the hottest giant planets, enabling new constraints on the composition of the interiors of giant planets outside our solar system.
\item The gas and dust emission from protoplanetary disks inform our understanding of their mass distribution, thermal, and chemical structure. The presence of planets may influence some or all of the observed continuum emission substructures in young disks however the gas emission is not as strongly tied to these substructures. Instead some gas species are more symmetrically distributed across the disk, while others show chemistry-induced structure in the disk.
\item Dust evolution is critical for understanding both pebble and planetesimal formation and the evolution of solids in disks.  While there is general agreement that pebble accretion may dominate the building of giant planets at large disk radii, it is still unclear as to how important this is compared to planetesimal accretion in the inner regions of disks. 
\item The C/O ratio of a planetary atmosphere is, on its own, likely to be a limited probe of planet formation.  Additional information from elemental ratios that involve nitrogen as an example, or other refractory elements may sharpen the constraints. That said, there is a predicted main sequence for the C/O ratio in the sense that Saturn and Neptune-like planets have low C/O ratios being completely dominated by solid accretion, while high mass giants will have high C/O ratios indicative of more gas dominated accretion. Although the spectral information is still very limited, this agrees with observations. 
\item Population synthesis simulations have become increasingly sophisticated over the last decade. Many additional developments in the physics of PPDs including updates on EOS of planetary building materials and a more general treatment of migration that includes MHD and disk wind effects are now being developed.  No synthesis to date has entirely succeeded in accounting for all the features of planetary populations. 
\item Planetary migration from beyond the ice line can bring water-rich mini-Neptunes into short-period orbits, with ice fractions as high as 50\%.  These populations may provide an explanation for the radius valley separating super-Earths from mini-Neptunes beyond the need for photoevaporation effects alone.  Because disk winds create more compact disks, it is expected that larger number of planets will migrate into short period orbits than in the case of viscous disks. 
\end{itemize}

The acceleration of discoveries about PPD physics and properties, outflows, and the composition of exoplanetary atmospheres and planets gives good reason for optimism that this emerging paradigm can lead to new insights into how the properties and composition of planets are connected to their formation. 

\section{Cross References}

\begin{itemize}
\item Internal Structure of Giant and Icy Planets: Importance of Heavy Elements and Mixing
\item Exoplanet Populations and Their Dependence on Host Star Properties
\item A Brief Overview of Planet Formation
\item Dust Evolution in Protoplanetary Disks
\item Chemistry During the Gas-Rich Stage of Planet Formation
\item Instabilities and Flow Structures in Protoplanetary Disks: Setting the Stage for
Planetesimal Formation
\item Pebble Accretion
\item Planetary Migration in Protoplanetary Disks
\item Formation of Giant Planets
\item Planetary Population Synthesis
\item Dynamical Evolution of Planetary Systems
\item Tightly Packed Planetary Systems
\end{itemize}

\section{Acknowledgements}  We thank Phil Armitage for his thoughtful referee report that helped to improve the review.  We also thank Christoph Mordasini, Remo Burn, Caroline Dorn, Illaria Pascucci, Ryan Cloutier, Jess Speedie, Paul Mollière, Bertram Bitsch, Mario Flock, Thomas Henning, Tom Ray, Caroline Kissig, Kees Dullemond, Yasuhiro Hasegawa, Bennett Skinner, Timmy Delage, and Atesh Goksu for the informative discussions during the course of this project.  REP is supported by a Discovery grant from the Natural Sciences and Engineering Research Council of Canada (NSERC). AJC is employed as a Staff Scientist in the Theoretical Astrophysics for Extra-Solar Planets Research chair at the LMU.  JI is supported as a PhD candidate in Planetary Sciences in the Division of Geological and Planetary Sciences at the California Institute of Technology by her advisor, Heather A. Knutson.  Finally,  Ralph Pudritz and Phil Armitage thank Saskia Ellis for her excellent work, patience, and guidance as our Editorial Assistant for the entire section on "Formation and Evolution of Planets and Planetary Systems" in the 2nd Edition of Springer's, Handbook of Exoplanets.

\bibliographystyle{spbasicHBexo}  
\bibliography{mybib} 

\begin{thebibliography}{331}
\providecommand{\natexlab}[1]{#1}
\providecommand{\url}[1]{{#1}}
\providecommand{\urlprefix}{URL }
\expandafter\ifx\csname urlstyle\endcsname\relax
  \providecommand{\doi}[1]{DOI~\discretionary{}{}{}#1}\else
  \providecommand{\doi}{DOI~\discretionary{}{}{}\begingroup \urlstyle{rm}\Url}\fi
\providecommand{\eprint}[2][]{\url{#2}}

\bibitem[{{Ag{\'u}ndez} et~al.(2014){Ag{\'u}ndez}, {Parmentier}, {Venot}, {Hersant}, and {Selsis}}]{Agundez2014}
{Ag{\'u}ndez} M, {Parmentier} V, {Venot} O, {Hersant} F {Selsis} F (2014) {Pseudo 2D chemical model of hot-Jupiter atmospheres: application to HD 209458b and HD 189733b}. \aap 564:A73

\bibitem[{{Aikawa} and {Herbst}(1999)}]{Aikawa_Herbst1999}
{Aikawa} Y {Herbst} E (1999) {Molecular evolution in protoplanetary disks. Two-dimensional distributions and column densities of gaseous molecules}. \aap 351:233--246

\bibitem[{Akinsanmi et~al.(2023)Akinsanmi, Lendl, Boue, and Barros}]{akinsanmi_effect_2023}
Akinsanmi B, Lendl M, Boue G Barros SCC (2023) On the effect of tidal deformation on planetary phase curves. \doi{10.48550/arXiv.2310.03553}, \urlprefix\url{https://ui.adsabs.harvard.edu/abs/2023arXiv231003553A}, publication Title: arXiv e-prints ADS Bibcode: 2023arXiv231003553A

\bibitem[{Akinsanmi et~al.(2024)Akinsanmi, Barros, Lendl, Carone, Cubillos, Bekkelien, Fortier, Florén, Cameron, Boué, Bruno, Demory, Brandeker, Sousa, Wilson, Deline, Bonfanti, Scandariato, Hooton, Correia, Demangeon, Smith, Singh, Alibert, Alonso, Asquier, Bárczy, Navascues, Baumjohann, Beck, Beck, Benz, Billot, Bonfils, Borsato, Broeg, Buder, Charnoz, Csizmadia, Davies, Deleuil, Delrez, Ehrenreich, Erikson, Farinato, Fossati, Fridlund, Gandolfi, Gillon, Güdel, Günther, Heitzmann, Helling, Hoyer, Isaak, Kiss, Lam, Laskar, Etangs, Magrin, Maxted, Mecina, Mordasini, Nascimbeni, Olofsson, Ottensamer, Pagano, Pallé, Peter, Piazza, Piotto, Pollacco, Queloz, Ragazzoni, Rando, Rauer, Ribas, Santos, Ségransan, Simon, Stalport, Szabó, Thomas, Udry, Grootel, Venturini, Villaver, and Walton}]{akinsanmi_tidal_2024}
Akinsanmi B, Barros SCC, Lendl M et~al. (2024) The tidal deformation and atmosphere of {WASP}-12 b from its phase curve. Astronomy \& Astrophysics 685:A63, \urlprefix\url{https://www.aanda.org/articles/aa/abs/2024/05/aa48502-23/aa48502-23.html}, publisher: EDP Sciences

\bibitem[{{Alderson} et~al.(2023){Alderson}, {Wakeford}, {Alam}, {Batalha}, {Lothringer}, {Adams Redai}, {Barat}, {Brande}, {Damiano}, {Daylan}, {Espinoza}, {Flagg}, {Goyal}, {Grant}, {Hu}, {Inglis}, {Lee}, {Mikal-Evans}, {Ramos-Rosado}, {Roy}, {Wallack}, {Batalha}, {Bean}, {Benneke}, {Berta-Thompson}, {Carter}, {Changeat}, {Col{\'o}n}, {Crossfield}, {D{\'e}sert}, {Foreman-Mackey}, {Gibson}, {Kreidberg}, {Line}, {L{\'o}pez-Morales}, {Molaverdikhani}, {Moran}, {Morello}, {Moses}, {Mukherjee}, {Schlawin}, {Sing}, {Stevenson}, {Taylor}, {Aggarwal}, {Ahrer}, {Allen}, {Barstow}, {Bell}, {Blecic}, {Casewell}, {Chubb}, {Crouzet}, {Cubillos}, {Decin}, {Feinstein}, {Fortney}, {Harrington}, {Heng}, {Iro}, {Kempton}, {Kirk}, {Knutson}, {Krick}, {Leconte}, {Lendl}, {MacDonald}, {Mancini}, {Mansfield}, {May}, {Mayne}, {Miguel}, {Nikolov}, {Ohno}, {Palle}, {Parmentier}, {Petit dit de la Roche}, {Piaulet}, {Powell}, {Rackham}, {Redfield}, {Rogers}, {Rustamkulov}, {Tan}, {Tremblin}, {Tsai}, {Turner}, {de Val-Borro}, {Venot},
  {Welbanks}, {Wheatley}, and {Zhang}}]{Alderson2023}
{Alderson} L, {Wakeford} HR, {Alam} MK et~al. (2023) {Early Release Science of the exoplanet WASP-39b with JWST NIRSpec G395H}. \nat 614(7949):664--669

\bibitem[{{Alessi} and {Pudritz}(2022)}]{Alessi_Pudritz2022}
{Alessi} M {Pudritz} RE (2022) {Combined effects of disc winds and turbulence-driven accretion on planet populations}. \mnras 515(2):2548--2577

\bibitem[{{Alessi} et~al.(2017){Alessi}, {Pudritz}, and {Cridland}}]{APC16a}
{Alessi} M, {Pudritz} RE {Cridland} AJ (2017) {On the formation and chemical composition of super Earths}. \mnras 464:428--452

\bibitem[{{Alessi} et~al.(2020){Alessi}, {Pudritz}, and {Cridland}}]{APC2020}
{Alessi} M, {Pudritz} RE {Cridland} AJ (2020) {Formation of planetary populations - II. Effects of initial disc size and radial dust drift}. \mnras 493(1):1013--1033

\bibitem[{{Alibert} et~al.(2011){Alibert}, {Mordasini}, and {Benz}}]{Alibert2011}
{Alibert} Y, {Mordasini} C {Benz} W (2011) {Extrasolar planet population synthesis. III. Formation of planets around stars of different masses}. \aap 526:A63

\bibitem[{{ALMA Partnership} et~al.(2015){ALMA Partnership}, {Brogan}, {P{\'e}rez} et~al.}]{ALMA2015}
{ALMA Partnership}, {Brogan} CL, {P{\'e}rez} LM et~al. (2015) {The 2014 ALMA Long Baseline Campaign: First Results from High Angular Resolution Observations toward the HL Tau Region}. \apjl 808:L3

\bibitem[{{Andr{\'e}} et~al.(2010){Andr{\'e}}, {Men'shchikov}, {Bontemps}, {K{\"o}nyves}, {Motte}, {Schneider}, {Didelon}, {Minier}, {Saraceno}, {Ward-Thompson}, {di Francesco}, {White}, {Molinari}, {Testi}, {Abergel}, {Griffin}, {Henning}, {Royer}, {Mer{\'\i}n}, {Vavrek}, {Attard}, {Arzoumanian}, {Wilson}, {Ade}, {Aussel}, {Baluteau}, {Benedettini}, {Bernard}, {Blommaert}, {Cambr{\'e}sy}, {Cox}, {di Giorgio}, {Hargrave}, {Hennemann}, {Huang}, {Kirk}, {Krause}, {Launhardt}, {Leeks}, {Le Pennec}, {Li}, {Martin}, {Maury}, {Olofsson}, {Omont}, {Peretto}, {Pezzuto}, {Prusti}, {Roussel}, {Russeil}, {Sauvage}, {Sibthorpe}, {Sicilia-Aguilar}, {Spinoglio}, {Waelkens}, {Woodcraft}, and {Zavagno}}]{Andre2010}
{Andr{\'e}} P, {Men'shchikov} A, {Bontemps} S et~al. (2010) {From filamentary clouds to prestellar cores to the stellar IMF: Initial highlights from the Herschel Gould Belt Survey}. \aap 518:L102

\bibitem[{{Andr{\'e}} et~al.(2014){Andr{\'e}}, {Di Francesco}, {Ward-Thompson}, {Inutsuka}, {Pudritz}, and {Pineda}}]{Andre2014}
{Andr{\'e}} P, {Di Francesco} J, {Ward-Thompson} D et~al. (2014) {From Filamentary Networks to Dense Cores in Molecular Clouds: Toward a New Paradigm for Star Formation}. In: {Beuther} H, {Klessen} RS, {Dullemond} CP {Henning} T (eds) Protostars and Planets VI, pp 27--51, \doi{10.2458/azu_uapress_9780816531240-ch002}

\bibitem[{{Andrews} and {Williams}(2007)}]{AndrewsWilliams2007}
{Andrews} SM {Williams} JP (2007) {High-Resolution Submillimeter Constraints on Circumstellar Disk Structure}. \apj 659:705--728

\bibitem[{{Andrews} et~al.(2010){Andrews}, {Wilner}, {Hughes}, {Qi}, and {Dullemond}}]{Andrews2010}
{Andrews} SM, {Wilner} DJ, {Hughes} AM, {Qi} C {Dullemond} CP (2010) {Protoplanetary Disk Structures in Ophiuchus. II. Extension to Fainter Sources}. \apj 723:1241--1254

\bibitem[{{Andrews} et~al.(2018){Andrews}, {Huang}, {P{\'e}rez}, {Isella}, {Dullemond}, {Kurtovic}, {Guzm{\'a}n}, {Carpenter}, {Wilner}, {Zhang}, {Zhu}, {Birnstiel}, {Bai}, {Benisty}, {Hughes}, {{\"O}berg}, and {Ricci}}]{Andrews2018}
{Andrews} SM, {Huang} J, {P{\'e}rez} LM et~al. (2018) {The Disk Substructures at High Angular Resolution Project (DSHARP). I. Motivation, Sample, Calibration, and Overview}. \apjl 869(2):L41

\bibitem[{{Ansdell} et~al.(2016){Ansdell}, {Williams}, {van der Marel}, {Carpenter}, {Guidi}, {Hogerheijde}, {Mathews}, {Manara}, {Miotello}, {Natta}, {Oliveira}, {Tazzari}, {Testi}, {van Dishoeck}, and {van Terwisga}}]{Ansdell2016}
{Ansdell} M, {Williams} JP, {van der Marel} N et~al. (2016) {ALMA Survey of Lupus Protoplanetary Disks. I. Dust and Gas Masses}. \apj 828:46

\bibitem[{{Aoyama} and {Bai}(2023)}]{Aoyama_Bai2023}
{Aoyama} Y {Bai} XN (2023) {Three-dimensional Global Simulations of Type-II Planet-Disk Interaction with a Magnetized Disk Wind. I. Magnetic Flux Concentration and Gap Properties}. \apj 946(1):5

\bibitem[{{Armitage}(2020)}]{Armitage2020}
{Armitage} PJ (2020) {Astrophysics of planet formation, Second Edition}

\bibitem[{{Atreya} et~al.(2016){Atreya}, {Crida}, {Guillot}, {Lunine}, {Madhusudhan}, and {Mousis}}]{Atreya2016}
{Atreya} SK, {Crida} A, {Guillot} T et~al. (2016) {The Origin and Evolution of Saturn, with Exoplanet Perspective}. arXiv e-prints arXiv:1606.04510

\bibitem[{{Avenhaus} et~al.(2018){Avenhaus}, {Quanz}, {Garufi}, {Perez}, {Casassus}, {Pinte}, {Bertrang}, {Caceres}, {Benisty}, and {Dominik}}]{Avenhaus2018}
{Avenhaus} H, {Quanz} SP, {Garufi} A et~al. (2018) {Disks around T Tauri Stars with SPHERE (DARTTS-S). I. SPHERE/IRDIS Polarimetric Imaging of Eight Prominent T Tauri Disks}. \apj 863(1):44

\bibitem[{{Bacciotti} et~al.(2025){Bacciotti}, {Nony}, {Podio}, {Dougados}, {Garufi}, {Cabrit}, {Codella}, {Zimniak}, and {Ferreira}}]{Bacciotti2025}
{Bacciotti} F, {Nony} T, {Podio} L et~al. (2025) {ALMA chemical survey of disk-outflow sources in Taurus (ALMA-DOT) VII: the layered molecular outflow from HL Tau and its relationship with the ringed disk}. arXiv e-prints arXiv:2501.03920

\bibitem[{{Bae} et~al.(2017){Bae}, {Zhu}, and {Hartmann}}]{Bae2017}
{Bae} J, {Zhu} Z {Hartmann} L (2017) {On the Formation of Multiple Concentric Rings and Gaps in Protoplanetary Disks}. \apj 850(2):201

\bibitem[{{Bae} et~al.(2023){Bae}, {Isella}, {Zhu}, {Martin}, {Okuzumi}, and {Suriano}}]{Bae2023}
{Bae} J, {Isella} A, {Zhu} Z et~al. (2023) {Structured Distributions of Gas and Solids in Protoplanetary Disks}. In: {Inutsuka} S, {Aikawa} Y, {Muto} T, {Tomida} K {Tamura} M (eds) Protostars and Planets VII, Astronomical Society of the Pacific Conference Series, vol 534, p 423, \doi{10.48550/arXiv.2210.13314}

\bibitem[{{Bai}(2014)}]{Bai2014}
{Bai} XN (2014) {Hall-effect-Controlled Gas Dynamics in Protoplanetary Disks. I. Wind Solutions at the Inner Disk}. \apj 791:137

\bibitem[{{Bai}(2016)}]{Bai2016}
{Bai} XN (2016) {Towards a Global Evolutionary Model of Protoplanetary Disks}. \apj 821:80

\bibitem[{{Bai} and {Stone}(2013)}]{BaiStone2013}
{Bai} XN {Stone} JM (2013) {Wind-driven Accretion in Protoplanetary Disks. I. Suppression of the Magnetorotational Instability and Launching of the Magnetocentrifugal Wind}. \apj 769:76

\bibitem[{{Bai} and {Stone}(2017)}]{BaiStone2017}
{Bai} XN {Stone} JM (2017) {Hall Effect-Mediated Magnetic Flux Transport in Protoplanetary Disks}. \apj 836:46

\bibitem[{{Balbus} and {Hawley}(1991)}]{BH1991}
{Balbus} SA {Hawley} JF (1991) {A powerful local shear instability in weakly magnetized disks. I - Linear analysis. II - Nonlinear evolution}. \apj 376:214--233

\bibitem[{{Banerjee} and {Pudritz}(2006)}]{BanerjeePudritz2006}
{Banerjee} R {Pudritz} RE (2006) {Outflows and Jets from Collapsing Magnetized Cloud Cores}. \apj 641:949--960

\bibitem[{{Banzatti} et~al.(2023){Banzatti}, {Pontoppidan}, {Carr}, {Jellison}, {Pascucci}, {Najita}, {Romero-Mirza}, {{\"O}berg}, {Kalyaan}, {Pinilla}, {Krijt}, {Long}, {Lambrechts}, {Rosotti}, {Herczeg}, {Salyk}, {Zhang}, {Bergin}, {Ballering}, {Meyer}, {Bruderer}, and {Jdiscs Collaboration}}]{Banzatti2023}
{Banzatti} A, {Pontoppidan} KM, {Carr} JS et~al. (2023) {JWST Reveals Excess Cool Water near the Snow Line in Compact Disks, Consistent with Pebble Drift}. \apjl 957(2):L22

\bibitem[{{Baraffe} et~al.(2014){Baraffe}, {Chabrier}, {Fortney}, and {Sotin}}]{Baraffe2014}
{Baraffe} I, {Chabrier} G, {Fortney} J {Sotin} C (2014) {Planetary Internal Structures}. Protostars and Planets VI pp 763--786

\bibitem[{Batalha et~al.(2019)Batalha, Marley, Lewis, and Fortney}]{batalha_exoplanet_2019}
Batalha NE, Marley MS, Lewis NK Fortney JJ (2019) Exoplanet {Reflected}-light {Spectroscopy} with {PICASO}. The Astrophysical Journal 878:70, \urlprefix\url{https://ui.adsabs.harvard.edu/abs/2019ApJ...878...70B}, publisher: IOP ADS Bibcode: 2019ApJ...878...70B

\bibitem[{{Batalha}(2014)}]{Batalha2014}
{Batalha} NM (2014) {Exploring exoplanet populations with NASA's Kepler Mission}. Proceedings of the National Academy of Science 111:12,647--12,654

\bibitem[{{Bate}(2012)}]{Bate2012}
{Bate} MR (2012) {Stellar, brown dwarf and multiple star properties from a radiation hydrodynamical simulation of star cluster formation}. \mnras 419:3115--3146

\bibitem[{{Bate}(2018)}]{Bate2018}
{Bate} MR (2018) {On the diversity and statistical properties of protostellar discs}. \mnras 475:5618--5658

\bibitem[{{Benz} et~al.(2014){Benz}, {Ida}, {Alibert}, {Lin}, and {Mordasini}}]{Benz2014}
{Benz} W, {Ida} S, {Alibert} Y, {Lin} D {Mordasini} C (2014) {Planet Population Synthesis}. Protostars and Planets VI pp 691--713

\bibitem[{{Bergin} et~al.(2015){Bergin}, {Blake}, {Ciesla}, {Hirschmann}, and {Li}}]{Bergin2015}
{Bergin} EA, {Blake} GA, {Ciesla} F, {Hirschmann} MM {Li} J (2015) {Tracing the ingredients for a habitable earth from interstellar space through planet formation}. Proceedings of the National Academy of Science 112:8965--8970

\bibitem[{{B{\'e}thune} et~al.(2017){B{\'e}thune}, {Lesur}, and {Ferreira}}]{Bethune2017}
{B{\'e}thune} W, {Lesur} G {Ferreira} J (2017) {Global simulations of protoplanetary disks with net magnetic flux. I. Non-ideal MHD case}. \aap 600:A75

\bibitem[{{Birnstiel}(2024)}]{Birnstiel2024}
{Birnstiel} T (2024) {Dust Growth and Evolution in Protoplanetary Disks}. \araa 62(1):157--202

\bibitem[{{Birnstiel} et~al.(2012){Birnstiel}, {Klahr}, and {Ercolano}}]{Birnstiel2012}
{Birnstiel} T, {Klahr} H {Ercolano} B (2012) {A simple model for the evolution of the dust population in protoplanetary disks}. \aap 539:A148

\bibitem[{{Birnstiel} et~al.(2016){Birnstiel}, {Fang}, and {Johansen}}]{Birnstiel2016}
{Birnstiel} T, {Fang} M {Johansen} A (2016) {Dust Evolution and the Formation of Planetesimals}. \ssr 205(1-4):41--75

\bibitem[{{Bitsch} et~al.(2013){Bitsch}, {Crida}, {Morbidelli}, {Kley}, and {Dobbs-Dixon}}]{Bitsch2013}
{Bitsch} B, {Crida} A, {Morbidelli} A, {Kley} W {Dobbs-Dixon} I (2013) {Stellar irradiated discs and implications on migration of embedded planets. I. Equilibrium discs}. \aap 549:A124

\bibitem[{{Bitsch} et~al.(2015){Bitsch}, {Lambrechts}, and {Johansen}}]{Bitsch2015}
{Bitsch} B, {Lambrechts} M {Johansen} A (2015) {The growth of planets by pebble accretion in evolving protoplanetary discs}. \aap 582:A112

\bibitem[{{Bitsch} et~al.(2018){Bitsch}, {Morbidelli}, {Johansen}, {Lega}, {Lambrechts}, and {Crida}}]{Bitsch2018}
{Bitsch} B, {Morbidelli} A, {Johansen} A et~al. (2018) {Pebble-isolation mass: Scaling law and implications for the formation of super-Earths and gas giants}. \aap 612:A30

\bibitem[{{Bitsch} et~al.(2019){Bitsch}, {Izidoro}, {Johansen}, {Raymond}, {Morbidelli}, {Lambrechts}, and {Jacobson}}]{Bitsch2019}
{Bitsch} B, {Izidoro} A, {Johansen} A et~al. (2019) {Formation of planetary systems by pebble accretion and migration: growth of gas giants}. \aap 623:A88

\bibitem[{{Blandford} and {Payne}(1982)}]{BlandfordPayne1982}
{Blandford} RD {Payne} DG (1982) {Hydromagnetic flows from accretion discs and the production of radio jets}. \mnras 199:883--903

\bibitem[{{Bodenheimer} and {Pollack}(1986)}]{BodenheimerPollack1986}
{Bodenheimer} P {Pollack} JB (1986) {Calculations of the accretion and evolution of giant planets The effects of solid cores}. \icarus 67:391--408

\bibitem[{{Bolton} et~al.(2017){Bolton}, {Lunine}, {Stevenson}, {Connerney}, {Levin}, {Owen}, {Bagenal}, {Gautier}, {Ingersoll}, {Orton}, {Guillot}, {Hubbard}, {Bloxham}, {Coradini}, {Stephens}, {Mokashi}, {Thorne}, and {Thorpe}}]{Bolton2017}
{Bolton} SJ, {Lunine} J, {Stevenson} D et~al. (2017) {The Juno Mission}. \ssr 213:5--37

\bibitem[{{Bond} et~al.(2010){Bond}, {O'Brien}, and {Lauretta}}]{Bond2010}
{Bond} JC, {O'Brien} DP {Lauretta} DS (2010) {The Compositional Diversity of Extrasolar Terrestrial Planets. I. In Situ Simulations}. \apj 715:1050--1070

\bibitem[{{Booth} et~al.(2021){Booth}, {van der Marel}, {Leemker}, {van Dishoeck}, and {Ohashi}}]{Booth2021}
{Booth} AS, {van der Marel} N, {Leemker} M, {van Dishoeck} EF {Ohashi} S (2021) {A major asymmetric ice trap in a planet-forming disk. II. Prominent SO and SO$_{2}$ pointing to C/O < 1}. \aap 651:L6

\bibitem[{{Booth} and {Ilee}(2019)}]{BoothIlee2019}
{Booth} RA {Ilee} JD (2019) {Planet-forming material in a protoplanetary disc: the interplay between chemical evolution and pebble drift}. \mnras 487(3):3998--4011

\bibitem[{{Booth} et~al.(2017){Booth}, {Clarke}, {Madhusudhan}, and {Ilee}}]{Booth2017}
{Booth} RA, {Clarke} CJ, {Madhusudhan} N {Ilee} JD (2017) {Chemical enrichment of giant planets and discs due to pebble drift}. \mnras 469:3994--4011

\bibitem[{{Bosman} et~al.(2017){Bosman}, {Tielens}, and {van Dishoeck}}]{Bosman2017b}
{Bosman} AD, {Tielens} AGGM {van Dishoeck} EF (2017) {Efficiency of radial transport of ices in protoplanetary disks probed with infrared observations: the case of CO$\_2$}. ArXiv e-prints

\bibitem[{{Boss}(1997)}]{Boss1997}
{Boss} AP (1997) {Giant planet formation by gravitational instability.} Science 276:1836--1839

\bibitem[{{Bowler}(2016)}]{Bowler2016}
{Bowler} BP (2016) {Imaging Extrasolar Giant Planets}. \pasp 128(10):102,001

\bibitem[{{Brewer} et~al.(2017){Brewer}, {Fischer}, and {Madhusudhan}}]{Brewer16}
{Brewer} JM, {Fischer} DA {Madhusudhan} N (2017) {C/O and O/H Ratios Suggest Some Hot Jupiters Originate Beyond the Snow Line}. \aj 153:83

\bibitem[{{Br{\"u}gger} et~al.(2020){Br{\"u}gger}, {Burn}, {Coleman}, {Alibert}, and {Benz}}]{Brügger2020}
{Br{\"u}gger} N, {Burn} R, {Coleman} GAL, {Alibert} Y {Benz} W (2020) {Pebbles versus planetesimals. The outcomes of population synthesis models}. \aap 640:A21

\bibitem[{{Bryan} and {Lee}(2024)}]{Bryan_Lee2024}
{Bryan} ML {Lee} EJ (2024) {Friends Not Foes: Strong Correlation between Inner Super-Earths and Outer Gas Giants}. \apjl 968(2):L25

\bibitem[{{Burn} et~al.(2024){Burn}, {Mordasini}, {Mishra}, {Haldemann}, {Venturini}, {Emsenhuber}, and {Henning}}]{Burn2024}
{Burn} R, {Mordasini} C, {Mishra} L et~al. (2024) {A radius valley between migrated steam worlds and evaporated rocky cores}. Nature Astronomy 8:463--471

\bibitem[{Burrows and Sharp(1999)}]{burrows_chemical_1999}
Burrows A Sharp CM (1999) Chemical {Equilibrium} {Abundances} in {Brown} {Dwarf} and {Extrasolar} {Giant} {Planet} {Atmospheres}. The Astrophysical Journal 512:843--863, \urlprefix\url{https://ui.adsabs.harvard.edu/abs/1999ApJ...512..843B}, publisher: IOP ADS Bibcode: 1999ApJ...512..843B

\bibitem[{{Butscher} et~al.(2015){Butscher}, {Duvernay}, {Theule}, {Danger}, {Carissan}, {Hagebaum-Reignier}, and {Chiavassa}}]{Butscher2015}
{Butscher} T, {Duvernay} F, {Theule} P et~al. (2015) {Formation mechanism of glycolaldehyde and ethylene glycol in astrophysical ices from HCO$^{•}$ and $^{•}$CH$_{2}$OH recombination: an experimental study}. \mnras 453:1587--1596

\bibitem[{Calamari et~al.(2022)Calamari, Faherty, Burningham, Gonzales, Bardalez-Gagliuffi, Vos, Gemma, Whiteford, and Gaarn}]{calamari_atmospheric_2022}
Calamari E, Faherty JK, Burningham B et~al. (2022) An {Atmospheric} {Retrieval} of the {Brown} {Dwarf} {Gliese} {229B}. The Astrophysical Journal 940:164, \urlprefix\url{https://ui.adsabs.harvard.edu/abs/2022ApJ...940..164C}, publisher: IOP ADS Bibcode: 2022ApJ...940..164C

\bibitem[{Calamari et~al.(2024)Calamari, Faherty, Visscher, Gemma, Burningham, and Rothermich}]{calamari_predicting_2024}
Calamari E, Faherty JK, Visscher C et~al. (2024) Predicting {Cloud} {Conditions} in {Substellar} {Mass} {Objects} {Using} {Ultracool} {Dwarf} {Companions}. The Astrophysical Journal 963(1):67, \urlprefix\url{https://dx.doi.org/10.3847/1538-4357/ad1f6d}, publisher: The American Astronomical Society

\bibitem[{{Chabrier}(2005)}]{Chabrier2005}
{Chabrier} G (2005) {The Initial Mass Function: From Salpeter 1955 to 2005}. In: {Corbelli} E, {Palla} F {Zinnecker} H (eds) The Initial Mass Function 50 Years Later, Astrophysics and Space Science Library, vol 327, p~41, \doi{10.1007/978-1-4020-3407-7_5}

\bibitem[{{Chabrier} and {Baraffe}(2007)}]{ChabrierBaraffe2007}
{Chabrier} G {Baraffe} I (2007) {Heat Transport in Giant (Exo)planets: A New Perspective}. \apjl 661:L81--L84

\bibitem[{Chachan et~al.(2023)Chachan, Knutson, Lothringer, and Blake}]{chachan_breaking_2023}
Chachan Y, Knutson HA, Lothringer J Blake GA (2023) Breaking {Degeneracies} in {Formation} {Histories} by {Measuring} {Refractory} {Content} in {Gas} {Giants}. The Astrophysical Journal 943(2):112, \urlprefix\url{https://dx.doi.org/10.3847/1538-4357/aca614}, publisher: The American Astronomical Society

\bibitem[{{Chambers}(2019)}]{Chambers2019}
{Chambers} J (2019) {An Analytic Model for an Evolving Protoplanetary Disk with a Disk Wind}. \apj 879(2):98

\bibitem[{{Chambers}(2009)}]{Chambers2009}
{Chambers} JE (2009) {An Analytic Model for the Evolution of a Viscous, Irradiated Disk}. \apj 705:1206--1214

\bibitem[{{Chatterjee} and {Ford}(2015)}]{ChatterjeeFord2015}
{Chatterjee} S {Ford} EB (2015) {Planetesimal Interactions Can Explain the Mysterious Period Ratios of Small Near-Resonant Planets}. \apj 803:33

\bibitem[{{Chatterjee} et~al.(2008){Chatterjee}, {Ford}, {Matsumura}, and {Rasio}}]{Chatterjee2008}
{Chatterjee} S, {Ford} EB, {Matsumura} S {Rasio} FA (2008) {Dynamical Outcomes of Planet-Planet Scattering}. \apj 686:580-602

\bibitem[{{Chen} and {Kipping}(2017)}]{ChenKipping2017}
{Chen} J {Kipping} D (2017) {Probabilistic Forecasting of the Masses and Radii of Other Worlds}. \apj 834:17

\bibitem[{{Chiang} and {Laughlin}(2013)}]{ChiangLaughlin2013}
{Chiang} E {Laughlin} G (2013) {The minimum-mass extrasolar nebula: in situ formation of close-in super-Earths}. \mnras 431:3444--3455

\bibitem[{{Chiang} and {Goldreich}(1997)}]{ChiangGoldreich1997}
{Chiang} EI {Goldreich} P (1997) {Spectral Energy Distributions of T Tauri Stars with Passive Circumstellar Disks}. \apj 490:368--376

\bibitem[{Chuang et~al.(2018)Chuang, Fedoseev, Qasim, Ioppolo, van Dishoeck, and Linnartz}]{Chuang2018}
Chuang KJ, Fedoseev G, Qasim D et~al. (2018) Reactive desorption of co hydrogenation products under cold pre-stellar core conditions. The Astrophysical Journal 853(2):102, \urlprefix\url{http://stacks.iop.org/0004-637X/853/i=2/a=102}

\bibitem[{{Cleeves} et~al.(2013){Cleeves}, {Adams}, and {Bergin}}]{Cleeves2013}
{Cleeves} LI, {Adams} FC {Bergin} EA (2013) {Exclusion of Cosmic Rays in Protoplanetary Disks: Stellar and Magnetic Effects}. \apj 772:5

\bibitem[{{Cleeves} et~al.(2014){Cleeves}, {Bergin}, {Alexander}, {Du}, {Graninger}, {{\"O}berg}, and {Harries}}]{Cleeves2014}
{Cleeves} LI, {Bergin} EA, {Alexander} CMO et~al. (2014) {The ancient heritage of water ice in the solar system}. Science 345:1590--1593

\bibitem[{{Cloutier} and {Menou}(2020)}]{Cloutier2020}
{Cloutier} R {Menou} K (2020) {Evolution of the Radius Valley around Low-mass Stars from Kepler and K2}. \aj 159(5):211

\bibitem[{{Coleman} and {Nelson}(2014)}]{ColemanNelson2014}
{Coleman} GAL {Nelson} RP (2014) {On the formation of planetary systems via oligarchic growth in thermally evolving viscous discs}. \mnras 445:479--499

\bibitem[{{Coleman} and {Nelson}(2016)}]{ColemaneNelson2016}
{Coleman} GAL {Nelson} RP (2016) {Giant planet formation in radially structured protoplanetary discs}. \mnras 460:2779--2795

\bibitem[{{Cooper} and {Showman}(2006)}]{Cooper2006}
{Cooper} CS {Showman} AP (2006) {Dynamics and Disequilibrium Carbon Chemistry in Hot Jupiter Atmospheres, with Application to HD 209458b}. \apj 649:1048--1063

\bibitem[{{Cordwell} and {Rafikov}(2024)}]{Cordwell_Rafikov2024}
{Cordwell} AJ {Rafikov} RR (2024) {Early stages of gap opening by planets in protoplanetary discs}. \mnras 534(2):1394--1413

\bibitem[{{Crida} and {Morbidelli}(2007)}]{CridaMorbidelli2007}
{Crida} A {Morbidelli} A (2007) {Cavity opening by a giant planet in a protoplanetary disc and effects on planetary migration}. \mnras 377:1324--1336

\bibitem[{{Cridland} et~al.(2016){Cridland}, {Pudritz}, and {Alessi}}]{Crid16a}
{Cridland} AJ, {Pudritz} RE {Alessi} M (2016) {Composition of early planetary atmospheres - I. Connecting disc astrochemistry to the formation of planetary atmospheres}. \mnras 461:3274--3295

\bibitem[{{Cridland} et~al.(2017{\natexlab{a}}){Cridland}, {Pudritz}, and {Birnstiel}}]{Crid16b}
{Cridland} AJ, {Pudritz} RE {Birnstiel} T (2017{\natexlab{a}}) {Radial drift of dust in protoplanetary discs: the evolution of ice lines and dead zones}. \mnras 465:3865--3878

\bibitem[{{Cridland} et~al.(2017{\natexlab{b}}){Cridland}, {Pudritz}, {Birnstiel}, {Cleeves}, and {Bergin}}]{Crid17}
{Cridland} AJ, {Pudritz} RE, {Birnstiel} T, {Cleeves} LI {Bergin} EA (2017{\natexlab{b}}) {Composition of early planetary atmospheres - II. Coupled Dust and chemical evolution in protoplanetary discs}. \mnras 469:3910--3927

\bibitem[{{Cridland} et~al.(2019{\natexlab{a}}){Cridland}, {Pudritz}, and {Alessi}}]{Crid2019a}
{Cridland} AJ, {Pudritz} RE {Alessi} M (2019{\natexlab{a}}) {Physics of planet trapping with applications to HL Tau}. \mnras 484(1):345--363

\bibitem[{{Cridland} et~al.(2019{\natexlab{b}}){Cridland}, {van Dishoeck}, {Alessi}, and {Pudritz}}]{Crid2019c}
{Cridland} AJ, {van Dishoeck} EF, {Alessi} M {Pudritz} RE (2019{\natexlab{b}}) {Connecting planet formation and astrochemistry. A main sequence for C/O in hot exoplanetary atmospheres}. \aap 632:A63

\bibitem[{{Cridland} et~al.(2020){Cridland}, {van Dishoeck}, {Alessi}, and {Pudritz}}]{Crid2020b}
{Cridland} AJ, {van Dishoeck} EF, {Alessi} M {Pudritz} RE (2020) {Connecting planet formation and astrochemistry. C/Os and N/Os of warm giant planets and Jupiter analogues}. \aap 642:A229

\bibitem[{{Cuzzi} and {Zahnle}(2004)}]{CuzziZahnle2004}
{Cuzzi} JN {Zahnle} KJ (2004) {Material Enhancement in Protoplanetary Nebulae by Particle Drift through Evaporation Fronts}. \apj 614:490--496

\bibitem[{Damasceno et~al.(2024)Damasceno, Seidel, Prinoth, Psaridi, Esparza-Borges, Stangret, Santos, Zapatero-Osorio, Alibert, Allart, Silva, Cointepas, Silva, Cristo, Marcantonio, Ehrenreich, Hernández, Herrero-Cisneros, Lendl, Lillo-Box, Martins, Micela, Pallé, Sousa, Steiner, Vaulato, Zhao, and Pepe}]{damasceno_atmospheric_2024}
Damasceno YC, Seidel JV, Prinoth B et~al. (2024) The atmospheric composition of the ultra-hot {Jupiter} {WASP}-178 b observed with {ESPRESSO}. Astronomy \& Astrophysics 689:A54, \urlprefix\url{https://www.aanda.org/articles/aa/abs/2024/09/aa50119-24/aa50119-24.html}, publisher: EDP Sciences

\bibitem[{{Dittkrist} et~al.(2014){Dittkrist}, {Mordasini}, {Klahr}, {Alibert}, and {Henning}}]{Dittkrist2014}
{Dittkrist} KM, {Mordasini} C, {Klahr} H, {Alibert} Y {Henning} T (2014) {Impacts of planet migration models on planetary populations. Effects of saturation, cooling and stellar irradiation}. \aap 567:A121

\bibitem[{{Dong} et~al.(2015){Dong}, {Zhu}, and {Whitney}}]{Dong2015}
{Dong} R, {Zhu} Z {Whitney} B (2015) {Observational Signatures of Planets in Protoplanetary Disks I. Gaps Opened by Single and Multiple Young Planets in Disks}. \apj 809(1):93

\bibitem[{{Dorn} et~al.(2017){Dorn}, {Venturini}, {Khan}, {Heng}, {Alibert}, {Helled}, {Rivoldini}, and {Benz}}]{Dorn2017}
{Dorn} C, {Venturini} J, {Khan} A et~al. (2017) {A generalized Bayesian inference method for constraining the interiors of super Earths and sub-Neptunes}. \aap 597:A37

\bibitem[{{Drazkowska} et~al.(2023){Drazkowska}, {Bitsch}, {Lambrechts}, {Mulders}, {Harsono}, {Vazan}, {Liu}, {Ormel}, {Kretke}, and {Morbidelli}}]{Drazkowska2023}
{Drazkowska} J, {Bitsch} B, {Lambrechts} M et~al. (2023) {Planet Formation Theory in the Era of ALMA and Kepler: from Pebbles to Exoplanets}. In: {Inutsuka} S, {Aikawa} Y, {Muto} T, {Tomida} K {Tamura} M (eds) Protostars and Planets VII, Astronomical Society of the Pacific Conference Series, vol 534, p 717, \doi{10.48550/arXiv.2203.09759}

\bibitem[{{Drozdovskaya} et~al.(2016){Drozdovskaya}, {Walsh}, {van Dishoeck}, {Furuya}, {Marboeuf}, {Thiabaud}, {Harsono}, and {Visser}}]{Drozdovskaya2016}
{Drozdovskaya} MN, {Walsh} C, {van Dishoeck} EF et~al. (2016) {Cometary ices in forming protoplanetary disc midplanes}. \mnras 462(1):977--993

\bibitem[{{Duffell} et~al.(2014){Duffell}, {Haiman}, {MacFadyen}, {D'Orazio}, and {Farris}}]{Duffell2014}
{Duffell} PC, {Haiman} Z, {MacFadyen} AI, {D'Orazio} DJ {Farris} BD (2014) {The Migration of Gap-opening Planets is Not Locked to Viscous Disk Evolution}. \apjl 792:L10

\bibitem[{{Dullemond} et~al.(2018){Dullemond}, {Birnstiel}, {Huang}, {Kurtovic}, {Andrews}, {Guzm{\'a}n}, {P{\'e}rez}, {Isella}, {Zhu}, {Benisty}, {Wilner}, {Bai}, {Carpenter}, {Zhang}, and {Ricci}}]{Dullemond2018}
{Dullemond} CP, {Birnstiel} T, {Huang} J et~al. (2018) {The Disk Substructures at High Angular Resolution Project (DSHARP). VI. Dust Trapping in Thin-ringed Protoplanetary Disks}. \apjl 869(2):L46

\bibitem[{{Edgar}(2008)}]{Edgar2008}
{Edgar} RG (2008) {Type II Migration: Varying Planet Mass and Disc Viscosity}. ArXiv e-prints

\bibitem[{{Eistrup} et~al.(2016){Eistrup}, {Walsh}, and {van Dishoeck}}]{Eistrup2016}
{Eistrup} C, {Walsh} C {van Dishoeck} EF (2016) {Setting the volatile composition of (exo)planet-building material. Does chemical evolution in disk midplanes matter?} \aap 595:A83

\bibitem[{{Eistrup} et~al.(2018){Eistrup}, {Walsh}, and {van Dishoeck}}]{Eistrup2018}
{Eistrup} C, {Walsh} C {van Dishoeck} EF (2018) {Molecular abundances and C/O ratios in chemically evolving planet-forming disk midplanes}. \aap 613:A14

\bibitem[{{Elser} et~al.(2012){Elser}, {Meyer}, and {Moore}}]{Elser2012}
{Elser} S, {Meyer} MR {Moore} B (2012) {On the origin of elemental abundances in the terrestrial planets}. \icarus 221:859--874

\bibitem[{{Emsenhuber} et~al.(2021){Emsenhuber}, {Mordasini}, {Burn}, {Alibert}, {Benz}, and {Asphaug}}]{Emsenhuber2021}
{Emsenhuber} A, {Mordasini} C, {Burn} R et~al. (2021) {The New Generation Planetary Population Synthesis (NGPPS). I. Bern global model of planet formation and evolution, model tests, and emerging planetary systems}. \aap 656:A69

\bibitem[{{Ercolano} et~al.(2021){Ercolano}, {Picogna}, {Monsch}, {Drake}, and {Preibisch}}]{Ercolano2021}
{Ercolano} B, {Picogna} G, {Monsch} K, {Drake} JJ {Preibisch} T (2021) {The dispersal of protoplanetary discs - II: photoevaporation models with observationally derived irradiating spectra}. \mnras 508(2):1675--1685

\bibitem[{{Fabrycky} and {Tremaine}(2007)}]{FabryckyTremaine2007}
{Fabrycky} D {Tremaine} S (2007) {Shrinking Binary and Planetary Orbits by Kozai Cycles with Tidal Friction}. \apj 669:1298--1315

\bibitem[{{Fabrycky} et~al.(2014){Fabrycky}, {Lissauer}, {Ragozzine}, {Rowe}, {Steffen}, {Agol}, {Barclay}, {Batalha}, {Borucki}, {Ciardi}, {Ford}, {Gautier}, {Geary}, {Holman}, {Jenkins}, {Li}, {Morehead}, {Morris}, {Shporer}, {Smith}, {Still}, and {Van Cleve}}]{Fabrycky2014}
{Fabrycky} DC, {Lissauer} JJ, {Ragozzine} D et~al. (2014) {Architecture of Kepler's Multi-transiting Systems. II. New Investigations with Twice as Many Candidates}. \apj 790:146

\bibitem[{Faedi et~al.(2011)Faedi, Barros, Anderson, Brown, Collier~Cameron, Pollacco, Boisse, Hébrard, Lendl, Lister, Smalley, Street, Triaud, Bento, Bouchy, Butters, Enoch, Haswell, Hellier, Keenan, Miller, Moulds, Moutou, Norton, Queloz, Santerne, Simpson, Skillen, Smith, Udry, Watson, West, and Wheatley}]{faedi_wasp-39b_2011}
Faedi F, Barros SCC, Anderson DR et~al. (2011) {WASP}-39b: a highly inflated {Saturn}-mass planet orbiting a late {G}-type star. Astronomy and Astrophysics 531:A40, \urlprefix\url{https://ui.adsabs.harvard.edu/abs/2011A&A...531A..40F}, publisher: EDP ADS Bibcode: 2011A\&A...531A..40F

\bibitem[{{Fang} and {Margot}(2012)}]{RangMargot2012}
{Fang} J {Margot} JL (2012) {Architecture of Planetary Systems Based on Kepler Data: Number of Planets and Coplanarity}. \apj 761:92

\bibitem[{{Fedele} et~al.(2013){Fedele}, {Bruderer}, {van Dishoeck}, {Hogerheijde}, {Panic}, {Brown}, and {Henning}}]{Fedele2013}
{Fedele} D, {Bruderer} S, {van Dishoeck} EF et~al. (2013) {Probing the Radial Temperature Structure of Protoplanetary Disks with Herschel/HIFI}. \apjl 776:L3

\bibitem[{{Fischer} and {Valenti}(2005)}]{FischerValenti2005}
{Fischer} DA {Valenti} J (2005) {The Planet-Metallicity Correlation}. \apj 622:1102--1117

\bibitem[{{Flaherty} et~al.(2017){Flaherty}, {Hughes}, {Rose}, {Simon}, {Qi}, {Andrews}, {K{\'o}sp{\'a}l}, {Wilner}, {Chiang}, {Armitage}, and {Bai}}]{Flaherty2017}
{Flaherty} KM, {Hughes} AM, {Rose} SC et~al. (2017) {A Three-dimensional View of Turbulence: Constraints on Turbulent Motions in the HD 163296 Protoplanetary Disk Using DCO$^{+}$}. \apj 843(2):150

\bibitem[{{Flaherty} et~al.(2018){Flaherty}, {Hughes}, {Teague}, {Simon}, {Andrews}, and {Wilner}}]{Flaherty2018}
{Flaherty} KM, {Hughes} AM, {Teague} R et~al. (2018) {Turbulence in the TW Hya Disk}. \apj 856(2):117

\bibitem[{{Flock} et~al.(2015){Flock}, {Ruge}, {Dzyurkevich}, {Henning}, {Klahr}, and {Wolf}}]{Flock2015}
{Flock} M, {Ruge} JP, {Dzyurkevich} N et~al. (2015) {Gaps, rings, and non-axisymmetric structures in protoplanetary disks. From simulations to ALMA observations}. \aap 574:A68

\bibitem[{{Fogel} et~al.(2011){Fogel}, {Bethell}, {Bergin}, {Calvet}, and {Semenov}}]{Fogel2011}
{Fogel} JKJ, {Bethell} TJ, {Bergin} EA, {Calvet} N {Semenov} D (2011) {Chemistry of a Protoplanetary Disk with Grain Settling and Ly{$\alpha$} Radiation}. \apj 726:29

\bibitem[{{Franceschi} et~al.(2024){Franceschi}, {Henning}, {Tabone}, {Perotti}, {Caratti o Garatti}, {Bettoni}, {van Dishoeck}, {Kamp}, {Absil}, {G{\"u}del}, {Olofsson}, {Waters}, {Arabhavi}, {Christiaens}, {Gasman}, {Grant}, {Jang}, {Rodgers-Lee}, {Samland}, {Schwarz}, {Temmink}, {Barrado}, {Boccaletti}, {Geers}, {Lagage}, {Pantin}, {Ray}, {Scheithauer}, {Vandenbussche}, and {Wright}}]{Franceschi2024}
{Franceschi} R, {Henning} T, {Tabone} B et~al. (2024) {MINDS: Mid-infrared atomic and molecular hydrogen lines in the inner disk around a low-mass star}. \aap 687:A96

\bibitem[{{Frank} et~al.(2014){Frank}, {Ray}, {Cabrit}, {Hartigan}, {Arce}, {Bacciotti}, {Bally}, {Benisty}, {Eisl{\"o}ffel}, {G{\"u}del}, {Lebedev}, {Nisini}, and {Raga}}]{Frank2014}
{Frank} A, {Ray} TP, {Cabrit} S et~al. (2014) {Jets and Outflows from Star to Cloud: Observations Confront Theory}. Protostars and Planets VI pp 451--474

\bibitem[{{Fulton} et~al.(2017){Fulton}, {Petigura}, {Howard}, {Isaacson}, {Marcy}, {Cargile}, {Hebb}, {Weiss}, {Johnson}, {Morton}, {Sinukoff}, {Crossfield}, and {Hirsch}}]{Fulton2017}
{Fulton} BJ, {Petigura} EA, {Howard} AW et~al. (2017) {The California-Kepler Survey. III. A Gap in the Radius Distribution of Small Planets}. \aj 154(3):109

\bibitem[{{Gammie}(1996)}]{Gammie1996}
{Gammie} CF (1996) {Linear Theory of Magnetized, Viscous, Self-gravitating Gas Disks}. \apj 462:725

\bibitem[{{Gammie}(2001)}]{Gammie2001}
{Gammie} CF (2001) {Nonlinear Outcome of Gravitational Instability in Cooling, Gaseous Disks}. \apj 553(1):174--183

\bibitem[{{Garufi} et~al.(2020){Garufi}, {Avenhaus}, {P{\'e}rez}, {Quanz}, {van Holstein}, {Bertrang}, {Casassus}, {Cieza}, {Principe}, {van der Plas}, and {Zurlo}}]{Garufi2020}
{Garufi} A, {Avenhaus} H, {P{\'e}rez} S et~al. (2020) {Disks Around T Tauri Stars with SPHERE (DARTTS-S). II. Twenty-one new polarimetric images of young stellar disks}. \aap 633:A82

\bibitem[{{Gasman} et~al.(2023){Gasman}, {van Dishoeck}, {Grant}, {Temmink}, {Tabone}, {Henning}, {Kamp}, {G{\"u}del}, {Lagage}, {Perotti}, {Christiaens}, {Samland}, {Arabhavi}, {Argyriou}, {Abergel}, {Absil}, {Barrado}, {Boccaletti}, {Bouwman}, {Caratti o Garatti}, {Geers}, {Glauser}, {Guadarrama}, {Jang}, {Kanwar}, {Lahuis}, {Morales-Calder{\'o}n}, {Mueller}, {Nehm{\'e}}, {Olofsson}, {Pantin}, {Pawellek}, {Ray}, {Rodgers-Lee}, {Scheithauer}, {Schreiber}, {Schwarz}, {Vandenbussche}, {Vlasblom}, {Waters}, {Wright}, {Colina}, {Greve}, and {{\"O}stlin}}]{Gasman2023}
{Gasman} D, {van Dishoeck} EF, {Grant} SL et~al. (2023) {MINDS. Abundant water and varying C/O across the disk of Sz 98 as seen by JWST/MIRI}. \aap 679:A117

\bibitem[{Gibson et~al.(2022)Gibson, Nugroho, Lothringer, Maguire, and Sing}]{gibson_relative_2022}
Gibson NP, Nugroho SK, Lothringer J, Maguire C Sing DK (2022) Relative abundance constraints from high-resolution optical transmission spectroscopy of {WASP}-121b, and a fast model-filtering technique for accelerating retrievals. Monthly Notices of the Royal Astronomical Society 512(3):4618--4638, \urlprefix\url{https://doi.org/10.1093/mnras/stac091}

\bibitem[{{Gillett} and {Forrest}(1973)}]{GillettForrest1973}
{Gillett} FC {Forrest} WJ (1973) {Spectra of the Becklin-Neugebauer point source and the Kleinmann-Low nebula from 2.8 to 13.5 microns.} \apj 179:483--491

\bibitem[{{Ginzburg} et~al.(2018){Ginzburg}, {Schlichting}, and {Sari}}]{Ginzburg2018}
{Ginzburg} S, {Schlichting} HE {Sari} R (2018) {Core-powered mass-loss and the radius distribution of small exoplanets}. \mnras 476(1):759--765

\bibitem[{{Glassgold} et~al.(2004){Glassgold}, {Najita}, and {Igea}}]{Glassgold2004}
{Glassgold} AE, {Najita} J {Igea} J (2004) {Heating Protoplanetary Disk Atmospheres}. \apj 615(2):972--990

\bibitem[{{Goldreich} and {Tremaine}(1979)}]{GoldreichTremaine1979}
{Goldreich} P {Tremaine} S (1979) {The excitation of density waves at the Lindblad and corotation resonances by an external potential}. \apj 233:857--871

\bibitem[{{Gonz{\'a}lez-Cataldo} et~al.(2014){Gonz{\'a}lez-Cataldo}, {Wilson}, and {Militzer}}]{Gonzalez2014}
{Gonz{\'a}lez-Cataldo} F, {Wilson} HF {Militzer} B (2014) {Ab Initio Free Energy Calculations of the Solubility of Silica in Metallic Hydrogen and Application to Giant Planet Cores}. \apj 787:79

\bibitem[{{Gorti} et~al.(2016){Gorti}, {Liseau}, {S{\'a}ndor}, and {Clarke}}]{Gorti2016}
{Gorti} U, {Liseau} R, {S{\'a}ndor} Z {Clarke} C (2016) {Disk Dispersal: Theoretical Understanding and Observational Constraints}. \ssr 205:125--152

\bibitem[{Grant et~al.(2023)Grant, Lewis, Wakeford, Batalha, Glidden, Goyal, Mullens, MacDonald, May, Seager, Stevenson, Valenti, Visscher, Alderson, Allen, Cañas, Colón, Clampin, Espinoza, Gressier, Huang, Lin, Long, Louie, Peña-Guerrero, Ranjan, Sotzen, Valentine, Anderson, Balmer, Bellini, Hoch, Kammerer, Libralato, Mountain, Perrin, Pueyo, Rickman, Rebollido, Sohn, Marel, and Watkins}]{grant_jwst-tst_2023}
Grant D, Lewis NK, Wakeford HR et~al. (2023) {JWST}-{TST} {DREAMS}: {Quartz} {Clouds} in the {Atmosphere} of {WASP}-17b. The Astrophysical Journal Letters 956(2):L32, \urlprefix\url{https://dx.doi.org/10.3847/2041-8213/acfc3b}, publisher: The American Astronomical Society

\bibitem[{{Grant} et~al.(2023){Grant}, {van Dishoeck}, {Tabone}, {Gasman}, {Henning}, {Kamp}, {G{\"u}del}, {Lagage}, {Bettoni}, {Perotti}, {Christiaens}, {Samland}, {Arabhavi}, {Argyriou}, {Abergel}, {Absil}, {Barrado}, {Boccaletti}, {Bouwman}, {Caratti o Garatti}, {Geers}, {Glauser}, {Guadarrama}, {Jang}, {Kanwar}, {Lahuis}, {Morales-Calder{\'o}n}, {Mueller}, {Nehm{\'e}}, {Olofsson}, {Pantin}, {Pawellek}, {Ray}, {Rodgers-Lee}, {Scheithauer}, {Schreiber}, {Schwarz}, {Temmink}, {Vandenbussche}, {Vlasblom}, {Waters}, {Wright}, {Colina}, {Greve}, {Justannont}, and {{\"O}stlin}}]{Grant2023}
{Grant} SL, {van Dishoeck} EF, {Tabone} B et~al. (2023) {MINDS. The Detection of $^{13}$CO$_{2}$ with JWST-MIRI Indicates Abundant CO$_{2}$ in a Protoplanetary Disk}. \apjl 947(1):L6

\bibitem[{{Grasset} et~al.(2009){Grasset}, {Schneider}, and {Sotin}}]{Grasset2009}
{Grasset} O, {Schneider} J {Sotin} C (2009) {A Study of the Accuracy of Mass-Radius Relationships for Silicate-Rich and Ice-Rich Planets up to 100 Earth Masses}. \apj 693:722--733

\bibitem[{{Gressel} et~al.(2015){Gressel}, {Turner}, {Nelson}, and {McNally}}]{Gressel2015}
{Gressel} O, {Turner} NJ, {Nelson} RP {McNally} CP (2015) {Global Simulations of Protoplanetary Disks With Ohmic Resistivity and Ambipolar Diffusion}. \apj 801:84

\bibitem[{{Haffert} et~al.(2019){Haffert}, {Bohn}, {de Boer}, {Snellen}, {Brinchmann}, {Girard}, {Keller}, and {Bacon}}]{Haffert2019}
{Haffert} SY, {Bohn} AJ, {de Boer} J et~al. (2019) {Two accreting protoplanets around the young star PDS 70}. Nature Astronomy 3:749--754

\bibitem[{{Haisch} et~al.(2001){Haisch}, {Lada}, and {Lada}}]{Haisch2001}
{Haisch} KE Jr, {Lada} EA {Lada} CJ (2001) {Disk Frequencies and Lifetimes in Young Clusters}. \apjl 553:L153--L156

\bibitem[{{Haldemann} et~al.(2020){Haldemann}, {Alibert}, {Mordasini}, and {Benz}}]{Haldemann2020}
{Haldemann} J, {Alibert} Y, {Mordasini} C {Benz} W (2020) {AQUA: a collection of H$_{2}$O equations of state for planetary models}. \aap 643:A105

\bibitem[{{Haldemann} et~al.(2024){Haldemann}, {Dorn}, {Venturini}, {Alibert}, and {Benz}}]{Haldemann2024}
{Haldemann} J, {Dorn} C, {Venturini} J, {Alibert} Y {Benz} W (2024) {BICEPS: An improved characterization model for low- and intermediate-mass exoplanets}. \aap 681:A96

\bibitem[{{Hansen} and {Murray}(2013)}]{HansenMurray2013}
{Hansen} BMS {Murray} N (2013) {Testing in Situ Assembly with the Kepler Planet Candidate Sample}. \apj 775:53

\bibitem[{{Hartmann}(2008)}]{Hartmann2008}
{Hartmann} L (2008) {Masses and mass distributions of protoplanetary disks}. Physica Scripta Volume T 130(1):014012

\bibitem[{{Hartmann} and {Kenyon}(1987)}]{HartmannKenyon1987}
{Hartmann} L {Kenyon} SJ (1987) {High spectral resolution infrared observations of V1057 Cygni}. \apj 322:393--398

\bibitem[{{Hasegawa} and {Pudritz}(2011)}]{HP11}
{Hasegawa} Y {Pudritz} RE (2011) {The origin of planetary system architectures - I. Multiple planet traps in gaseous discs}. \mnras 417:1236--1259

\bibitem[{{Hasegawa} and {Pudritz}(2012)}]{HP12}
{Hasegawa} Y {Pudritz} RE (2012) {Evolutionary Tracks of Trapped, Accreting Protoplanets: The Origin of the Observed Mass-Period Relation}. \apj 760(2):117

\bibitem[{{Hasegawa} and {Pudritz}(2013)}]{HP13}
{Hasegawa} Y {Pudritz} RE (2013) {Planetary Populations in the Mass-Period Diagram: A Statistical Treatment of Exoplanet Formation and the Role of Planet Traps}. \apj 778:78

\bibitem[{{Hasegawa} and {Pudritz}(2014)}]{HP14}
{Hasegawa} Y {Pudritz} RE (2014) {Planet Traps and Planetary Cores: Origins of the Planet-Metallicity Correlation}. \apj 794:25

\bibitem[{{Hasegawa} et~al.(2017){Hasegawa}, {Okuzumi}, {Flock}, and {Turner}}]{Hasegawa2017}
{Hasegawa} Y, {Okuzumi} S, {Flock} M {Turner} NJ (2017) {Magnetically Induced Disk Winds and Transport in the HL Tau Disk}. \apj 845(1):31

\bibitem[{{Helled} et~al.(2014){Helled}, {Bodenheimer}, {Podolak}, {Boley}, {Meru}, {Nayakshin}, {Fortney}, {Mayer}, {Alibert}, and {Boss}}]{Helled2014}
{Helled} R, {Bodenheimer} P, {Podolak} M et~al. (2014) {Giant Planet Formation, Evolution, and Internal Structure}. Protostars and Planets VI pp 643--665

\bibitem[{{Helling} et~al.(2014){Helling}, {Woitke}, {Rimmer}, {Kamp}, {Thi}, and {Meijerink}}]{Helling2014}
{Helling} C, {Woitke} P, {Rimmer} PB et~al. (2014) {Disk Evolution, Element Abundances and Cloud Properties of Young Gas Giant Planets}. Life 4

\bibitem[{{Henning} and {Semenov}(2013)}]{HenningSemenov2013}
{Henning} T {Semenov} D (2013) {Chemistry in Protoplanetary Disks}. Chemical Reviews 113:9016--9042

\bibitem[{{Henning} et~al.(2024){Henning}, {Kamp}, {Samland}, {Arabhavi}, {Kanwar}, {van Dishoeck}, {G{\"u}del}, {Lagage}, {Waelkens}, {Abergel}, {Absil}, {Barrado}, {Boccaletti}, {Bouwman}, {Caratti o Garatti}, {Geers}, {Glauser}, {Lahuis}, {Mueller}, {Nehm{\'e}}, {Olofsson}, {Pantin}, {Ray}, {Scheithauer}, {Vandenbussche}, {Waters}, {Wright}, {Argyriou}, {Christiaens}, {Franceschi}, {Gasman}, {Grant}, {Guadarrama}, {Jang}, {Morales-Calder{\'o}n}, {Pawellek}, {Perotti}, {Rodgers-Lee}, {Schreiber}, {Schwarz}, {Tabone}, {Temmink}, {Vlasblom}, {Colina}, {Greve}, and {{\"O}stlin}}]{Henning_MINDS2024}
{Henning} T, {Kamp} I, {Samland} M et~al. (2024) {MINDS: The JWST MIRI Mid-INfrared Disk Survey}. \pasp 136(5):054302

\bibitem[{{Hern{\'a}ndez} et~al.(2007){Hern{\'a}ndez}, {Calvet}, {Brice{\~n}o}, {Hartmann}, {Vivas}, {Muzerolle}, {Downes}, {Allen}, and {Gutermuth}}]{Hernandez2007}
{Hern{\'a}ndez} J, {Calvet} N, {Brice{\~n}o} C et~al. (2007) {Spitzer Observations of the Orion OB1 Association: Disk Census in the Low-Mass Stars}. \apj 671:1784--1799

\bibitem[{Hinkel et~al.(2014)Hinkel, Timmes, Young, Pagano, and Turnbull}]{hinkel2014}
Hinkel NR, Timmes FX, Young PA, Pagano MD Turnbull MC (2014) Stellar {Abundances} in the {Solar} {Neighborhood}: {The} {Hypatia} {Catalog}. The Astronomical Journal 148:54, \urlprefix\url{https://ui.adsabs.harvard.edu/abs/2014AJ....148...54H}, publisher: IOP ADS Bibcode: 2014AJ....148...54H

\bibitem[{{Hollenbach} et~al.(1994){Hollenbach}, {Johnstone}, {Lizano}, and {Shu}}]{Hollenbach1994}
{Hollenbach} D, {Johnstone} D, {Lizano} S {Shu} F (1994) {Photoevaporation of Disks around Massive Stars and Application to Ultracompact H II Regions}. \apj 428:654

\bibitem[{{Howard} et~al.(2010){Howard}, {Marcy}, {Johnson}, {Fischer}, {Wright}, {Isaacson}, {Valenti}, {Anderson}, {Lin}, and {Ida}}]{Howard2010}
{Howard} AW, {Marcy} GW, {Johnson} JA et~al. (2010) {The Occurrence and Mass Distribution of Close-in Super-Earths, Neptunes, and Jupiters}. Science 330:653

\bibitem[{{Howard} et~al.(2012){Howard}, {Marcy}, {Bryson}, {Jenkins}, {Rowe}, {Batalha}, {Borucki}, {Koch}, {Dunham}, {Gautier}, {Van Cleve}, {Cochran}, {Latham}, {Lissauer}, {Torres}, {Brown}, {Gilliland}, {Buchhave}, {Caldwell}, {Christensen-Dalsgaard}, {Ciardi}, {Fressin}, {Haas}, {Howell}, {Kjeldsen}, {Seager}, {Rogers}, {Sasselov}, {Steffen}, {Basri}, {Charbonneau}, {Christiansen}, {Clarke}, {Dupree}, {Fabrycky}, {Fischer}, {Ford}, {Fortney}, {Tarter}, {Girouard}, {Holman}, {Johnson}, {Klaus}, {Machalek}, {Moorhead}, {Morehead}, {Ragozzine}, {Tenenbaum}, {Twicken}, {Quinn}, {Isaacson}, {Shporer}, {Lucas}, {Walkowicz}, {Welsh}, {Boss}, {Devore}, {Gould}, {Smith}, {Morris}, {Prsa}, {Morton}, {Still}, {Thompson}, {Mullally}, {Endl}, and {MacQueen}}]{Howard2012}
{Howard} AW, {Marcy} GW, {Bryson} ST et~al. (2012) {Planet Occurrence within 0.25 AU of Solar-type Stars from Kepler}. \apjs 201:15

\bibitem[{{Howard} et~al.(2013){Howard}, {Sanchis-Ojeda}, {Marcy}, {Johnson}, {Winn}, {Isaacson}, {Fischer}, {Fulton}, {Sinukoff}, and {Fortney}}]{Howard2013}
{Howard} AW, {Sanchis-Ojeda} R, {Marcy} GW et~al. (2013) {A rocky composition for an Earth-sized exoplanet}. \nat 503:381--384

\bibitem[{{Huang} et~al.(2018){Huang}, {Andrews}, {Dullemond}, {Isella}, {P{\'e}rez}, {Guzm{\'a}n}, {{\"O}berg}, {Zhu}, {Zhang}, {Bai}, {Benisty}, {Birnstiel}, {Carpenter}, {Hughes}, {Ricci}, {Weaver}, and {Wilner}}]{Huang2018}
{Huang} J, {Andrews} SM, {Dullemond} CP et~al. (2018) {The Disk Substructures at High Angular Resolution Project (DSHARP). II. Characteristics of Annular Substructures}. \apjl 869(2):L42

\bibitem[{{Huang} et~al.(2025){Huang}, {Yu}, {Lee}, {Dong}, and {Bai}}]{Huang2025}
{Huang} P, {Yu} F, {Lee} EJ, {Dong} R {Bai} XN (2025) {Leaky Dust Traps in Planet-Embedded Protoplanetary Disks}. arXiv e-prints arXiv:2503.19026

\bibitem[{{Hughes} et~al.(2018){Hughes}, {Duch{\^e}ne}, and {Matthews}}]{Hughes2018}
{Hughes} AM, {Duch{\^e}ne} G {Matthews} BC (2018) {Debris Disks: Structure, Composition, and Variability}. \araa 56:541--591

\bibitem[{{Ida} and {Lin}(2004{\natexlab{a}})}]{IdaLin2004}
{Ida} S {Lin} DNC (2004{\natexlab{a}}) {Toward a Deterministic Model of Planetary Formation. I. A Desert in the Mass and Semimajor Axis Distributions of Extrasolar Planets}. \apj 604:388--413

\bibitem[{{Ida} and {Lin}(2004{\natexlab{b}})}]{IdaLin2004b}
{Ida} S {Lin} DNC (2004{\natexlab{b}}) {Toward a Deterministic Model of Planetary Formation. II. The Formation and Retention of Gas Giant Planets around Stars with a Range of Metallicities}. \apj 616:567--572

\bibitem[{{Ida} and {Lin}(2005)}]{IdaLin2005}
{Ida} S {Lin} DNC (2005) {Toward a Deterministic Model of Planetary Formation. III. Mass Distribution of Short-Period Planets around Stars of Various Masses}. \apj 626:1045--1060

\bibitem[{{Ida} and {Lin}(2008{\natexlab{a}})}]{IdaLin2008}
{Ida} S {Lin} DNC (2008{\natexlab{a}}) {Toward a Deterministic Model of Planetary Formation. IV. Effects of Type I Migration}. \apj 673:487-501

\bibitem[{{Ida} and {Lin}(2008{\natexlab{b}})}]{IdaLin2008b}
{Ida} S {Lin} DNC (2008{\natexlab{b}}) {Toward a Deterministic Model of Planetary Formation. V. Accumulation Near the Ice Line and Super-Earths}. \apj 685:584-595

\bibitem[{{Ikoma} et~al.(2000){Ikoma}, {Nakazawa}, and {Emori}}]{Ikoma2000}
{Ikoma} M, {Nakazawa} K {Emori} H (2000) {Formation of Giant Planets: Dependences on Core Accretion Rate and Grain Opacity}. \apj 537:1013--1025

\bibitem[{Inglis et~al.(2024)Inglis, Batalha, Lewis, Kataria, Knutson, Kilpatrick, Gagnebin, Mukherjee, Pettyjohn, Crossfield, Foote, Grant, Henry, Lally, McKemmish, Sing, Wakeford, Zapata~Trujillo, and Zellem}]{inglis_quartz_2024}
Inglis J, Batalha NE, Lewis NK et~al. (2024) Quartz {Clouds} in the {Dayside} {Atmosphere} of the {Quintessential} {Hot} {Jupiter} {HD} 189733 b. The Astrophysical Journal 973:L41, \urlprefix\url{https://ui.adsabs.harvard.edu/abs/2024ApJ...973L..41I}, publisher: IOP ADS Bibcode: 2024ApJ...973L..41I

\bibitem[{{Javoy}(1995)}]{Javoy1995}
{Javoy} M (1995) {The integral enstatite chondrite model of the Earth}. \grl 22:2219--2222

\bibitem[{{Jin} et~al.(2016){Jin}, {Li}, {Isella}, {Li}, and {Ji}}]{Jin2016}
{Jin} S, {Li} S, {Isella} A, {Li} H {Ji} J (2016) {Modeling Dust Emission of HL Tau Disk Based on Planet-Disk Interactions}. \apj 818(1):76

\bibitem[{{Johansen} and {Lambrechts}(2017)}]{Johansen2017}
{Johansen} A {Lambrechts} M (2017) {Forming Planets via Pebble Accretion}. Annual Review of Earth and Planetary Sciences 45(1):359--387

\bibitem[{{Johansen} et~al.(2007){Johansen}, {Oishi}, {Mac Low}, {Klahr}, {Henning}, and {Youdin}}]{Johansen2007}
{Johansen} A, {Oishi} JS, {Mac Low} MM et~al. (2007) {Rapid planetesimal formation in turbulent circumstellar disks}. \nat 448:1022--1025

\bibitem[{{Johnson} et~al.(2010){Johnson}, {Aller}, {Howard}, and {Crepp}}]{Johnson2010}
{Johnson} JA, {Aller} KM, {Howard} AW {Crepp} JR (2010) {Giant Planet Occurrence in the Stellar Mass-Metallicity Plane}. \pasp 122:905

\bibitem[{{J{\o}rgensen} et~al.(2009){J{\o}rgensen}, {van Dishoeck}, {Visser}, {Bourke}, {Wilner}, {Lommen}, {Hogerheijde}, and {Myers}}]{Jorgensen2009}
{J{\o}rgensen} JK, {van Dishoeck} EF, {Visser} R et~al. (2009) {PROSAC: a submillimeter array survey of low-mass protostars. II. The mass evolution of envelopes, disks, and stars from the Class 0 through I stages}. \aap 507:861--879

\bibitem[{{J{\o}rgensen} et~al.(2012){J{\o}rgensen}, {Favre}, {Bisschop}, {Bourke}, {van Dishoeck}, and {Schmalzl}}]{Jorgensen2012}
{J{\o}rgensen} JK, {Favre} C, {Bisschop} SE et~al. (2012) {Detection of the Simplest Sugar, Glycolaldehyde, in a Solar-type Protostar with ALMA}. \apjl 757:L4

\bibitem[{{Juri{\'c}} and {Tremaine}(2008)}]{JuricTremaine2008}
{Juri{\'c}} M {Tremaine} S (2008) {Dynamical Origin of Extrasolar Planet Eccentricity Distribution}. \apj 686:603-620

\bibitem[{{Kamp} et~al.(2023){Kamp}, {Henning}, {Arabhavi}, {Bettoni}, {Christiaens}, {Gasman}, {Grant}, {Morales-Calder{\'o}n}, {Tabone}, {Abergel}, {Absil}, {Argyriou}, {Barrado}, {Boccaletti}, {Bouwman}, {Caratti o Garatti}, {van Dishoeck}, {Geers}, {Glauser}, {G{\"u}del}, {Guadarrama}, {Jang}, {Kanwar}, {Lagage}, {Lahuis}, {Mueller}, {Nehm{\'e}}, {Olofsson}, {Pantin}, {Pawellek}, {Perotti}, {Ray}, {Rodgers-Lee}, {Samland}, {Scheithauer}, {Schreiber}, {Schwarz}, {Temmink}, {Vandenbussche}, {Vlasblom}, {Waelkens}, {Waters}, and {Wright}}]{Kamp2023}
{Kamp} I, {Henning} T, {Arabhavi} AM et~al. (2023) {The chemical inventory of the inner regions of planet-forming disks {\textendash} the JWST/MINDS program}. Faraday Discussions 245:112--137

\bibitem[{{Kanagawa} et~al.(2015{\natexlab{a}}){Kanagawa}, {Muto}, {Tanaka}, {Tanigawa}, {Takeuchi}, {Tsukagoshi}, and {Momose}}]{Kanagawa2015b}
{Kanagawa} KD, {Muto} T, {Tanaka} H et~al. (2015{\natexlab{a}}) {Mass Estimates of a Giant Planet in a Protoplanetary Disk from the Gap Structures}. \apjl 806(1):L15

\bibitem[{{Kanagawa} et~al.(2015{\natexlab{b}}){Kanagawa}, {Tanaka}, {Muto}, {Tanigawa}, and {Takeuchi}}]{Kanagawa2015a}
{Kanagawa} KD, {Tanaka} H, {Muto} T, {Tanigawa} T {Takeuchi} T (2015{\natexlab{b}}) {Formation of a disc gap induced by a planet: effect of the deviation from Keplerian disc rotation}. \mnras 448(1):994--1006

\bibitem[{{Kanwar} et~al.(2024){Kanwar}, {Kamp}, {Jang}, {Waters}, {van Dishoeck}, {Christiaens}, {Arabhavi}, {Henning}, {G{\"u}del}, {Woitke}, {Absil}, {Barrado}, {Caratti o Garatti}, {Glauser}, {Lahuis}, {Scheithauer}, {Vandenbussche}, {Gasman}, {Grant}, {Kurtovic}, {Perotti}, {Tabone}, and {Temmink}}]{Kanwar2024}
{Kanwar} J, {Kamp} I, {Jang} H et~al. (2024) {MINDS. Hydrocarbons detected by JWST/MIRI in the inner disk of Sz28 consistent with a high C/O gas-phase chemistry}. \aap 689:A231

\bibitem[{{Kempton} and {Knutson}(2024)}]{Kempton_Knutson2024}
{Kempton} EMR {Knutson} HA (2024) {Transiting Exoplanet Atmospheres in the Era of JWST}. Reviews in Mineralogy and Geochemistry 90(1):411--464

\bibitem[{{Keppler} et~al.(2018){Keppler}, {Benisty}, {M{\"u}ller}, {Henning}, {van Boekel}, {Cantalloube}, {Ginski}, {van Holstein}, {Maire}, {Pohl}, {Samland}, {Avenhaus}, {Baudino}, {Boccaletti}, {de Boer}, {Bonnefoy}, {Chauvin}, {Desidera}, {Langlois}, {Lazzoni}, {Marleau}, {Mordasini}, {Pawellek}, {Stolker}, {Vigan}, {Zurlo}, {Birnstiel}, {Brandner}, {Feldt}, {Flock}, {Girard}, {Gratton}, {Hagelberg}, {Isella}, {Janson}, {Juhasz}, {Kemmer}, {Kral}, {Lagrange}, {Launhardt}, {Matter}, {M{\'e}nard}, {Milli}, {Molli{\`e}re}, {Olofsson}, {P{\'e}rez}, {Pinilla}, {Pinte}, {Quanz}, {Schmidt}, {Udry}, {Wahhaj}, {Williams}, {Buenzli}, {Cudel}, {Dominik}, {Galicher}, {Kasper}, {Lannier}, {Mesa}, {Mouillet}, {Peretti}, {Perrot}, {Salter}, {Sissa}, {Wildi}, {Abe}, {Antichi}, {Augereau}, {Baruffolo}, {Baudoz}, {Bazzon}, {Beuzit}, {Blanchard}, {Brems}, {Buey}, {De Caprio}, {Carbillet}, {Carle}, {Cascone}, {Cheetham}, {Claudi}, {Costille}, {Delboulb{\'e}}, {Dohlen}, {Fantinel}, {Feautrier}, {Fusco}, {Giro}, {Gluck},
  {Gry}, {Hubin}, {Hugot}, {Jaquet}, {Le Mignant}, {Llored}, {Madec}, {Magnard}, {Martinez}, {Maurel}, {Meyer}, {M{\"o}ller-Nilsson}, {Moulin}, {Mugnier}, {Orign{\'e}}, {Pavlov}, {Perret}, {Petit}, {Pragt}, {Puget}, {Rabou}, {Ramos}, {Rigal}, {Rochat}, {Roelfsema}, {Rousset}, {Roux}, {Salasnich}, {Sauvage}, {Sevin}, {Soenke}, {Stadler}, {Suarez}, {Turatto}, and {Weber}}]{Keppler2018}
{Keppler} M, {Benisty} M, {M{\"u}ller} A et~al. (2018) {Discovery of a planetary-mass companion within the gap of the transition disk around PDS 70}. \aap 617:A44

\bibitem[{{Kimmig} et~al.(2020){Kimmig}, {Dullemond}, and {Kley}}]{Kimmig2020}
{Kimmig} CN, {Dullemond} CP {Kley} W (2020) {Effect of wind-driven accretion on planetary migration}. \aap 633:A4

\bibitem[{{Kimura} and {Ikoma}(2022)}]{Kimura_Ikoma2022}
{Kimura} T {Ikoma} M (2022) {Predicted diversity in water content of terrestrial exoplanets orbiting M dwarfs}. Nature Astronomy 6:1296--1307

\bibitem[{{Klassen} et~al.(2016){Klassen}, {Pudritz}, {Kuiper}, {Peters}, and {Banerjee}}]{Klassen2016}
{Klassen} M, {Pudritz} RE, {Kuiper} R, {Peters} T {Banerjee} R (2016) {Simulating the Formation of Massive Protostars. I. Radiative Feedback and Accretion Disks}. \apj 823:28

\bibitem[{{Kley} and {Nelson}(2012)}]{KleyNelson2012}
{Kley} W {Nelson} RP (2012) {Planet-Disk Interaction and Orbital Evolution}. \araa 50:211--249

\bibitem[{{Kokubo} and {Ida}(2002)}]{KokiboIda2002}
{Kokubo} E {Ida} S (2002) {Formation of Protoplanet Systems and Diversity of Planetary Systems}. \apj 581:666--680

\bibitem[{{Konigl} and {Pudritz}(2000)}]{Konigl_Pudritz2000}
{Konigl} A {Pudritz} RE (2000) {Disk Winds and the Accretion-Outflow Connection}. In: {Mannings} V, {Boss} AP {Russell} SS (eds) Protostars and Planets IV, p 759, \doi{10.48550/arXiv.astro-ph/9903168}

\bibitem[{{Kratter} et~al.(2008){Kratter}, {Matzner}, and {Krumholz}}]{Kratter2008}
{Kratter} KM, {Matzner} CD {Krumholz} MR (2008) {Global Models for the Evolution of Embedded, Accreting Protostellar Disks}. \apj 681:375-390

\bibitem[{{Krijt} and {Ciesla}(2016)}]{Krijt2016}
{Krijt} S {Ciesla} FJ (2016) {Dust Diffusion and Settling in the Presence of Collisions: Trapping (sub)micron Grains in the Midplane}. \apj 822:111

\bibitem[{{Krijt} et~al.(2020){Krijt}, {Bosman}, {Zhang}, {Schwarz}, {Ciesla}, and {Bergin}}]{Krijt2020}
{Krijt} S, {Bosman} AD, {Zhang} K et~al. (2020) {CO Depletion in Protoplanetary Disks: A Unified Picture Combining Physical Sequestration and Chemical Processing}. \apj 899(2):134

\bibitem[{{Kuffmeier}(2024)}]{Kuffmeier2024}
{Kuffmeier} M (2024) {Magnetohydrodynamical modeling of star-disk formation: from isolated spherical collapse towards incorporation of external dynamics}. Frontiers in Astronomy and Space Sciences 11:1403075

\bibitem[{{Kuffmeier} et~al.(2020){Kuffmeier}, {Goicovic}, and {Dullemond}}]{Kuffmeier2020}
{Kuffmeier} M, {Goicovic} FG {Dullemond} CP (2020) {Late encounter events as source of disks and spiral structures. Forming second generation disks}. \aap 633:A3

\bibitem[{{Lambrechts} and {Johansen}(2014)}]{Lambrechts2014}
{Lambrechts} M {Johansen} A (2014) {Forming the cores of giant planets from the radial pebble flux in protoplanetary discs}. \aap 572:A107

\bibitem[{{Lau} et~al.(2022){Lau}, {Dr{\k{a}}zkowska}, {Stammler}, {Birnstiel}, and {Dullemond}}]{Lau2022}
{Lau} TCH, {Dr{\k{a}}zkowska} J, {Stammler} SM, {Birnstiel} T {Dullemond} CP (2022) {Rapid formation of massive planetary cores in a pressure bump}. \aap 668:A170

\bibitem[{{Leconte} and {Chabrier}(2012)}]{LeconteChabrier2012}
{Leconte} J {Chabrier} G (2012) {A new vision of giant planet interiors: Impact of double diffusive convection}. \aap 540:A20

\bibitem[{{Leconte} and {Chabrier}(2013)}]{LeconteChabrier2013}
{Leconte} J {Chabrier} G (2013) {Layered convection as the origin of Saturn's luminosity anomaly}. Nature Geoscience 6:347--350

\bibitem[{{Lesur} et~al.(2014){Lesur}, {Kunz}, and {Fromang}}]{Lesur2014}
{Lesur} G, {Kunz} MW {Fromang} S (2014) {Thanatology in protoplanetary discs. The combined influence of Ohmic, Hall, and ambipolar diffusion on dead zones}. \aap 566:A56

\bibitem[{{Lesur} et~al.(2023){Lesur}, {Flock}, {Ercolano}, {Lin}, {Yang}, {Barranco}, {Benitez-Llambay}, {Goodman}, {Johansen}, {Klahr}, {Laibe}, {Lyra}, {Marcus}, {Nelson}, {Squire}, {Simon}, {Turner}, {Umurhan}, and {Youdin}}]{Lesur2023}
{Lesur} G, {Flock} M, {Ercolano} B et~al. (2023) {Hydro-, Magnetohydro-, and Dust-Gas Dynamics of Protoplanetary Disks}. In: {Inutsuka} S, {Aikawa} Y, {Muto} T, {Tomida} K {Tamura} M (eds) Protostars and Planets VII, Astronomical Society of the Pacific Conference Series, vol 534, p 465, \doi{10.48550/arXiv.2203.09821}

\bibitem[{{Li} et~al.(2014){Li}, {Banerjee}, {Pudritz}, {J{\o}rgensen}, {Shang}, {Krasnopolsky}, and {Maury}}]{Li2014}
{Li} ZY, {Banerjee} R, {Pudritz} RE et~al. (2014) {The Earliest Stages of Star and Planet Formation: Core Collapse, and the Formation of Disks and Outflows}. Protostars and Planets VI pp 173--194

\bibitem[{{Ligterink} et~al.(2017){Ligterink}, {Coutens}, {Kofman}, {M{\"u}ller}, {Garrod}, {Calcutt}, {Wampfler}, {J{\o}rgensen}, {Linnartz}, and {van Dishoeck}}]{Ligterink2017}
{Ligterink} NFW, {Coutens} A, {Kofman} V et~al. (2017) {The ALMA-PILS survey: detection of CH$_{3}$NCO towards the low-mass protostar IRAS 16293-2422 and laboratory constraints on its formation}. \mnras 469:2219--2229

\bibitem[{{Lin} and {Papaloizou}(1986)}]{LinPapaloizou1986}
{Lin} DNC {Papaloizou} J (1986) {On the tidal interaction between protoplanets and the primordial solar nebula. II - Self-consistent nonlinear interaction}. \apj 307:395--409

\bibitem[{{Lin} and {Papaloizou}(1993)}]{LinPapaloizou1993}
{Lin} DNC {Papaloizou} JCB (1993) {On the tidal interaction between protostellar disks and companions}. In: {Levy} EH {Lunine} JI (eds) Protostars and Planets III, pp 749--835

\bibitem[{Line et~al.(2015)Line, Teske, Burningham, Fortney, and Marley}]{line_uniform_2015}
Line MR, Teske J, Burningham B, Fortney JJ Marley MS (2015) Uniform {Atmospheric} {Retrieval} {Analysis} of {Ultracool} {Dwarfs}. {I}. {Characterizing} {Benchmarks}, {Gl} {570D} and {HD} {3651B}. The Astrophysical Journal 807:183, \urlprefix\url{https://ui.adsabs.harvard.edu/abs/2015ApJ...807..183L}, publisher: IOP ADS Bibcode: 2015ApJ...807..183L

\bibitem[{{Lissauer} et~al.(2011){Lissauer}, {Ragozzine}, {Fabrycky}, {Steffen}, {Ford}, {Jenkins}, {Shporer}, {Holman}, {Rowe}, {Quintana}, {Batalha}, {Borucki}, {Bryson}, {Caldwell}, {Carter}, {Ciardi}, {Dunham}, {Fortney}, {Gautier}, {Howell}, {Koch}, {Latham}, {Marcy}, {Morehead}, and {Sasselov}}]{Lissauer2011}
{Lissauer} JJ, {Ragozzine} D, {Fabrycky} DC et~al. (2011) {Architecture and Dynamics of Kepler's Candidate Multiple Transiting Planet Systems}. \apjs 197:8

\bibitem[{Lothringer et~al.(2021)Lothringer, Rustamkulov, Sing, Gibson, Wilson, and Schlaufman}]{lothringer2021}
Lothringer JD, Rustamkulov Z, Sing DK et~al. (2021) A {New} {Window} into {Planet} {Formation} and {Migration}: {Refractory}-to-{Volatile} {Elemental} {Ratios} in {Ultra}-hot {Jupiters}. The Astrophysical Journal 914(1):12, \urlprefix\url{http://arxiv.org/abs/2011.10626}, arXiv:2011.10626 [astro-ph]

\bibitem[{{Lozovsky} et~al.(2017){Lozovsky}, {Helled}, {Rosenberg}, and {Bodenheimer}}]{Lozovsky2017}
{Lozovsky} M, {Helled} R, {Rosenberg} ED {Bodenheimer} P (2017) {Jupiter's Formation and Its Primordial Internal Structure}. \apj 836:227

\bibitem[{{Luo} et~al.(2024){Luo}, {Dorn}, and {Deng}}]{Luo2024}
{Luo} H, {Dorn} C {Deng} J (2024) {The interior as the dominant water reservoir in super-Earths and sub-Neptunes}. Nature Astronomy 8:1399--1407

\bibitem[{{Lynden-Bell} and {Pringle}(1974)}]{LBP1974}
{Lynden-Bell} D {Pringle} JE (1974) {The evolution of viscous discs and the origin of the nebular variables.} \mnras 168:603--637

\bibitem[{{Lyra} et~al.(2010){Lyra}, {Paardekooper}, and {Mac Low}}]{Lyra2010}
{Lyra} W, {Paardekooper} SJ {Mac Low} MM (2010) {Orbital Migration of Low-mass Planets in Evolutionary Radiative Models: Avoiding Catastrophic Infall}. \apjl 715:L68--L73

\bibitem[{{Madhusudhan} et~al.(2014){Madhusudhan}, {Amin}, and {Kennedy}}]{Madu2014}
{Madhusudhan} N, {Amin} MA {Kennedy} GM (2014) {Toward Chemical Constraints on Hot Jupiter Migration}. \apjl 794:L12

\bibitem[{{Madhusudhan} et~al.(2017){Madhusudhan}, {Bitsch}, {Johansen}, and {Eriksson}}]{Madhusudhan2017}
{Madhusudhan} N, {Bitsch} B, {Johansen} A {Eriksson} L (2017) {Atmospheric signatures of giant exoplanet formation by pebble accretion}. \mnras 469:4102--4115

\bibitem[{{Mansfield} et~al.(2018){Mansfield}, {Bean}, {Line}, {Parmentier}, {Kreidberg}, {D{\'e}sert}, {Fortney}, {Stevenson}, {Arcangeli}, and {Dragomir}}]{Mansfield2018}
{Mansfield} M, {Bean} JL, {Line} MR et~al. (2018) {An HST/WFC3 Thermal Emission Spectrum of the Hot Jupiter HAT-P-7b}. \aj 156(1):10

\bibitem[{{Masset}(2017)}]{Masset2017}
{Masset} FS (2017) {Coorbital thermal torques on low-mass protoplanets}. \mnras 472(4):4204--4219

\bibitem[{{Masset} et~al.(2006){Masset}, {Morbidelli}, {Crida}, and {Ferreira}}]{Masset2006}
{Masset} FS, {Morbidelli} A, {Crida} A {Ferreira} J (2006) {Disk Surface Density Transitions as Protoplanet Traps}. \apj 642(1):478--487

\bibitem[{{Mayor} and {Queloz}(1995)}]{Mayor1995}
{Mayor} M {Queloz} D (1995) {A Jupiter-mass companion to a solar-type star}. \nat 378:355--359

\bibitem[{{McNally} et~al.(2017){McNally}, {Nelson}, {Paardekooper}, {Gressel}, and {Lyra}}]{McNally2017}
{McNally} CP, {Nelson} RP, {Paardekooper} SJ, {Gressel} O {Lyra} W (2017) {Low mass planet migration in magnetically torqued dead zones - I. Static migration torque}. \mnras 472:1565--1575

\bibitem[{{Mignone} et~al.(2007){Mignone}, {Bodo}, {Massaglia}, {Matsakos}, {Tesileanu}, {Zanni}, and {Ferrari}}]{Mignone2007}
{Mignone} A, {Bodo} G, {Massaglia} S et~al. (2007) {PLUTO: A Numerical Code for Computational Astrophysics}. \apjs 170(1):228--242

\bibitem[{{Militzer} and {Hubbard}(2013)}]{MilitzerHubbard2013}
{Militzer} B {Hubbard} WB (2013) {Ab Initio Equation of State for Hydrogen-Helium Mixtures with Recalibration of the Giant-planet Mass-Radius Relation}. \apj 774:148

\bibitem[{{Miotello} et~al.(2016){Miotello}, {van Dishoeck}, {Kama}, and {Bruderer}}]{Miotello2016}
{Miotello} A, {van Dishoeck} EF, {Kama} M {Bruderer} S (2016) {Determining protoplanetary disk gas masses from CO isotopologues line observations}. \aap 594:A85

\bibitem[{{Miotello} et~al.(2023){Miotello}, {Kamp}, {Birnstiel}, {Cleeves}, and {Kataoka}}]{Miotello2023}
{Miotello} A, {Kamp} I, {Birnstiel} T, {Cleeves} LC {Kataoka} A (2023) {Setting the Stage for Planet Formation: Measurements and Implications of the Fundamental Disk Properties}. In: {Inutsuka} S, {Aikawa} Y, {Muto} T, {Tomida} K {Tamura} M (eds) Protostars and Planets VII, Astronomical Society of the Pacific Conference Series, vol 534, p 501, \doi{10.48550/arXiv.2203.09818}

\bibitem[{{Miyake} and {Nakagawa}(1993)}]{Miyake1993}
{Miyake} K {Nakagawa} Y (1993) {Effects of particle size distribution on opacity curves of protoplanetary disks around T Tauri stars}. \icarus 106:20

\bibitem[{{Mizuno} et~al.(1978){Mizuno}, {Nakazawa}, and {Hayashi}}]{Mizuno1978}
{Mizuno} H, {Nakazawa} K {Hayashi} C (1978) {Instability of a gaseous envelope surrounding a planetary core and formation of giant planets}. Progress of Theoretical Physics 60:699--710

\bibitem[{{Molli{\`e}re} et~al.(2017){Molli{\`e}re}, {van Boekel}, {Bouwman}, {Henning}, {Lagage}, and {Min}}]{Molliere2017}
{Molli{\`e}re} P, {van Boekel} R, {Bouwman} J et~al. (2017) {Observing transiting planets with JWST. Prime targets and their synthetic spectral observations}. \aap 600:A10

\bibitem[{{Molli{\`e}re} et~al.(2022){Molli{\`e}re}, {Molyarova}, {Bitsch}, {Henning}, {Schneider}, {Kreidberg}, {Eistrup}, {Burn}, {Nasedkin}, {Semenov}, {Mordasini}, {Schlecker}, {Schwarz}, {Lacour}, {Nowak}, and {Schulik}}]{molliere2022}
{Molli{\`e}re} P, {Molyarova} T, {Bitsch} B et~al. (2022) {Interpreting the Atmospheric Composition of Exoplanets: Sensitivity to Planet Formation Assumptions}. \apj 934(1):74

\bibitem[{{Mordasini} et~al.(2014){Mordasini}, {Klahr}, {Alibert}, {Miller}, and {Henning}}]{Mordasini2014}
{Mordasini} C, {Klahr} H, {Alibert} Y, {Miller} N {Henning} T (2014) {Grain opacity and the bulk composition of extrasolar planets. I. Results from scaling the ISM opacity}. \aap 566:A141

\bibitem[{{Mordasini} et~al.(2016){Mordasini}, {van Boekel}, {Molli{\`e}re}, {Henning}, and {Benneke}}]{Mordasini16}
{Mordasini} C, {van Boekel} R, {Molli{\`e}re} P, {Henning} T {Benneke} B (2016) {The Imprint of Exoplanet Formation History on Observable Present-day Spectra of Hot Jupiters}. \apj 832:41

\bibitem[{{Moriarty} et~al.(2014){Moriarty}, {Madhusudhan}, and {Fischer}}]{Moriarty14}
{Moriarty} J, {Madhusudhan} N {Fischer} D (2014) {Chemistry in an Evolving Protoplanetary Disk: Effects on Terrestrial Planet Composition}. \apj 787:81

\bibitem[{{Moscadelli} et~al.(2022){Moscadelli}, {Sanna}, {Beuther}, {Oliva}, and {Kuiper}}]{Moscadelli2022}
{Moscadelli} L, {Sanna} A, {Beuther} H, {Oliva} A {Kuiper} R (2022) {Snapshot of a magnetohydrodynamic disk wind traced by water maser observations}. Nature Astronomy 6:1068--1076

\bibitem[{Mukherjee et~al.(2023)Mukherjee, Batalha, Fortney, and Marley}]{mukherjee_picaso_2023}
Mukherjee S, Batalha NE, Fortney JJ Marley MS (2023) {PICASO} 3.0: {A} {One}-dimensional {Climate} {Model} for {Giant} {Planets} and {Brown} {Dwarfs}. The Astrophysical Journal 942:71, \urlprefix\url{https://ui.adsabs.harvard.edu/abs/2023ApJ...942...71M}, publisher: IOP ADS Bibcode: 2023ApJ...942...71M

\bibitem[{{Mulders} et~al.(2020){Mulders}, {O'Brien}, {Ciesla}, {Apai}, and {Pascucci}}]{Mulders2020}
{Mulders} GD, {O'Brien} DP, {Ciesla} FJ, {Apai} D {Pascucci} I (2020) {Earths in Other Solar Systems' N-body Simulations: The Role of Orbital Damping in Reproducing the Kepler Planetary Systems}. \apj 897(1):72

\bibitem[{{Nakatani} et~al.(2018){Nakatani}, {Hosokawa}, {Yoshida}, {Nomura}, and {Kuiper}}]{Nakatani2018}
{Nakatani} R, {Hosokawa} T, {Yoshida} N, {Nomura} H {Kuiper} R (2018) {Radiation Hydrodynamics Simulations of Photoevaporation of Protoplanetary Disks. II. Metallicity Dependence of UV and X-Ray Photoevaporation}. \apj 865(1):75

\bibitem[{{{\"O}berg} and {Bergin}(2021)}]{ObergBergin2021}
{{\"O}berg} KI {Bergin} EA (2021) {Astrochemistry and compositions of planetary systems}. \physrep 893:1--48

\bibitem[{{{\"O}berg} and {Wordsworth}(2019)}]{Oberg2019}
{{\"O}berg} KI {Wordsworth} R (2019) {Jupiter's Composition Suggests its Core Assembled Exterior to the N$_{2}$ Snowline}. \aj 158(5):194

\bibitem[{{{\"O}berg} et~al.(2011{\natexlab{a}}){{\"O}berg}, {Boogert}, {Pontoppidan}, {van den Broek}, {van Dishoeck}, {Bottinelli}, {Blake}, and {Evans}}]{Oberg2011a}
{{\"O}berg} KI, {Boogert} ACA, {Pontoppidan} KM et~al. (2011{\natexlab{a}}) {Ices in Starless and Starforming Cores}. In: {Cernicharo} J {Bachiller} R (eds) The Molecular Universe, IAU Symposium, vol 280, pp 65--78, \doi{10.1017/S1743921311024872}

\bibitem[{{{\"O}berg} et~al.(2011{\natexlab{b}}){{\"O}berg}, {Boogert}, {Pontoppidan}, {van den Broek}, {van Dishoeck}, {Bottinelli}, {Blake}, and {Evans}}]{Oberg2011b}
{{\"O}berg} KI, {Boogert} ACA, {Pontoppidan} KM et~al. (2011{\natexlab{b}}) {The Spitzer Ice Legacy: Ice Evolution from Cores to Protostars}. \apj 740:109

\bibitem[{{{\"O}berg} et~al.(2011{\natexlab{c}}){{\"O}berg}, {Murray-Clay}, and {Bergin}}]{Oberg11}
{{\"O}berg} KI, {Murray-Clay} R {Bergin} EA (2011{\natexlab{c}}) {The Effects of Snowlines on C/O in Planetary Atmospheres}. \apjl 743:L16

\bibitem[{{{\"O}berg} et~al.(2021){{\"O}berg}, {Guzm{\'a}n}, {Walsh}, {Aikawa}, {Bergin}, {Law}, {Loomis}, {Alarc{\'o}n}, {Andrews}, {Bae}, {Bergner}, {Boehler}, {Booth}, {Bosman}, {Calahan}, {Cataldi}, {Cleeves}, {Czekala}, {Furuya}, {Huang}, {Ilee}, {Kurtovic}, {Le Gal}, {Liu}, {Long}, {M{\'e}nard}, {Nomura}, {P{\'e}rez}, {Qi}, {Schwarz}, {Sierra}, {Teague}, {Tsukagoshi}, {Yamato}, {van't Hoff}, {Waggoner}, {Wilner}, and {Zhang}}]{Oberg2021MAPS}
{{\"O}berg} KI, {Guzm{\'a}n} VV, {Walsh} C et~al. (2021) {Molecules with ALMA at Planet-forming Scales (MAPS). I. Program Overview and Highlights}. \apjs 257(1):1

\bibitem[{{{\"O}berg} et~al.(2023){{\"O}berg}, {Facchini}, and {Anderson}}]{Oberg2023}
{{\"O}berg} KI, {Facchini} S {Anderson} DE (2023) {Protoplanetary Disk Chemistry}. \araa 61:287--328

\bibitem[{{Ormel} et~al.(2009){Ormel}, {Paszun}, {Dominik}, and {Tielens}}]{Ormel2009}
{Ormel} CW, {Paszun} D, {Dominik} C {Tielens} AGGM (2009) {Dust coagulation and fragmentation in molecular clouds. I. How collisions between dust aggregates alter the dust size distribution}. \aap 502:845--869

\bibitem[{{Otegi} et~al.(2020){Otegi}, {Bouchy}, and {Helled}}]{Otegi2020}
{Otegi} JF, {Bouchy} F {Helled} R (2020) {Revisited mass-radius relations for exoplanets below 120 M$_{{\ensuremath{\oplus}}}$}. \aap 634:A43

\bibitem[{{Ouyed} and {Pudritz}(1997)}]{Ouyed_Pudritz1997}
{Ouyed} R {Pudritz} RE (1997) {Numerical Simulations of Astrophysical Jets from Keplerian Disks. I. Stationary Models}. \apj 482(2):712--732

\bibitem[{{Owen} and {Wu}(2017)}]{Owen2017}
{Owen} JE {Wu} Y (2017) {The Evaporation Valley in the Kepler Planets}. \apj 847(1):29

\bibitem[{{Owen} et~al.(2011){Owen}, {Ercolano}, and {Clarke}}]{Owen2011}
{Owen} JE, {Ercolano} B {Clarke} CJ (2011) {Protoplanetary disc evolution and dispersal: the implications of X-ray photoevaporation}. \mnras 412:13--25

\bibitem[{{Paardekooper} et~al.(2010){Paardekooper}, {Baruteau}, {Crida}, and {Kley}}]{Paardekooper2010}
{Paardekooper} SJ, {Baruteau} C, {Crida} A {Kley} W (2010) {A torque formula for non-isothermal type I planetary migration - I. Unsaturated horseshoe drag}. \mnras 401:1950--1964

\bibitem[{{Paardekooper} et~al.(2011){Paardekooper}, {Baruteau}, and {Kley}}]{Paardekooper2011}
{Paardekooper} SJ, {Baruteau} C {Kley} W (2011) {A torque formula for non-isothermal Type I planetary migration - II. Effects of diffusion}. \mnras 410(1):293--303

\bibitem[{{Papaloizou} and {Lin}(1984)}]{PapaloizouLin1984}
{Papaloizou} J {Lin} DNC (1984) {On the tidal interaction between protoplanets and the primordial solar nebula. I - Linear calculation of the role of angular momentum exchange}. \apj 285:818--834

\bibitem[{{Pascucci} and {Sterzik}(2009)}]{Pascucci2009}
{Pascucci} I {Sterzik} M (2009) {Evidence for Disk Photoevaporation Driven by the Central Star}. \apj 702:724--732

\bibitem[{{Pascucci} et~al.(2023){Pascucci}, {Cabrit}, {Edwards}, {Gorti}, {Gressel}, and {Suzuki}}]{Pascucci2023}
{Pascucci} I, {Cabrit} S, {Edwards} S et~al. (2023) {The Role of Disk Winds in the Evolution and Dispersal of Protoplanetary Disks}. In: {Inutsuka} S, {Aikawa} Y, {Muto} T, {Tomida} K {Tamura} M (eds) Protostars and Planets VII, Astronomical Society of the Pacific Conference Series, vol 534, p 567, \doi{10.48550/arXiv.2203.10068}

\bibitem[{{Pascucci} et~al.(2025){Pascucci}, {Beck}, {Cabrit}, {Bajaj}, {Edwards}, {Louvet}, {Najita}, {Skinner}, {Gorti}, {Salyk}, {Brittain}, {Krijt}, {Muzerolle Page}, {Ruaud}, {Schwarz}, {Semenov}, {Duch{\^e}ne}, and {Villenave}}]{Pascucci2025}
{Pascucci} I, {Beck} TL, {Cabrit} S et~al. (2025) {The nested morphology of disk winds from young stars revealed by JWST/NIRSpec observations}. Nature Astronomy 9:81--89

\bibitem[{{Pasek} et~al.(2005){Pasek}, {Milsom}, {Ciesla}, {Lauretta}, {Sharp}, and {Lunine}}]{Pasek2005}
{Pasek} MA, {Milsom} JA, {Ciesla} FJ et~al. (2005) {Sulfur chemistry with time-varying oxygen abundance during Solar System formation}. \icarus 175:1--14

\bibitem[{{Pelletier} and {Pudritz}(1992)}]{PelletierPudritz1992}
{Pelletier} G {Pudritz} RE (1992) {Hydromagnetic disk winds in young stellar objects and active galactic nuclei}. \apj 394:117--138

\bibitem[{{Pepe} et~al.(2004){Pepe}, {Mayor}, {Queloz}, {Benz}, {Bonfils}, {Bouchy}, {Lo Curto}, {Lovis}, {M{\'e}gevand}, {Moutou}, {Naef}, {Rupprecht}, {Santos}, {Sivan}, {Sosnowska}, and {Udry}}]{Pepe2004}
{Pepe} F, {Mayor} M, {Queloz} D et~al. (2004) {The HARPS search for southern extra-solar planets. I. HD 330075 b: A new ``hot Jupiter''}. \aap 423:385--389

\bibitem[{{Perotti} et~al.(2023){Perotti}, {Christiaens}, {Henning}, {Tabone}, {Waters}, {Kamp}, {Olofsson}, {Grant}, {Gasman}, {Bouwman}, {Samland}, {Franceschi}, {van Dishoeck}, {Schwarz}, {G{\"u}del}, {Lagage}, {Ray}, {Vandenbussche}, {Abergel}, {Absil}, {Arabhavi}, {Argyriou}, {Barrado}, {Boccaletti}, {Caratti o Garatti}, {Geers}, {Glauser}, {Justannont}, {Lahuis}, {Mueller}, {Nehm{\'e}}, {Pantin}, {Scheithauer}, {Waelkens}, {Guadarrama}, {Jang}, {Kanwar}, {Morales-Calder{\'o}n}, {Pawellek}, {Rodgers-Lee}, {Schreiber}, {Colina}, {Greve}, {{\"O}stlin}, and {Wright}}]{Perotti2023}
{Perotti} G, {Christiaens} V, {Henning} T et~al. (2023) {Water in the terrestrial planet-forming zone of the PDS 70 disk}. \nat 620(7974):516--520

\bibitem[{{Petigura} et~al.(2018){Petigura}, {Marcy}, {Winn}, {Weiss}, {Fulton}, {Howard}, {Sinukoff}, {Isaacson}, {Morton}, and {Johnson}}]{Petigura2018}
{Petigura} EA, {Marcy} GW, {Winn} JN et~al. (2018) {The California-Kepler Survey. IV. Metal-rich Stars Host a Greater Diversity of Planets}. \aj 155(2):89

\bibitem[{{Pignatale} et~al.(2011){Pignatale}, {Maddison}, {Taquet}, {Brooks}, and {Liffman}}]{Pignatale2011}
{Pignatale} FC, {Maddison} ST, {Taquet} V, {Brooks} G {Liffman} K (2011) {The effect of the regular solution model in the condensation of protoplanetary dust}. \mnras 414:2386--2405

\bibitem[{{Pineda} et~al.(2023){Pineda}, {Arzoumanian}, {Andre}, {Friesen}, {Zavagno}, {Clarke}, {Inoue}, {Chen}, {Lee}, {Soler}, and {Kuffmeier}}]{Pineda2023}
{Pineda} JE, {Arzoumanian} D, {Andre} P et~al. (2023) {From Bubbles and Filaments to Cores and Disks: Gas Gathering and Growth of Structure Leading to the Formation of Stellar Systems}. In: {Inutsuka} S, {Aikawa} Y, {Muto} T, {Tomida} K {Tamura} M (eds) Protostars and Planets VII, Astronomical Society of the Pacific Conference Series, vol 534, p 233, \doi{10.48550/arXiv.2205.03935}

\bibitem[{{Pinte} et~al.(2016){Pinte}, {Dent}, {M{\'e}nard}, {Hales}, {Hill}, {Cortes}, and {de Gregorio-Monsalvo}}]{Pinte2016}
{Pinte} C, {Dent} WRF, {M{\'e}nard} F et~al. (2016) {Dust and Gas in the Disk of HL Tauri: Surface Density, Dust Settling, and Dust-to-gas Ratio}. \apj 816(1):25

\bibitem[{{Pinte} et~al.(2023){Pinte}, {Teague}, {Flaherty}, {Hall}, {Facchini}, and {Casassus}}]{Pinte2023}
{Pinte} C, {Teague} R, {Flaherty} K et~al. (2023) {Kinematic Structures in Planet-Forming Disks}. In: {Inutsuka} S, {Aikawa} Y, {Muto} T, {Tomida} K {Tamura} M (eds) Protostars and Planets VII, Astronomical Society of the Pacific Conference Series, vol 534, p 645, \doi{10.48550/arXiv.2203.09528}

\bibitem[{{Pollack} et~al.(1996){Pollack}, {Hubickyj}, {Bodenheimer}, {Lissauer}, {Podolak}, and {Greenzweig}}]{Pollack1996}
{Pollack} JB, {Hubickyj} O, {Bodenheimer} P et~al. (1996) {Formation of the Giant Planets by Concurrent Accretion of Solids and Gas}. \icarus 124:62--85

\bibitem[{{Pontoppidan} et~al.(2014){Pontoppidan}, {Salyk}, {Bergin}, {Brittain}, {Marty}, {Mousis}, and {{\"O}berg}}]{Pontoppidan2014}
{Pontoppidan} KM, {Salyk} C, {Bergin} EA et~al. (2014) {Volatiles in Protoplanetary Disks}. Protostars and Planets VI pp 363--385

\bibitem[{{Pudritz} and {Norman}(1986)}]{PudritzNorman1986}
{Pudritz} RE {Norman} CA (1986) {Bipolar hydromagnetic winds from disks around protostellar objects}. \apj 301:571--586

\bibitem[{{Pudritz} and {Ray}(2019)}]{pudritz_Ray2019}
{Pudritz} RE {Ray} TP (2019) {The Role of Magnetic Fields in Protostellar Outflows and Star Formation}. Frontiers in Astronomy and Space Sciences 6:54

\bibitem[{{Pudritz} et~al.(2007){Pudritz}, {Ouyed}, {Fendt}, and {Brandenburg}}]{Pudritz2007}
{Pudritz} RE, {Ouyed} R, {Fendt} C {Brandenburg} A (2007) {Disk Winds, Jets, and Outflows: Theoretical and Computational Foundations}. Protostars and Planets V pp 277--294

\bibitem[{{Queloz} et~al.(2000){Queloz}, {Mayor}, {Weber}, {Bl{\'e}cha}, {Burnet}, {Confino}, {Naef}, {Pepe}, {Santos}, and {Udry}}]{Queloz2000}
{Queloz} D, {Mayor} M, {Weber} L et~al. (2000) {The CORALIE survey for southern extra-solar planets. I. A planet orbiting the star Gliese 86}. \aap 354:99--102

\bibitem[{{Raettig} et~al.(2015){Raettig}, {Klahr}, and {Lyra}}]{Raettig2015}
{Raettig} N, {Klahr} H {Lyra} W (2015) {Particle Trapping and Streaming Instability in Vortices in Protoplanetary Disks}. \apj 804:35

\bibitem[{{Ray} et~al.(2007){Ray}, {Dougados}, {Bacciotti}, {Eisl{\"o}ffel}, and {Chrysostomou}}]{Ray2007}
{Ray} T, {Dougados} C, {Bacciotti} F, {Eisl{\"o}ffel} J {Chrysostomou} A (2007) {Toward Resolving the Outflow Engine: An Observational Perspective}. Protostars and Planets V pp 231--244

\bibitem[{{Raymond} et~al.(2014){Raymond}, {Kokubo}, {Morbidelli}, {Morishima}, and {Walsh}}]{Raymond2014}
{Raymond} SN, {Kokubo} E, {Morbidelli} A, {Morishima} R {Walsh} KJ (2014) {Terrestrial Planet Formation at Home and Abroad}. Protostars and Planets VI pp 595--618

\bibitem[{{Rice} et~al.(2003){Rice}, {Armitage}, {Bate}, and {Bonnell}}]{Rice2003}
{Rice} WKM, {Armitage} PJ, {Bate} MR {Bonnell} IA (2003) {The effect of cooling on the global stability of self-gravitating protoplanetary discs}. \mnras 339(4):1025--1030

\bibitem[{{Riols} and {Lesur}(2019{\natexlab{a}})}]{Riols_Lesur2019}
{Riols} A {Lesur} G (2019{\natexlab{a}}) {Spontaneous ring formation in wind-emitting accretion discs}. \aap 625:A108

\bibitem[{{Riols} and {Lesur}(2019{\natexlab{b}})}]{Riols2019}
{Riols} A {Lesur} G (2019{\natexlab{b}}) {Spontaneous ring formation in wind-emitting accretion discs}. \aap 625:A108

\bibitem[{{Riols} et~al.(2020){Riols}, {Lesur}, and {Menard}}]{Riols2020}
{Riols} A, {Lesur} G {Menard} F (2020) {Ring formation and dust dynamics in wind-driven protoplanetary discs: global simulations}. \aap 639:A95

\bibitem[{{Rivilla} et~al.(2017){Rivilla}, {Beltr{\'a}n}, {Cesaroni}, {Fontani}, {Codella}, and {Zhang}}]{Rivilla2017}
{Rivilla} VM, {Beltr{\'a}n} MT, {Cesaroni} R et~al. (2017) {Formation of ethylene glycol and other complex organic molecules in star-forming regions}. \aap 598:A59

\bibitem[{{Rogers} et~al.(2021){Rogers}, {Gupta}, {Owen}, and {Schlichting}}]{Rogers2021}
{Rogers} JG, {Gupta} A, {Owen} JE {Schlichting} HE (2021) {Photoevaporation versus core-powered mass-loss: model comparison with the 3D radius gap}. \mnras 508(4):5886--5902

\bibitem[{{Rogers}(2014)}]{Rogers2014}
{Rogers} LA (2014) {Glimpsing the Compositions of Sub-Neptune-Size Exoplanets}. In: {Booth} M, {Matthews} BC {Graham} JR (eds) Exploring the Formation and Evolution of Planetary Systems, IAU Symposium, vol 299, pp 247--251, \doi{10.1017/S1743921313008491}

\bibitem[{{Rogers} and {Wadsley}(2012)}]{Rogers2012}
{Rogers} PD {Wadsley} J (2012) {The fragmentation of protostellar discs: the Hill criterion for spiral arms}. \mnras 423(2):1896--1908

\bibitem[{{Ros} and {Johansen}(2013)}]{RosJohansen2013}
{Ros} K {Johansen} A (2013) {Ice condensation as a planet formation mechanism}. \aap 552:A137

\bibitem[{{Ruden}(2004)}]{Ruden2004}
{Ruden} SP (2004) {Evolution of Photoevaporating Protoplanetary Disks}. \apj 605:880--891

\bibitem[{{Rustamkulov} et~al.(2023){Rustamkulov}, {Sing}, {Mukherjee}, {May}, {Kirk}, {Schlawin}, {Line}, {Piaulet}, {Carter}, {Batalha}, {Goyal}, {L{\'o}pez-Morales}, {Lothringer}, {MacDonald}, {Moran}, {Stevenson}, {Wakeford}, {Espinoza}, {Bean}, {Batalha}, {Benneke}, {Berta-Thompson}, {Crossfield}, {Gao}, {Kreidberg}, {Powell}, {Cubillos}, {Gibson}, {Leconte}, {Molaverdikhani}, {Nikolov}, {Parmentier}, {Roy}, {Taylor}, {Turner}, {Wheatley}, {Aggarwal}, {Ahrer}, {Alam}, {Alderson}, {Allen}, {Banerjee}, {Barat}, {Barrado}, {Barstow}, {Bell}, {Blecic}, {Brande}, {Casewell}, {Changeat}, {Chubb}, {Crouzet}, {Daylan}, {Decin}, {D{\'e}sert}, {Mikal-Evans}, {Feinstein}, {Flagg}, {Fortney}, {Harrington}, {Heng}, {Hong}, {Hu}, {Iro}, {Kataria}, {Kempton}, {Krick}, {Lendl}, {Lillo-Box}, {Louca}, {Lustig-Yaeger}, {Mancini}, {Mansfield}, {Mayne}, {Miguel}, {Morello}, {Ohno}, {Palle}, {Petit dit de la Roche}, {Rackham}, {Radica}, {Ramos-Rosado}, {Redfield}, {Rogers}, {Shkolnik}, {Southworth}, {Teske}, {Tremblin},
  {Tucker}, {Venot}, {Waalkes}, {Welbanks}, {Zhang}, and {Zieba}}]{Rustamkulov2023}
{Rustamkulov} Z, {Sing} DK, {Mukherjee} S et~al. (2023) {Early Release Science of the exoplanet WASP-39b with JWST NIRSpec PRISM}. \nat 614(7949):659--663

\bibitem[{{Salmeron} and {Wardle}(2003)}]{SalmeronWardle2003}
{Salmeron} R {Wardle} M (2003) {Magnetorotational instability in stratified, weakly ionized accretion discs}. \mnras 345:992--1008

\bibitem[{{Salyk} et~al.(2008){Salyk}, {Pontoppidan}, {Blake}, {Lahuis}, {van Dishoeck}, and {Evans}}]{Salyk2008}
{Salyk} C, {Pontoppidan} KM, {Blake} GA et~al. (2008) {H$_{2}$O and OH Gas in the Terrestrial Planet-forming Zones of Protoplanetary Disks}. \apjl 676:L49

\bibitem[{{Sch{\"a}fer} et~al.(2017){Sch{\"a}fer}, {Yang}, and {Johansen}}]{Schafer2017}
{Sch{\"a}fer} U, {Yang} CC {Johansen} A (2017) {Initial mass function of planetesimals formed by the streaming instability}. \aap 597:A69

\bibitem[{{Schneider} and {Bitsch}(2021)}]{Schneider2021}
{Schneider} AD {Bitsch} B (2021) {How drifting and evaporating pebbles shape giant planets. I. Heavy element content and atmospheric C/O}. \aap 654:A71

\bibitem[{{Schwarz} et~al.(2016){Schwarz}, {Bergin}, {Cleeves}, {Blake}, {Zhang}, {{\"O}berg}, {van Dishoeck}, and {Qi}}]{Schwarz2016}
{Schwarz} KR, {Bergin} EA, {Cleeves} LI et~al. (2016) {The Radial Distribution of H$_{2}$ and CO in TW Hya as Revealed by Resolved ALMA Observations of CO Isotopologues}. \apj 823(2):91

\bibitem[{{Schwarz} et~al.(2025){Schwarz}, {Samland}, {Olofsson}, {Henning}, {Sellek}, {G{\"u}del}, {Tabone}, {Kamp}, {Lagage}, {van Dishoeck}, {Caratti o Garatti}, {Glauser}, {Ray}, {Arabhavi}, {Christiaens}, {Franceschi}, {Gasman}, {Grant}, {Kanwar}, {Kaeufer}, {Kurtovic}, {Perotti}, {Temmink}, and {Vlasblom}}]{Schwarz2025}
{Schwarz} KR, {Samland} M, {Olofsson} G et~al. (2025) {MINDS. JWST-MIRI Observations of a Spatially Resolved Atomic Jet and Polychromatic Molecular Wind toward SY Cha}. \apj 980(1):148

\bibitem[{{Seager} et~al.(2007){Seager}, {Kuchner}, {Hier-Majumder}, and {Militzer}}]{Seager2007}
{Seager} S, {Kuchner} M, {Hier-Majumder} CA {Militzer} B (2007) {Mass-Radius Relationships for Solid Exoplanets}. \apj 669(2):1279--1297

\bibitem[{{Seifried} et~al.(2014){Seifried}, {Banerjee}, {Pudritz}, and {Klessen}}]{Seifried2014}
{Seifried} D, {Banerjee} R, {Pudritz} RE {Klessen} RS (2014) {Disc Formation in Turbulent Cloud Cores: Circumventing the Magnetic Braking Catastrophe}. In: {Stamatellos} D, {Goodwin} S {Ward-Thompson} D (eds) The Labyrinth of Star Formation, Astrophysics and Space Science Proceedings, vol~36, p~75, \doi{10.1007/978-3-319-03041-8_13}

\bibitem[{{Seifried} et~al.(2015){Seifried}, {Banerjee}, {Pudritz}, and {Klessen}}]{Seifried2015}
{Seifried} D, {Banerjee} R, {Pudritz} RE {Klessen} RS (2015) {Accretion and magnetic field morphology around Class 0 stage protostellar discs}. \mnras 446:2776--2788

\bibitem[{{Shakura} and {Sunyaev}(1973)}]{SS1973}
{Shakura} NI {Sunyaev} RA (1973) {Black holes in binary systems. Observational appearance.} \aap 24:337--355

\bibitem[{{Showman} and {Guillot}(2002)}]{ShowmanGuillot2002}
{Showman} AP {Guillot} T (2002) {Atmospheric circulation and tides of ``51 Pegasus b-like'' planets}. \aap 385:166--180

\bibitem[{{Simon} et~al.(2016){Simon}, {Armitage}, {Li}, and {Youdin}}]{Simon2016}
{Simon} JB, {Armitage} PJ, {Li} R {Youdin} AN (2016) {The Mass and Size Distribution of Planetesimals Formed by the Streaming Instability. I. The Role of Self-gravity}. \apj 822:55

\bibitem[{{Speedie} et~al.(2022){Speedie}, {Pudritz}, {Cridland}, {Meru}, and {Booth}}]{Speedie2022}
{Speedie} J, {Pudritz} RE, {Cridland} AJ, {Meru} F {Booth} RA (2022) {Turbulent disc viscosity and the bifurcation of planet formation histories}. \mnras 510(4):6059--6084

\bibitem[{{Speedie} et~al.(2024){Speedie}, {Dong}, {Hall}, {Longarini}, {Veronesi}, {Paneque-Carre{\~n}o}, {Lodato}, {Tang}, {Teague}, and {Hashimoto}}]{Speedie2024}
{Speedie} J, {Dong} R, {Hall} C et~al. (2024) {Gravitational instability in a planet-forming disk}. \nat 633(8028):58--62

\bibitem[{{Spezzano} et~al.(2017){Spezzano}, {Caselli}, {Bizzocchi}, {Giuliano}, and {Lattanzi}}]{Spezzano2017}
{Spezzano} S, {Caselli} P, {Bizzocchi} L, {Giuliano} BM {Lattanzi} V (2017) {The observed chemical structure of L1544}. \aap 606:A82

\bibitem[{{Stammler} and {Birnstiel}(2022)}]{Stammler2022}
{Stammler} SM {Birnstiel} T (2022) {DustPy: A Python Package for Dust Evolution in Protoplanetary Disks}. \apj 935(1):35

\bibitem[{{Stammler} et~al.(2017){Stammler}, {Birnstiel}, {Pani{\'c}}, {Dullemond}, and {Dominik}}]{Stammler2017}
{Stammler} SM, {Birnstiel} T, {Pani{\'c}} O, {Dullemond} CP {Dominik} C (2017) {Redistribution of CO at the location of the CO ice line in evolving gas and dust disks}. \aap 600:A140

\bibitem[{{Stevenson}(1985)}]{Stevenson1985}
{Stevenson} DJ (1985) {Cosmochemistry and structure of the giant planets and their satellites}. \icarus 62:4--15

\bibitem[{{Stevenson} and {Lunine}(1988)}]{StevensonLunine1988}
{Stevenson} DJ {Lunine} JI (1988) {Rapid formation of Jupiter by diffuse redistribution of water vapor in the solar nebula}. \icarus 75:146--155

\bibitem[{{Suriano} et~al.(2019){Suriano}, {Li}, {Krasnopolsky}, {Suzuki}, and {Shang}}]{Suriano2019}
{Suriano} SS, {Li} ZY, {Krasnopolsky} R, {Suzuki} TK {Shang} H (2019) {The formation of rings and gaps in wind-launching non-ideal MHD discs: three-dimensional simulations}. \mnras 484(1):107--124

\bibitem[{{Suzuki} et~al.(2016){Suzuki}, {Ogihara}, {Morbidelli}, {Crida}, and {Guillot}}]{Suzuki2016}
{Suzuki} TK, {Ogihara} M, {Morbidelli} A, {Crida} A {Guillot} T (2016) {Evolution of protoplanetary discs with magnetically driven disc winds}. \aap 596:A74

\bibitem[{{Tabone} et~al.(2023){Tabone}, {Bettoni}, {van Dishoeck}, {Arabhavi}, {Grant}, {Gasman}, {Henning}, {Kamp}, {G{\"u}del}, {Lagage}, {Ray}, {Vandenbussche}, {Abergel}, {Absil}, {Argyriou}, {Barrado}, {Boccaletti}, {Bouwman}, {Caratti o Garatti}, {Geers}, {Glauser}, {Justannont}, {Lahuis}, {Mueller}, {Nehm{\'e}}, {Olofsson}, {Pantin}, {Scheithauer}, {Waelkens}, {Waters}, {Black}, {Christiaens}, {Guadarrama}, {Morales-Calder{\'o}n}, {Jang}, {Kanwar}, {Pawellek}, {Perotti}, {Perrin}, {Rodgers-Lee}, {Samland}, {Schreiber}, {Schwarz}, {Colina}, {{\"O}stlin}, and {Wright}}]{Tabone2023}
{Tabone} B, {Bettoni} G, {van Dishoeck} EF et~al. (2023) {A rich hydrocarbon chemistry and high C to O ratio in the inner disk around a very low-mass star}. Nature Astronomy 7:805--814

\bibitem[{{Teague} et~al.(2022){Teague}, {Bae}, {Andrews}, {Benisty}, {Bergin}, {Facchini}, {Huang}, {Longarini}, and {Wilner}}]{Teague2022}
{Teague} R, {Bae} J, {Andrews} SM et~al. (2022) {Mapping the Complex Kinematic Substructure in the TW Hya Disk}. \apj 936(2):163

\bibitem[{{Terebey} et~al.(1984){Terebey}, {Shu}, and {Cassen}}]{Terebey+1984}
{Terebey} S, {Shu} FH {Cassen} P (1984) {The collapse of the cores of slowly rotating isothermal clouds}. \apj 286:529--551

\bibitem[{{Teske}(2024)}]{Teske2024}
{Teske} JK (2024) {The Star{\textendash}Planet Composition Connection}. \araa 62(1):333--368

\bibitem[{{Testi} et~al.(2014){Testi}, {Birnstiel}, {Ricci}, {Andrews}, {Blum}, {Carpenter}, {Dominik}, {Isella}, {Natta}, {Williams}, and {Wilner}}]{Testi2014}
{Testi} L, {Birnstiel} T, {Ricci} L et~al. (2014) {Dust Evolution in Protoplanetary Disks}. Protostars and Planets VI pp 339--361

\bibitem[{{Toppani} et~al.(2006){Toppani}, {Libourel}, {Robert}, and {Ghanbaja}}]{Toppani2006}
{Toppani} A, {Libourel} G, {Robert} F {Ghanbaja} J (2006) {Laboratory condensation of refractory dust in protosolar and circumstellar conditions}. \gca 70:5035--5060

\bibitem[{{Turner} et~al.(2014){Turner}, {Fromang}, {Gammie}, {Klahr}, {Lesur}, {Wardle}, and {Bai}}]{Turner2014}
{Turner} NJ, {Fromang} S, {Gammie} C et~al. (2014) {Transport and Accretion in Planet-Forming Disks}. Protostars and Planets VI pp 411--432

\bibitem[{{Udry} and {Santos}(2007)}]{UdrySantos2007}
{Udry} S {Santos} NC (2007) {Statistical Properties of Exoplanets}. \araa 45:397--439

\bibitem[{{Umebayashi} and {Nakano}(2009)}]{UmebayashiNakano2009}
{Umebayashi} T {Nakano} T (2009) {Effects of Radionuclides on the Ionization State of Protoplanetary Disks and Dense Cloud Cores}. \apj 690:69--81

\bibitem[{{Valencia} et~al.(2007){Valencia}, {Sasselov}, and {O'Connell}}]{Valencia2007}
{Valencia} D, {Sasselov} DD {O'Connell} RJ (2007) {Detailed Models of Super-Earths: How Well Can We Infer Bulk Properties?} \apj 665:1413--1420

\bibitem[{{van der Marel} and {Mulders}(2021)}]{vanderMarel2021}
{van der Marel} N {Mulders} GD (2021) {A Stellar Mass Dependence of Structured Disks: A Possible Link with Exoplanet Demographics}. \doi{10.3847/1538-3881/ac0255}

\bibitem[{{Vasyunin} et~al.(2017){Vasyunin}, {Caselli}, {Dulieu}, and {Jim{\'e}nez-Serra}}]{Vasyunin2017}
{Vasyunin} AI, {Caselli} P, {Dulieu} F {Jim{\'e}nez-Serra} I (2017) {Formation of Complex Molecules in Prestellar Cores: A Multilayer Approach}. \apj 842:33

\bibitem[{{Villenave} et~al.(2025){Villenave}, {Rosotti}, {Lambrechts}, {Ziampras}, {Pinte}, {Menard}, {Stapelfeldt}, {Duchene}, {Baylock}, and {Doi}}]{Villenave2025}
{Villenave} M, {Rosotti} G, {Lambrechts} M et~al. (2025) {Turbulence in protoplanetary disks: A systematic analysis of dust settling in 33 disks}. arXiv e-prints arXiv:2503.05872

\bibitem[{Visscher and Fegley(2005)}]{visscher_chemical_2005}
Visscher C Fegley B Jr (2005) Chemical {Constraints} on the {Water} and {Total} {Oxygen} {Abundances} in the {Deep} {Atmosphere} of {Saturn}. The Astrophysical Journal 623:1221--1227, \urlprefix\url{https://ui.adsabs.harvard.edu/abs/2005ApJ...623.1221V}, publisher: IOP ADS Bibcode: 2005ApJ...623.1221V

\bibitem[{Visscher et~al.(2010)Visscher, Moses, and Saslow}]{visscher_deep_2010}
Visscher C, Moses JI Saslow SA (2010) The deep water abundance on {Jupiter}: {New} constraints from thermochemical kinetics and diffusion modeling. Icarus 209:602--615, \urlprefix\url{https://ui.adsabs.harvard.edu/abs/2010Icar..209..602V}, aDS Bibcode: 2010Icar..209..602V

\bibitem[{{Wafflard-Fernandez} and {Lesur}(2023)}]{Wafflard_Lesur2023}
{Wafflard-Fernandez} G {Lesur} G (2023) {Planet-disk-wind interaction: The magnetized fate of protoplanets}. \aap 677:A70

\bibitem[{{Wahl} et~al.(2017){Wahl}, {Hubbard}, {Militzer}, {Guillot}, {Miguel}, {Movshovitz}, {Kaspi}, {Helled}, {Reese}, {Galanti}, {Levin}, {Connerney}, and {Bolton}}]{Wahl2017}
{Wahl} SM, {Hubbard} WB, {Militzer} B et~al. (2017) {Comparing Jupiter interior structure models to Juno gravity measurements and the role of a dilute core}. \grl 44:4649--4659

\bibitem[{Wahl et~al.(2021)Wahl, Thorngren, Lu, and Militzer}]{wahl_tidal_2021}
Wahl SM, Thorngren D, Lu T Militzer B (2021) Tidal {Response} and {Shape} of {Hot} {Jupiters}. The Astrophysical Journal 921(2):105, \urlprefix\url{https://dx.doi.org/10.3847/1538-4357/ac1a72}, publisher: The American Astronomical Society

\bibitem[{{Walsh} et~al.(2014){Walsh}, {Millar}, {Nomura}, {Herbst}, {Widicus Weaver}, {Aikawa}, {Laas}, and {Vasyunin}}]{Walsh2014}
{Walsh} C, {Millar} TJ, {Nomura} H et~al. (2014) {Complex organic molecules in protoplanetary disks}. \aap 563:A33

\bibitem[{{Wang} and {Fischer}(2015)}]{WangFischer2015}
{Wang} J {Fischer} DA (2015) {Revealing a Universal Planet-Metallicity Correlation for Planets of Different Sizes Around Solar-type Stars}. \aj 149:14

\bibitem[{{Ward}(1986)}]{Ward1986}
{Ward} WR (1986) {Density waves in the solar nebula - Differential Lindblad torque}. \icarus 67:164--180

\bibitem[{{Ward}(1997)}]{Ward1997}
{Ward} WR (1997) {Protoplanet Migration by Nebula Tides}. \icarus 126:261--281

\bibitem[{{Watkins} et~al.(2023){Watkins}, {Barnes}, {Henny}, {Kim}, {Kreckel}, {Meidt}, {Klessen}, {Glover}, {Williams}, {Keller}, {Leroy}, {Rosolowsky}, {Lee}, {Anand}, {Belfiore}, {Bigiel}, {Blanc}, {Boquien}, {Cao}, {Chandar}, {Chen}, {Chevance}, {Congiu}, {Dale}, {Deger}, {Egorov}, {Emsellem}, {Faesi}, {Grasha}, {Groves}, {Hassani}, {Henshaw}, {Herrera}, {Hughes}, {Jeffreson}, {Jim{\'e}nez-Donaire}, {Koch}, {Kruijssen}, {Larson}, {Liu}, {Lopez}, {Pessa}, {Pety}, {Querejeta}, {Saito}, {Sandstrom}, {Scheuermann}, {Schinnerer}, {Sormani}, {Stuber}, {Thilker}, {Usero}, and {Whitmore}}]{Watkins2023}
{Watkins} EJ, {Barnes} AT, {Henny} K et~al. (2023) {PHANGS-JWST First Results: A Statistical View on Bubble Evolution in NGC 628}. \apjl 944(2):L24

\bibitem[{{Watson} et~al.(2016){Watson}, {Calvet}, {Fischer}, {Forrest}, {Manoj}, {Megeath}, {Melnick}, {Najita}, {Neufeld}, {Sheehan}, {Stutz}, and {Tobin}}]{Watson2016}
{Watson} DM, {Calvet} NP, {Fischer} WJ et~al. (2016) {Evolution of Mass Outflow in Protostars}. \apj 828(1):52

\bibitem[{{Weidenschilling}(1977)}]{Weidenschilling1977}
{Weidenschilling} SJ (1977) {Aerodynamics of solid bodies in the solar nebula}. \mnras 180:57--70

\bibitem[{{Weiss} et~al.(2013){Weiss}, {Marcy}, {Rowe}, {Howard}, {Isaacson}, {Fortney}, {Miller}, {Demory}, {Fischer}, {Adams}, {Dupree}, {Howell}, {Kolbl}, {Johnson}, {Horch}, {Everett}, {Fabrycky}, and {Seager}}]{Weiss2013}
{Weiss} LM, {Marcy} GW, {Rowe} JF et~al. (2013) {The Mass of KOI-94d and a Relation for Planet Radius, Mass, and Incident Flux}. \apj 768:14

\bibitem[{{Weiss} et~al.(2023){Weiss}, {Millholland}, {Petigura}, {Adams}, {Batygin}, {Block}, and {Mordasini}}]{Weiss2023}
{Weiss} LM, {Millholland} SC, {Petigura} EA et~al. (2023) {Architectures of Compact Multi-Planet Systems: Diversity and Uniformity}. In: {Inutsuka} S, {Aikawa} Y, {Muto} T, {Tomida} K {Tamura} M (eds) Protostars and Planets VII, Astronomical Society of the Pacific Conference Series, vol 534, p 863, \doi{10.48550/arXiv.2203.10076}

\bibitem[{{Xu} et~al.(2017){Xu}, {Bai}, and {{\"O}berg}}]{Xu2017}
{Xu} R, {Bai} XN {{\"O}berg} K (2017) {Turbulent-diffusion Mediated CO Depletion in Weakly Turbulent Protoplanetary Disks}. \apj 835:162

\bibitem[{{Yan} and {Lazarian}(2002)}]{YanLazarian2002}
{Yan} H {Lazarian} A (2002) {Scattering of Cosmic Rays by Magnetohydrodynamic Interstellar Turbulence}. Physical Review Letters 89:281102

\bibitem[{{Youdin} and {Goodman}(2005)}]{YoudinGoodman2005}
{Youdin} AN {Goodman} J (2005) {Streaming Instabilities in Protoplanetary Disks}. \apj 620:459--469

\bibitem[{{Youdin} and {Shu}(2002)}]{YoudinShu2002}
{Youdin} AN {Shu} FH (2002) {Planetesimal Formation by Gravitational Instability}. \apj 580:494--505

\bibitem[{{Zapolsky} and {Salpeter}(1969)}]{ZapolskySalpeter1969}
{Zapolsky} HS {Salpeter} EE (1969) {The Mass-Radius Relation for Cold Spheres of Low Mass}. \apj 158:809

\bibitem[{{Zhang} et~al.(2018){Zhang}, {Zhu}, {Huang}, {Guzm{\'a}n}, {Andrews}, {Birnstiel}, {Dullemond}, {Carpenter}, {Isella}, {P{\'e}rez}, {Benisty}, {Wilner}, {Baruteau}, {Bai}, and {Ricci}}]{Zhang_S2018}
{Zhang} S, {Zhu} Z, {Huang} J et~al. (2018) {The Disk Substructures at High Angular Resolution Project (DSHARP). VII. The Planet-Disk Interactions Interpretation}. \apjl 869(2):L47

\bibitem[{{Zhao} et~al.(2016){Zhao}, {Caselli}, {Li}, {Krasnopolsky}, {Shang}, and {Nakamura}}]{Zhao2016}
{Zhao} B, {Caselli} P, {Li} ZY et~al. (2016) {Protostellar disc formation enabled by removal of small dust grains}. \mnras 460(2):2050--2076

\bibitem[{{Zhao} et~al.(2018){Zhao}, {Caselli}, {Li}, and {Krasnopolsky}}]{Zhao2018}
{Zhao} B, {Caselli} P, {Li} ZY {Krasnopolsky} R (2018) {Decoupling of magnetic fields in collapsing protostellar envelopes and disc formation and fragmentation}. \mnras 473(4):4868--4889

\bibitem[{{Zhu} et~al.(2012){Zhu}, {Nelson}, {Dong}, {Espaillat}, and {Hartmann}}]{Zhu_Nelson2012}
{Zhu} Z, {Nelson} RP, {Dong} R, {Espaillat} C {Hartmann} L (2012) {Dust Filtration by Planet-induced Gap Edges: Implications for Transitional Disks}. \apj 755(1):6

\end{thebibliography}

\end{document}